\documentstyle[a4p,epsfig,12pt,rotating,multicol,cite]{article}
\begin{document}
%=======================================================================
%       Parameters for the title page
%=======================================================================
%
%  PR number, Date and Version
%
\newcommand{\PRnum}    {CERN-EP/2003-040}
\newcommand{\PNnum}     {OPAL Physics Note PN-406}
\newcommand{\TNnum}     {OPAL Technical Note TN-xxx}
\newcommand{\Date}      {\today}
\newcommand{\Authors}    {}
\newcommand{\MailAddr}  {gwwilson@ku.edu}
\newcommand{\EdBoard}   {}
\newcommand{\DraftVer}  {Final Draft}
\newcommand{\DraftDate} {\today}
\newcommand{\TimeLimit} {}

%=======================================================================
%       Parameters affecting the general appearance
%=======================================================================
\def\toprule{\noalign{\hrule \medskip}}
\def\midrule{\noalign{\medskip\hrule }}
\def\botrule{\noalign{\medskip\hrule }}
\setlength{\parskip}{\medskipamount}

%=======================================================================
%       Define symbols
%=======================================================================

\newcommand{\ee}{{\mathrm e}^+ {\mathrm e}^-}
\newcommand{\sq}{\tilde{\mathrm q}}
\newcommand{\seff}{\tilde{\mathrm f}}
\newcommand{\sele}{\tilde{\mathrm e}}
\newcommand{\sell}{\tilde{\ell}}
\newcommand{\snu}{\tilde{\nu}}
\newcommand{\smu}{\tilde{\mu}}
\newcommand{\stau}{\tilde{\tau}}
%*** \newcommand{\ch}{\tilde{\chi}}
\newcommand{\chp}{\tilde{\chi}^+_1}
%*** \newcommand{\chm}{\tilde{\chi}^-}
\newcommand{\chpm}{\tilde{\chi}^\pm_1}
\newcommand{\nt}{\tilde{\chi}^0}
\newcommand{\qq}{{\mathrm q}\bar{\mathrm q}}
\newcommand{\sleppair}{\sell^+ \sell^-}
\newcommand{\nunu}{\nu \bar{\nu}}
\newcommand{\mumu}{\mu^+ \mu^-}
\newcommand{\tautau}{\tau^+ \tau^-}
\newcommand{\ellell}{\ell^+ \ell^-}
\newcommand{\nulqq}{\nu \ell {\mathrm q} \bar{\mathrm q}'}
\newcommand{\MZ}{M_{\mathrm Z}}

\newcommand {\stopm}         {\tilde{\mathrm{t}}_{1}}
\newcommand {\stops}         {\tilde{\mathrm{t}}_{2}}
\newcommand {\stopbar}       {\bar{\tilde{\mathrm{t}}}_{1}}
\newcommand {\stopx}         {\tilde{\mathrm{t}}}
\newcommand {\sneutrino}     {\tilde{\nu}}
\newcommand {\slepton}       {\tilde{\ell}}
\newcommand {\stopl}         {\tilde{\mathrm{t}}_{\mathrm L}}
\newcommand {\stopr}         {\tilde{\mathrm{t}}_{\mathrm R}}
\newcommand {\stoppair}      {\tilde{\mathrm{t}}_{1}
\bar{\tilde{\mathrm{t}}}_{1}}
\newcommand {\gluino}        {\tilde{\mathrm g}}

\newcommand {\neutralino}    {\tilde{\chi }^{0}_{1}}
\newcommand {\neutrala}      {\tilde{\chi }^{0}_{2}}
\newcommand {\neutralb}      {\tilde{\chi }^{0}_{3}}
\newcommand {\neutralc}      {\tilde{\chi }^{0}_{4}}
\newcommand {\bino}          {\tilde{\mathrm B}^{0}}
\newcommand {\wino}          {\tilde{\mathrm W}^{0}}
\newcommand {\higginoa}      {\tilde{\rm H_{1}}^{0}}
\newcommand {\higginob}      {\tilde{\mathrm H_{1}}^{0}}
\newcommand {\chargino}      {\tilde{\chi }^{\pm}_{1}}
\newcommand {\charginop}     {\tilde{\chi }^{+}_{1}}
\newcommand {\KK}            {{\mathrm K}^{0}-\bar{\mathrm K}^{0}}
\newcommand {\ff}            {{\mathrm f} \bar{\mathrm f}}
\newcommand {\bstopm} {\mbox{$\boldmath {\tilde{\mathrm{t}}_{1}} $}}
\newcommand {\Mt}            {M_{\mathrm t}}
\newcommand {\mscalar}       {m_{0}}
\newcommand {\Mgaugino}      {M_{1/2}}
\newcommand {\rs}            {\sqrt{s}}
\newcommand {\WW}            {{\mathrm W}^+{\mathrm W}^-}
%*** \newcommand {\eetautau}      {\ee-\rightarrow {\tau^+}{\tau^-}}
\newcommand {\MGUT}          {M_{\mathrm {GUT}}}
\newcommand {\Zboson}        {${\mathrm Z}^{0}$}
\newcommand {\Wpm}           {{\mathrm W}^{\pm}}
\newcommand {\allqq}         {\sum_{q \neq t} q \bar{q}}
\newcommand {\mixang}        {\theta _{\mathrm {mix}}}
\newcommand {\thacop}        {\theta _{\mathrm {Acop}}}
\newcommand {\cosjet}        {\cos\thejet}
\newcommand {\costhr}        {\cos\thethr}
\newcommand {\djoin}         {d_{\mathrm{join}}}
\newcommand {\mstop}         {m_{\stopm}}
\newcommand {\msell}         {m_{\sell}}
\newcommand {\mchi}          {m_{\neutralino}}
\newcommand {\pp}{p \bar{p}}

\newcommand{\dWW}{\mbox{\wpair$\rightarrow$ \llnunu}}

\newcommand{\epair}{\mbox{${\mathrm e}^+{\mathrm e}^-$}}
\newcommand{\mupair}{\mbox{$\mu^+\mu^-$}}
\newcommand{\taupair}{\mbox{$\tau^+\tau^-$}}
\newcommand{\qpair}{\mbox{${\mathrm q}\overline{\mathrm q}$}}
\newcommand{\eeee}{\mbox{\epair\epair}}
\newcommand{\eemumu}{\mbox{\epair\mupair}}
\newcommand{\eetautau}{\mbox{\epair\taupair}}
\newcommand{\eeqq}{\mbox{\epair\qpair}}
\newcommand{\llqq}{\mbox{$\ell\ell$\qpair}}
\newcommand{\fs}{ final states}
\newcommand{\epairf}{\mbox{\epair\fs}}
\newcommand{\mupairf}{\mbox{\mupair\fs}}
\newcommand{\taupairf}{\mbox{\taupair\fs}}
\newcommand{\qpairf}{\mbox{\qpair\fs}}
\newcommand{\eeeef}{\mbox{\eeee\fs}}
\newcommand{\eemumuf}{\mbox{\eemumu\fs}}
\newcommand{\eetautauf}{\mbox{\eetautau\fs}}
\newcommand{\eeqqf}{\mbox{\eeqq\fs}}
\newcommand{\ffff}{four fermion final states}
\newcommand{\llnunu}{\mbox{$\ell^+\nu\,\ell^-\nbar$}}
\newcommand{\lnuqq}{\mbox{\lept\nubar\qpair}}
\newcommand{\zee}{\mbox{Zee}}
\newcommand{\zzg}{\mbox{ZZ/Z$\gamma$}}
\newcommand{\wenu}{\mbox{We$\nu$}}

\newcommand{\el}{\mbox{${\mathrm e}^-$}}
\newcommand{\selem}{\mbox{$\tilde{\mathrm e}^-$}}
\newcommand{\smum}{\mbox{$\tilde\mu^-$}}
\newcommand{\staum}{\mbox{$\tilde\tau^-$}}
\newcommand{\slept}{\mbox{$\tilde{\ell}^\pm$}}
\newcommand{\sleptm}{\mbox{$\tilde{\ell}^-$}}
\newcommand{\lept}{\mbox{$\ell^-$}}
\newcommand{\Hl}{\mbox{$\mathrm{L}^\pm$}}
\newcommand{\Hm}{\mbox{$\mathrm{L}^-$}}
\newcommand{\Hnu}{\mbox{$\nu_{\mathrm{L}}$}}
\newcommand{\nul}{\mbox{$\nu_\ell$}}
\newcommand{\nubar}{\mbox{$\overline{\nu}_\ell$}}
\newcommand{\nbar}{\mbox{$\overline{\nu}$}}
\newcommand{\spair}{\mbox{$\tilde{\ell}^+\tilde{\ell}^-$}}
\newcommand{\lpair}{\mbox{$\ell^+\ell^-$}}
\newcommand{\staupair}{\mbox{$\tilde{\tau}^+\tilde{\tau}^-$}}
\newcommand{\smupair}{\mbox{$\tilde{\mu}^+\tilde{\mu}^-$}}
\newcommand{\selepair}{\mbox{$\tilde{\mathrm e}^+\tilde{\mathrm e}^-$}}
\newcommand{\ch}{\mbox{$\tilde{\chi}^\pm_1$}}
\newcommand{\chpair}{\mbox{$\tilde{\chi}^+_1\tilde{\chi}^-_1$}}
\newcommand{\chm}{\mbox{$\tilde{\chi}^-_1$}}
\newcommand{\chmp}{\mbox{$\tilde{\chi}^\pm_1$}}
\newcommand{\chz}{\mbox{$\tilde{\chi}^0_1$}}
\newcommand{\dch}{\mbox{\chm$\rightarrow$\chz\lept\nubar}}
\newcommand{\dslept}{\mbox{\sleptm$\rightarrow$\lept\chz}}
\newcommand{\dH}{\mbox{$\mathrm{H}^{\pm} \rightarrow \tau^\pm \nu_\tau$}}
\newcommand{\mch}{\mbox{$m_{\tilde{\chi}^\pm_1}$}}
\newcommand{\mslept}{\mbox{$m_{\tilde{\ell}}$}}
\newcommand{\mstau}{\mbox{$m_{\staum}$}}
\newcommand{\msmu}{\mbox{$m_{\smum}$}}
\newcommand{\msele}{\mbox{$m_{\selem}$}}
\newcommand{\mchz}{\mbox{$m_{\tilde{\chi}^0_1}$}}
\newcommand{\dm}{\mbox{$\Delta m$}}
\newcommand{\dmch}{\mbox{$\Delta m_{\ch-\chz}$}}
\newcommand{\dmslept}{\mbox{$\Delta m_{\slept-\chz}$}}
\newcommand{\dmhl}{\mbox{$\Delta m_{\Hl-\Hnu}$}}
\newcommand{\w}{\mbox{W$^\pm$}}
\newcommand{\dchtwo}{\mbox{$\chpm \rightarrow \ell^\pm {\tilde{\nu}_\ell}$}}
\newcommand{\dchthree}{\mbox{$\chmp\rightarrow{\mathrm W}^\pm\chz\rightarrow\ell^\pm\nu\chz$}}
\newcommand{\chargthree}{\chpair (3-body decays)}
\newcommand{\chargtwo}{\chpair (2-body decays)}

% my variables
\newcommand{\acopc}{\mbox{$\phi^{\mathrm{acop}}$}}
\newcommand{\acolc}{\mbox{$\theta^{\mathrm{acol}}$}}
\newcommand{\acop}{\mbox{$\phi_{\mathrm{acop}}$}}
\newcommand{\acol}{\mbox{$\theta_{\mathrm{acol}}$}}
%*** \newcommand{\pt}{\mbox{$p_{t}$}}
\newcommand{\gwwpt}{\mbox{$p_{\mathrm{t}}$}}
\newcommand{\pz}{\mbox{$p_{\mathrm{z}}^{\mathrm{miss}}$}}
\newcommand{\ptevt}{\mbox{$p_{\mathrm{t}}^{\mathrm{miss}}$}}
\newcommand{\ptaxic}{\mbox{$a_{\mathrm{t}}^{\mathrm{miss}}$}}
\newcommand{\stevt}{\mbox{$p_{\mathrm{t}}^{\mathrm{miss}}$/\Ebeam}}
\newcommand{\staxic}{\mbox{$a_{\mathrm{t}}^{\mathrm{miss}}$/\Ebeam}}
\newcommand{\dptaxic}{\mbox{missing $p_{t}$ wrt. event axis \ptaxic}}
\newcommand{\cosevt}{\mbox{$\mid\cos\theta_{\mathrm{p}}^{\mathrm{miss}}\mid$}}
\newcommand{\axicos}{\mbox{$\mid\cos\theta_{\mathrm{a}}^{\mathrm{miss}}\mid$}}
\newcommand{\pthet}{\mbox{$\theta_{\mathrm{p}}^{\mathrm{miss}}$}}
\newcommand{\athet}{\mbox{$\theta_{\mathrm{a}}^{\mathrm{miss}}$}}
\newcommand{\dcosevt}{\mbox{$\mid\cos\theta\mid$ of missing p$_{t}$}}
\newcommand{\daxicos}{\mbox{$\mid\cos\theta\mid$ of missing p$_{t}$ wrt. event
axis}}
\newcommand{\efdsw}{\mbox{$x_{\mathrm{FDSW}}$}}
\newcommand{\acopf}{\mbox{$\Delta\phi_{\mathrm{FDSW}}$}}
\newcommand{\acopm}{\mbox{$\Delta\phi_{\mathrm{MUON}}$}}
\newcommand{\acopt}{\mbox{$\Delta\phi_{\mathrm{trk}}$}}
\newcommand{\po}{\mbox{$E_{\mathrm{isol}}^\gamma$}}
\newcommand{\qprod}{\mbox{$q1$$*$$q2$}}
\newcommand{\lcode}{lepton identification code}
\newcommand{\nctro}{\mbox{$N_{\mathrm{trk}}^{\mathrm{out}}$}}
\newcommand{\necao}{\mbox{$N_{\mathrm{ecal}}^{\mathrm{out}}$}}
\newcommand{\mout}{\mbox{$m^{\mathrm{out}}$}}
\newcommand{\nctec}{\mbox{\nctro$+$\necao}}
\newcommand{\gfract}{\mbox{$F_{\mathrm{good}}$}}
\newcommand{\zz}       {\mbox{$|z_0|$}}
\newcommand{\dz}       {\mbox{$|d_0|$}}
\newcommand{\sint}      {\mbox{$\sin\theta$}}
\newcommand{\cost}      {\mbox{$\cos\theta$}}
\newcommand{\mcost}     {\mbox{$|\cos\theta|$}}
\newcommand{\dedx}     {\mbox{$dE/dx$}}
\newcommand{\wdedx}     {\mbox{$W_{dE/dx}$}}
\newcommand{\xe}     {\mbox{$x_E$}}

% Terry's new defs
\newcommand{\mH}{\mbox{$m_{\mathrm{H}^+}$}}
\newcommand{\p}     {\mbox{$\pm$}}
\newcommand{\ssix}     {\mbox{$\protect\sqrt{s}$~=~161~GeV}}
\newcommand{\sseven}     {\mbox{$\protect\sqrt{s}$~=~172~GeV}}
\newcommand{\seight}     {\mbox{$\protect\sqrt{s}$~=~183~GeV}}
\newcommand{\snine}      {\mbox{$\protect\sqrt{s}$~=~189~GeV}}
\newcommand{\sten}      {\mbox{$\protect\sqrt{s}$~=~208~GeV}}
\newcommand{\sthree}     {\mbox{$\protect\sqrt{s}$~=~130--136~GeV}}
\newcommand{\mrecoil}     {\mbox{$m_{\mathrm{recoil}}$}}
\newcommand{\llmass}     {\mbox{$m_{ll}$}}
\newcommand{\sml}{\mbox{Standard Model \llnunu\ events}}
\newcommand{\sme}{\mbox{Standard Model events}}
\newcommand{\sig}{events containing a lepton pair plus missing transverse momentum}
\newcommand{\wpair}{\mbox{$\mathrm{W}^+\mathrm{W}^-$}}
\newcommand{\dW}{\mbox{W$^-\rightarrow$\lept\nubar}}
\newcommand{\dsele}{\mbox{\selem$\rightarrow$ e$^-$\chz}}
\newcommand{\eeeell}{\mbox{\epair$\rightarrow$\epair\lpair}}
\newcommand{\eell}{\mbox{\epair\lpair}}
\newcommand{\llgam}{\mbox{$\ell^+\ell^-(\gamma)$}}
\newcommand{\nunugam}{\mbox{$\nu\bar{\nu}\gamma\gamma$}}
\newcommand{\nngbra}{\mbox{$\nu\bar{\nu}\gamma(\gamma)$}}
\newcommand{\acope}{\mbox{$\Delta\phi_{\mathrm{EE}}$}}
\newcommand{\nee}{\mbox{N$_{\mathrm{EE}}$}}
\newcommand{\eesum}{\mbox{$\Sigma_{\mathrm{EE}}$}}
\newcommand{\at}{\mbox{$a_{t}$}}
\newcommand{\spp}{\mbox{$p$/\Ebeam}}
\newcommand{\acoph}{\mbox{$\Delta\phi_{\mathrm{HCAL}}$}}
% Graham's definitions
\newcommand{\ACOP}{\mbox{$\phi_{\mathrm{acop}}$}}
\newcommand{\XT}{\mbox{$x_T$}}
\newcommand{\XONE}{\mbox{$x_1$}}
\newcommand{\XTWO}{\mbox{$x_2$}}
\newcommand{\MLL}{\mbox{$m_{\ell\ell}$}}
\newcommand{\MRECOIL}{\mbox{$m_{\mathrm{recoil}}$}}
\newcommand {\mm}       {\mu^+ \mu^-}
\newcommand {\emu}         {\mathrm{e}^{\pm} \mu^{\mp}}
\newcommand {\et}         {\mathrm{e}^{\pm} \tau^{\mp}}
\newcommand {\mt}         {\mu^{\pm} \tau^{\mp}}
\newcommand {\lemu}       {\ell=\mathrm{e},\mu}
\newcommand{\Zz}{\mbox{${\mathrm{Z}^0}$}}

% journals
\newcommand{\ZP}[3]    {Z. Phys. {\bf C#1} (#2) #3.}
\newcommand{\PL}[3]    {Phys. Lett. {\bf B#1} (#2) #3.}
\newcommand{\etal}     {{\it et al}.}

% miscellaneous
\newcommand{\Ecm}{\mbox{$E_{\mathrm{cm}}$}}
\newcommand{\Ebeam}{\mbox{$E_{\mathrm{beam}}$}}
\newcommand{\ipb}{\mbox{pb$^{-1}$}}
\newcommand{\wrt}{with respect to}
\newcommand{\sm}{Standard Model}
\newcommand{\smb}{Standard Model background}
\newcommand{\smp}{Standard Model processes}
\newcommand{\smc}{Standard Model Monte Carlo}
\newcommand{\mc}{Monte Carlo}
\newcommand{\btb}{back-to-back}
\newcommand{\tp}{two-photon}
\newcommand{\tpb}{two-photon background}
\newcommand{\tpp}{two-photon processes}
\newcommand{\lp}{lepton pairs}
\newcommand{\vto}{\mbox{$\tau$ veto}}

% Def. fuer groesser-ungefaehr:
\newcommand{\gsim}{\;\raisebox{-0.9ex}
           {$\textstyle\stackrel{\textstyle >}{\sim}$}\;}
% Def. fuer kleiner-ungefaehr:
\newcommand{\lsim}{\;\raisebox{-0.9ex}{$\textstyle\stackrel{\textstyle<}
           {\sim}$}\;}

\newcommand{\degree}    {^\circ}
%
%-----------------------------------------
%  Variables for machine; math mode only
%-----------------------------------------
%*** \newcommand{\Ecm}       {E_{\mathrm{cm}}}
%*** \newcommand{\Ebeam}     {E_{\mathrm{b}}}
\newcommand{\roots}     {\protect\sqrt{s}}
%----------------------------------------
%  Variables for events; math mode only
%----------------------------------------
%
%     Thrust
%
\newcommand{\thrust}    {T}
\newcommand{\nthrust}   {\hat{n}_{\mathrm{thrust}}}
\newcommand{\thethr}    {\theta_{\,\mathrm{thrust}}}
\newcommand{\phithr}    {\phi_{\mathrm{thrust}}}
\newcommand{\acosthr}   {|\cos\thethr|}
\newcommand{\thejet}    {\theta_{\,\mathrm{jet}}}
\newcommand{\acosjet}   {|\cos\thejet|}
\newcommand{\thmiss}    { \theta_{miss} }
\newcommand{\cosmiss}   {| \cos \thmiss |}
%
%     Energy, etc.
%
\newcommand{\Evis}      {E_{\mathrm{vis}}}
\newcommand{\Rvis}      {E_{\mathrm{vis}}\,/\roots}
\newcommand{\Mvis}      {M_{\mathrm{vis}}}
\newcommand{\Rbal}      {R_{\mathrm{bal}}}
%\newcommand{\pt}       {p_{\mathrm{t}}}
%
%     Acoplanarity
%
\newcommand{\phiacop}   {\phi_{\mathrm{acop}}}
%
%  David's new variables
%-------------------------
%
\newcommand{\LS}      {\mbox{$L_{\mathrm{S}}$}}
\newcommand{\LB}      {\mbox{$L_{\mathrm{B}}$}}
\newcommand{\LR}      {\mbox{$L_{\mathrm{R}}$}}
\newcommand{\PS}      {\mbox{$P_S(x_i)$}}
\newcommand{\PB}      {\mbox{$P_B(x_i)$}}
\newcommand{\signine}   {\mbox{$\sigma_{95}$}}
\newcommand{\expsig}   {\mbox{$\langle\signine\rangle$}}
\newcommand{\Lsigs}      {\mbox{$L_i(\sigma_\mathrm{{s}})$}}
\newcommand{\Lsigsn}      {\mbox{$L_i(\sigma_\mathrm{{s}}^{189})$}}
\newcommand{\sigs}     {\mbox{$\sigma_{\mathrm{{s}}}$}}
\newcommand{\lri}      {L_{R_i}}
\newcommand{\mf}     {\mbox{$(m-15)/2$}}
\newcommand{\dstau}{\mbox{$\staum\rightarrow\tau^-\chz$}}
%
%---------
%  Units
%---------
%----------------------------
%  Bibliographic references
%----------------------------
%
%     Journal names
%
\newcommand{\PhysLett}  {Phys.~Lett.}
\newcommand{\PRL} {Phys.~Rev.\ Lett.}
\newcommand{\PhysRep}   {Phys.~Rep.}
\newcommand{\PhysRev}   {Phys.~Rev.}
\newcommand{\NPhys}  {Nucl.~Phys.}
\newcommand{\NIM} {Nucl.~Instr.\ Meth.}
\newcommand{\CPC} {Comp.~Phys.\ Comm.}
\newcommand{\ZPhys}  {Z.~Phys.}
\newcommand{\IEEENS} {IEEE Trans.\ Nucl.~Sci.}
%
%     Collaboration names
%
\newcommand{\OPALColl}  {OPAL Collab.}
\newcommand{\JADEColl}  {JADE Collab.}
%
%*** \newcommand{\etal}      {{\it et~al.}}
%-------
%  etc
%-------
\newcommand{\onecol}[2] {\multicolumn{1}{#1}{#2}}
\newcommand{\ra}        {\rightarrow}   % \to can be used as well

%=======================================================================
%       Title Page
%=======================================================================

%-----------------------------------------------------------------------
\begin{titlepage}
%
%     Header
%
\begin{center}{\large EUROPEAN ORGANIZATION FOR NUCLEAR RESEARCH
}\end{center}\bigskip
\begin{flushright}
    \PRnum\\
    14$^{\mathrm{th}}$ July 2003
\end{flushright}
\bigskip
%
%     Main title
%
\begin{center}
    \LARGE\bf\boldmath
    Search for Anomalous Production of
    Di-lepton Events with Missing Transverse Momentum 
    in \epair\ Collisions
    at $\sqrt{s} = $ 183--209 GeV
\end{center}
\bigskip
%
%     Author names
%
\begin{center}
 \Large
The OPAL Collaboration
\bigskip
\bigskip
\end{center}
%
%     Abstract
%
\begin{abstract}%=======================================================

In total 1317 
di-lepton events with 
significant missing transverse momentum 
were identified
in a total data sample of 680~pb$^{-1}$ 
collected at \epair\ centre-of-mass energies ranging from 
183~GeV to 209~GeV.
%In total 1317 events were identified.
The number of di-lepton events, 
the dependence on centre-of-mass energy, and the event properties
are consistent with expectations from Standard Model processes, 
predominantly \wpair\ production with both W bosons 
decaying leptonically. 
This topology is also an experimental
signature for the pair production of new particles that decay
to a charged lepton accompanied by one or more
invisible particles.
%Discrimination techniques designed to optimise 
%the sensitivity to particular new physics channels
%are applied to the data.
No evidence for new phenomena is apparent.
Upper limits are presented on 
the production cross-section multiplied by the relevant 
branching ratio squared
for sleptons, leptonically decaying charginos and 
charged Higgs bosons. 
Mass limits are also given.
%in a manner intended to minimise the number of 
%model assumptions for the slepton, leptonically decaying charginos and 
%charged Higgs bosons searches.
%%%%% The following was in the original abstract
%Assuming a 100\% branching ratio for the decay
%$\sell^\pm_R \rightarrow  {\ell^\pm} \nt_1$, where $\nt_1$ is the
%lightest neutralino, we exclude at 95\% confidence level (CL) 
%right-handed smuons with masses below 94.0~GeV for 
%\mbox{$\msmu - \mchz > 4$}~GeV and
%right-handed staus with masses below 89.8~GeV for 
%\mbox{$\mstau - \mchz > 8$}~GeV.
%Right-handed selectrons are excluded at 95\% CL for 
%masses below 97.5~GeV for \mbox{$\msele - \mchz > 11$}~GeV,
%within the framework of the
%Minimal Supersymmetric Standard Model assuming
%$\mu < -100$~GeV and $\tan{\beta} = 1.5$.
%Charged Higgs bosons, H$^{\pm}$, are excluded at 95\% CL for masses
%below 92.0~GeV, assuming a 100\% branching ratio for the decay \dH .

\end{abstract}%=========================================================
 \bigskip
\bigskip\bigskip
\begin{center}
{\large (Submitted to Eur.~Phys.~J.~C.)}
\end{center}
\bigskip
\begin{center}
%{Authors: \Authors}
\end{center}
\begin{center}
%{Editorial Board: \EdBoard}
\end{center}
\begin{center}
{\bf \TimeLimit}
\end{center}
\end{titlepage}

\begin{center}{\Large        The OPAL Collaboration
}\end{center}\bigskip
\begin{center}{
%begin authorlist PLEASE DO NOT DELETE THIS COMMENT
G.\thinspace Abbiendi$^{  2}$,
C.\thinspace Ainsley$^{  5}$,
P.F.\thinspace {\AA}kesson$^{  3}$,
G.\thinspace Alexander$^{ 22}$,
J.\thinspace Allison$^{ 16}$,
P.\thinspace Amaral$^{  9}$, 
G.\thinspace Anagnostou$^{  1}$,
K.J.\thinspace Anderson$^{  9}$,
S.\thinspace Arcelli$^{  2}$,
S.\thinspace Asai$^{ 23}$
% ...
%end authorlist PLEASE DO NOT DELETE THIS COMMENT
}\end{center}\bigskip
\bigskip
%begin institutes
$^{  1}$School of Physics and Astronomy, University of Birmingham,
Birmingham B15 2TT, UK
\newline
$^{  2}$Dipartimento di Fisica dell' Universit\`a di Bologna and INFN,
I-40126 Bologna, Italy
\newline
$^{  3}$Physikalisches Institut, Universit\"at Bonn,
D-53115 Bonn, Germany
\newline
% ...
$^{  5}$Cavendish Laboratory, Cambridge CB3 0HE, UK
\newline
% ...
$^{  9}$Enrico Fermi Institute and Department of Physics,
University of Chicago, Chicago IL 60637, USA
\newline
% ...
$^{ 16}$Department of Physics, Schuster Laboratory, The University,
Manchester M13 9PL, UK
\newline
% ...
$^{ 22}$Department of Physics and Astronomy, Tel Aviv University,
Tel Aviv 69978, Israel
\newline
$^{ 23}$International Centre for Elementary Particle Physics and
Department of Physics, University of Tokyo, Tokyo 113-0033, and
Kobe University, Kobe 657-8501, Japan
\newline
% ...
%end notes

%=======================================================================
\section{Introduction}
%=======================================================================
\label{sec:intro}

We report the final results from an investigation of
events containing two oppositely
charged leptons and significant missing transverse momentum 
recorded with the OPAL detector at LEP.
%
%produced in the highest energy 
%e$^+$e$^-$ collisions
%achieved to date. 
These events were produced between 
% analysed data were collected from 
1997 and 2000 
in the highest energy 
e$^+$e$^-$ collisions achieved to date, 
at 
centre-of-mass energies ($\sqrt{s}$) ranging from 183 to 209~GeV.
The data used in this analysis amount to a 
total integrated luminosity 
of 680 \ipb .
The event topology which is studied, henceforth called
acoplanar di-lepton events, consists of low multiplicity events 
with two oppositely charged leptons, significant missing transverse 
momentum and the possible presence of additional photons.

The number of observed events and their 
properties are compared with the expectations for  \smp , which 
are dominated by 
the \llnunu\ final state ($\ell$ = e, $\mu, \tau$) arising from \wpair\
production in which both
W bosons decay leptonically.
%: $\dW$.
This topology is also an experimental
signature for the pair production of new particles that decay
to produce a charged lepton accompanied by one or more
invisible particles. 
The invisible particles may be neutrinos, or 
the lightest stable supersymmetric~\cite{SUSY} particle (LSP), 
or weakly interacting neutral
particles with long lifetimes, which decay outside the
detector volume.
The LSP may be the lightest neutralino, $\nt_1$, 
the lightest sneutrino, $\tilde{\nu}$, or the gravitino, $\tilde{G}$.
We present the results of 
searches for the pair production and stated decay mode of the following
new particles:
\begin{description}
\item[charged scalar leptons (sleptons):]
$\sell^\pm \rightarrow  {\ell^\pm} 
\nt_1$ (or $\sell^\pm \rightarrow  {\ell^\pm} \tilde{G}$),
where $\sell^\pm$ may be a selectron ($\sele$), smuon ($\smu$) or stau
($\stau$) and $\ell^\pm$ is the corresponding charged lepton.
\item[charged Higgs:] $\mathrm{H}^{\pm} \rightarrow \tau^\pm \nu_\tau$.
\item[charginos:] $\chpm \rightarrow \ell^\pm \snu$ (``2-body'' decays)
\ \   or \ \ 
$\chpm \rightarrow \ell^\pm \nu \chz$ (``3-body'' decays).
\end{description}

The slepton searches are also relevant to
interpreting the results of searches for chargino and neutralino
production since the chargino and neutralino production 
cross-sections and branching 
ratios depend on the 
sneutrino and charged slepton masses.
The search for
charged sleptons provides constraints on the
slepton masses, notably the    
selectron mass, while indirect limits on the sneutrino masses, notably the
electron-sneutrino mass, can be obtained in models where the
charged slepton and sneutrino masses
are related. The search results described here will also be used in
a separate publication regarding the search for sleptons with
non-negligible lifetime.

In most respects the analysis is similar to our published searches
at centre-of-mass 
energies of 161, 172, 183 and 189~GeV~\cite{paper172,paper183,paper189}. 
This paper supersedes the results of~\cite{paper183,paper189}.
The analysis is performed in two stages.
The first stage consists of a general event selection designed 
for all possible low multiplicity 
\sig\ (Section~\ref{sec:gen}).
In this context the \sm\ \llnunu\ events are considered as signal
in addition to possible new physics sources.
All \smp\ that do not lead to \llnunu\ final states, e.g. \eell\ and
\llgam\ , are considered as background and are reduced to a
rather low level ($\approx 3$\%) by the event selection.
The observed numbers of events and kinematic distributions for the data 
are compared with expectations from \smp\ in Section~\ref{sec:compare}.
In the second stage the detailed properties of
the events (e.g. the type of leptons observed and their momenta), which 
vary greatly depending on the type of new particles considered and on
free parameters within the models, are used to separate as far as  possible 
the events consistent with
potential new physics sources from  \wpair\ production and other \smp\
(Section~\ref{sec:search}). 
Constraints on new physics are discussed in Section~\ref{sec:results}.

Slepton search results from the ALEPH collaboration 
at $\roots\leq$209 GeV~\cite{ALEPH208} have been published recently.
The L3 and DELPHI collaborations have published searches for sleptons in 
this channel using data with $\roots\leq$189 GeV~\cite{olep189}.

%Searches for sleptons at LEP2 using this topology have been presented
%also by other collaborations~\cite{olep}, though at the time of writing, 
%this is the only publication to include an analysis of the substantial
%data-set taken at 189~GeV and therefore gives the most stringent
%results to date. 
%=======================================================================
\section{OPAL Detector and Monte Carlo Simulation}
%=======================================================================
\label{opaldet}

A detailed description of the  OPAL detector can be found 
elsewhere~\cite{ref:OPAL-detector,tenim} and only the
general features are described here.

The central detector consisted of a system of chambers providing
charged particle tracking over 96\% of the full solid angle
inside a 0.435~T uniform magnetic field parallel to the beam axis. 
It consisted of a two-layer
silicon micro-strip vertex detector, a high precision vertex 
drift chamber,
a large volume jet chamber and a set of $z$-chambers  that measured 
the track coordinates along the beam direction. 

A lead-glass electromagnetic
calorimeter (ECAL) was located outside the magnet coil.  It provided, 
in combination with the gamma-catchers (GC) and forward detectors (FD), 
which were lead-scintillator 
sandwich calorimeters and, at smaller angles,
silicon tungsten calorimeters (SW), geometrical acceptance with excellent
hermeticity down to approximately 25~mrad\footnote{For some polar angles, precision energy measurements were compromised 
by upstream material such as the synchrotron shield installed for LEP2 at
approximately 30 mrad. However, the ability to veto significant energy 
deposits with very high efficiency extended down to polar angles of 25 mrad.} 
from the beam direction.

The magnet return yoke was instrumented for hadron calorimetry 
and consisted of barrel and endcap sections along with pole tip detectors that
together covered the region $|\cos \theta |<0.99$.
Outside the hadron calorimeter (HCAL), four layers of muon chambers 
covered the polar angle range of $|\cos \theta |<0.98$. 
Arrays of thin scintillating tiles were installed in the
endcap region to improve trigger performance, time resolution and hermeticity
for operation at LEP~2~\cite{tenim}. 
Of particular relevance to this analysis are the four layers
of scintillating tiles (the MIP-plug)  installed at 
each end of the OPAL detector covering
the angular range $43 < \theta < 220$~mrad.
These tiles were commissioned in 1997 and became fully operational
for data taken from 1998 at $\sqrt{s} \geq 189$~GeV.
Time-of-flight (TOF) scintillators in the barrel region aided
cosmic ray rejection.

The following \smp\ are simulated.
Four-fermion production is simulated 
%using the grc4f~\cite{grc4f}
%generator at $\roots$=183~GeV, and 
using 
the {\sc Koralw}~\cite{koralw}
generator which uses the grc4f~\cite{grc4f}
matrix elements to calculate the four-fermion cross-sections including
interference effects and includes a detailed description of hard
radiation from initial and final state charged particles.
%4-fermion production is generated using {\sc Koralw}~\cite{koralw},
%grc4f~\cite{grc4f}, and {\sc Excalibur}~\cite{excalibur}.
%The {\sc Koralw} generator (which uses the matrix elements of grc4f)
%was not used in~\cite{paper172,paper183}, but is
%used for 4-fermion simulation at 189~GeV as it
%has a more thorough treatment of initial state radiation than grc4f.
Two-photon production of muon pairs and tau pairs
is generated with the BDK generator~\cite{BDK} 
and the 
program of Vermaseren~\cite{vermaseren} is used for the two-photon
production of electron pairs.
The grc4f event generator excluding multi-peripheral processes
is used for \eell and \eeqq .
The  
{\sc Phojet}~\cite{phojet} and {\sc Herwig}~\cite{herwig} 
event generators are used to 
study backgrounds from two-photon production of hadrons.
%Because of the large total cross-section for  \eeee and \eemumu\ ,
%soft cuts are applied at the generator level to preselect
%events that might possibly lead to background in the selection of
%\llnunu\ final states.
%No generator level cuts are applied to the \eetautau\ generation.
The production of lepton pairs is generated using 
{\sc Bhwide}~\cite{bhwide} and {\sc Teegg}~\cite{teegg} for $\ee(\gamma)$, 
 and {\sc KK2f}~\cite{KK2f} 
%and {\sc Koralz}~\cite{koralz} 
for $\mumu(\gamma)$ and $\tautau(\gamma)$.
%The production of quark pairs, \qpair(g), is generated using {\sc Pythia}~\cite{pythia}
The final state \nngbra\ is generated with
{\sc Nunugpv98}~\cite{NUNUGPV}.

Slepton pair production is generated using {\sc
  Susygen}~\cite{SUSYGEN}.
%Selectron samples are generated with $\mu$ = -200
%  GeV and $\tan{\beta}$ = 1.5.
Chargino pair production is generated using {\sc Dfgt}~\cite{DFGT} for
three-body decays, and  {\sc Susygen} for two-body decays.
Charged Higgs boson pair production  is generated using
{\sc Hzha}~\cite{HZHA}. 
%and {\sc Pythia}.

Backgrounds from the accelerator or 
cosmic-ray interactions can lead to additional hits, energy deposition and
even reconstructed tracks being superimposed on triggered data events. 
Such effects of detector occupancy are simulated for all Monte Carlo 
samples by
adding to the Monte Carlo events the hits, energy depositions 
and additional jets found in randomly triggered
beam-crossing data events (BXRSA trigger\cite{opal_trigger} with a constant 0.1 Hz rate) 
corresponding to the same centre-of-mass energy. 
Systematic checks showed that this constant rate model adequately described 
the average inefficiency during a LEP fill, despite the instantaneous 
luminosity decreasing typically by a factor of 2 or 3.

All \sm\ and new physics \mc\ samples are processed with a full simulation  
of the OPAL detector \cite{gopal} and subjected to the same
reconstruction and analysis programs as used for the OPAL data.

\section{General Event Selection}
\label{sec:gen}
\subsection{Introduction}
The general event selection is formed by 
requiring that an event 
be selected by either or both of two independent event selections, 
referred to here as selection I and selection II.  Selection I 
is designed to retain efficiency for events with low visible 
energy.  This is characteristic of slepton or chargino events  
where the mass difference, \dm\ , between the parent particle and 
the invisible daughter particle (e.g. \chz ) is small. 
Selection II 
is optimised to maximise the efficiency for \dWW\ events, 
while keeping other \sm\ background events to a minimum.

Both selections require evidence for two\footnote{This is not 
strictly true for the single lepton selection 
of selection I (Section~\ref{sec-1lept}).} 
charged leptons and an
invisible system carrying significant missing transverse momentum 
(\ptevt).  The maximum \ptevt\ which can be carried away by undetected
particles travelling close to the beam is set by the maximum angle to
the beam at which such a particle will not be detected.  This is
25~mrad -- the angle to which the silicon tungsten detector extends.  A
particle with energy of \Ebeam\ may thus carry away $\ptevt =
0.025 \: \Ebeam$ without being detected.

Some background processes containing secondary neutrinos (particularly
from tau decay) may have large values of \ptevt\ with the direction of
the missing momentum vector pointing away from the beam axis.  Such
events tend to be fairly coplanar, and the component of \ptevt\ which is
perpendicular to the event thrust axis in the transverse plane (called
\ptaxic) is much less sensitive than \ptevt\ to the presence of
neutrinos from tau decays or to poorly measured particles.  This can
be seen by considering electrons produced in tau decay.  Low momentum
electrons produced from this source can have a large angle relative to
the original tau direction, but their momentum transverse to the
original direction (and hence their contribution to \ptaxic) is
small. 
%In events with low acoplanarity (where the potential
%background is concentrated), cuts are applied to \ptaxic\ and the
%direction of the missing momentum is calculated using \ptaxic\ rather
%than \ptevt.  For events with high acoplanarity cuts on \ptaxic\ no
%longer discriminate well between signal and background, and the
%conventional cuts on \ptevt\ are used.

The event selection as it was first implemented 
was described in detail
in~\cite{paper172}.  
In~\cite{paper183} we made use of the improved
hermeticity for non-showering particles in the forward direction provided 
by the MIP-plug\footnote{
Note that the general event selection and the Monte Carlo simulation
samples used for the 183~GeV data-set are unchanged from the 183~GeV
publication. The poorer MIP-plug detector performance in 1997
would have necessitated a re-optimisation 
of the latest general event selection, and we judged such a re-analysis 
to be inappropriate.}.
Subsequent improvements have been made for the analysis of the data
taken at 189~GeV~\cite{paper189} by reducing the sensitivity to
mis-measurements of \ptevt\ and 
using the much improved background rejection capabilities of the MIP-plug
to increase substantially the efficiency of selection II.
For this paper, selection I has been consolidated and much refined.
In particular, a selection has been added for single lepton events
in which a single high transverse momentum charged lepton 
is observed in the central detector. These are usually 
events where the second charged lepton is produced at a polar angle
sufficiently close to the beam direction that it does not produce
a visible track in the central detector.
In the context of this paper, this ``single-lepton'' selection is used
to recuperate di-leptons where the second lepton was missed.
Various small modifications were made to selection II in order to
reduce sensitivity to poorly modelled backgrounds. An example of this
is using stricter track quality criteria to reduce the possibility of
tracks from overlaid beam-gas events affecting the measurement of
the event kinematics.

%A complete description of the current event selection 
%would be prohibitively long. 
%Instead we give below a reasonably detailed
%technical description which summarises the event selection.
The current event selection is summarised below.

\subsection{Selection I}
\label{sec-sele-terry}

Selection I 
%(see~\cite{terrytn} for a detailed description) 
is designed
to retain efficiency for events with very low
visible energy, but nevertheless significant \ptevt.  This is typical
of new physics signal events with small \dm.  The selection
requires evidence that a pair of leptons has been produced and of
significant missing transverse momentum.  Subsequent 
cuts reduce the probability
that the signature of missing transverse momentum is faked by events with
secondary neutrinos from tau decay or poorly measured particles.

At least one lepton in the event is required to be well identified and
to satisfy requirements on isolation and transverse momentum.  Much
looser requirements are made on the possible presence of a second
lepton in the event.

In order to be considered in the event selection, tracks in the central 
detector are required to satisfy $p_t > 0.1$~GeV. Clusters in the 
barrel ECAL are required to satisfy $E > 0.1$~GeV and clusters in the endcap 
ECAL are required to satisfy $E > 0.2$~GeV. 
Converting photons are identified and the tracks and clusters associated to 
the conversion are replaced by a single 4-vector representing the photon.

\subsubsection{Lepton Candidates}

The first stage is to look for lepton candidates in the event.
A track is identified as a lepton candidate if it has $p>1.5$~GeV and 
it is identified as an electron, muon or hadronic tau decay\footnote{
The lepton identification (ID) 
made at the event selection stage is used for the purpose
of the selection only.  A separate lepton ID is applied to selected
events and used by the search analysis (see 
section~\ref{sec:lept}).}.  
The electron ID is based on the ratio of ECAL
energy to track momentum ($E/p$), and d$E$/d$x$ information.  Muons are
identified using muon chamber or HCAL hits which match with a track in
the central detector, or from a high momentum track which matches with 
a low energy ECAL cluster.  To
identify a hadronic tau the following criteria are applied:
\begin{itemize}
\item Within a cone of half-opening angle 35$^\circ$ around 
      a track of $p_t > $1.5~GeV there are no more
      than three tracks in total.
\item The invariant mass of all tracks and clusters within the cone is
      less than the tau mass. This is calculated 
      assuming the pion mass for each track
      and correcting for double counting of tracks and clusters.
\end{itemize}

The lepton candidates are also required to be isolated.  There must be
no more than 2 additional tracks and no more than 2 additional 
clusters in an isolation cone defined
around the lepton candidate (half opening angle 20$^\circ$ for
electrons and muons, half opening angle 60$^\circ$ for hadronic tau decays),
and the energy sum of the tracks or of the clusters must be less than
2~GeV.

\subsubsection{Di-lepton Event Selection }

Firstly evidence for a pair of leptons is required:

\begin{itemize}
\item There must be at least one and no more than two isolated lepton
      candidates with $\gwwpt > 1.5$~GeV. 
\item If the event contains two isolated lepton candidates then
      all tracks in the event must be associated with the lepton candidates.
\item If there is only one isolated lepton candidate then the other
      tracks and clusters in the event are considered as a possible
      second lepton candidate provided the following:
      \begin{itemize}
	\item[(a)] There must be between 1 and 3 additional tracks, at
                   least one of which must have $\gwwpt >0.3$~GeV.
        \item[(b)] The invariant mass of the additional tracks must be
                   less than 3~GeV and the invariant mass of the
                   additional tracks and clusters must be less than
                   8~GeV. 
        \item[(c)] $\gamma\beta$, the net momentum of the additional
                   tracks and clusters divided by their invariant
                   mass, is required to be greater than 2.0. 
      \end{itemize}
\end{itemize}

Next, significant missing transverse momentum is required.  
For events with a large
acoplanarity angle ($\acop > \pi/2$ where \acop\ is the 
supplement of the azimuthal opening angle), the following cut is applied
\footnote{The effect of measurement errors is taken into account in
selection I by taking each lepton in turn and shifting its momentum
up and down by one standard deviation of its estimated measurement
error.  At each stage the values of \ptevt\ and \ptaxic\ are 
recalculated and the minimum of all the values obtained is the one used
for comparison to the cut values.}:
\begin{itemize}
\item $x_{\mathrm{t}} > 0.045$, where $x_{\mathrm{t}}$ is the scaled missing transverse momentum of the event ($\stevt$).
\end{itemize}
For events with small acoplanarity angle ($\acop < \pi/2$):
\begin{itemize}
\item $x_{\mathrm{t}} > 0.035$
\item At small acoplanarity, cuts on \staxic\ and \athet\ (where
      $\athet = \tan^{-1}[\ptaxic /\pz]$, and
      \pz\  is the total momentum of the observed particles in
      the z direction) are
      used to reduce background from processes such as \taupair\ and 
      \epair\ \taupair\ . Events are divided into subsets using variables 
      which help estimate how likely they are to originate from \taupair\ or
      \epair\ \taupair\ .
  The cut values vary according to the subset (in increasing order 
  of similarity to \taupair\ or \epair\ \taupair\ background):
      \begin{itemize}
        \item[(a)] $x_{\mathrm{t}} > 0.2$ and unassociated energy less than 
                  1.5 GeV and 
                  the lepton pair is identified as $\ee$ , $\mumu$ or $\emu$ :
                 $\staxic > 0.011$ and $\athet > 0.025$
	\item[(b)] $x_{\mathrm{t}} > 0.2$ or 
unassociated energy less than 1.5 GeV but not in subset (a) :  
                   $\staxic > 0.018$ and $\athet > 0.05$
	\item[(c)] $x_{\mathrm{t}} < 0.2$ and unassociated energy 
                   greater than 1.5 GeV: 
                   $\staxic > 0.025$ and $\athet > 0.1$
      \end{itemize}
\item  $x_{\mathrm{t}} + \staxic > 0.07$.
\end{itemize}

Further cuts are applied to reduce the effect of processes which may
fake the signature of missing transverse momentum.  These are mostly 
vetoes against energy in the forward region (GC, FD, SW or MIP-plug).  
However care is taken to ensure that the activity in the forward region 
is only used to veto the event if the activity could
possibly explain the apparent missing momentum.  Also different
requirements are made depending on the amount of missing transverse
momentum observed. 
%Full details are available in~\cite{terrytn}.

Radiative lepton pair events containing high energy isolated photons 
form a potentially serious source of background, because quantities 
such as \ptevt\ and \ptaxic\ may be poorly measured. However, in order 
to retain efficiency for potential signal events containing isolated 
photons, events are rejected only if the presence of the isolated 
photon could possibly have caused the observed \ptevt\ and 
\ptaxic\ .

\subsubsection{Single Lepton Selection}
\label{sec-1lept}

Selection I also includes a selection for events with only one lepton
visible in the central detector, and the other lepton travelling
sufficiently close to the beam axis that no track is produced in the
central tracking chambers.

The principal single lepton selection criteria are:
\begin{itemize}
\item The event must contain one and only one identified, isolated 
      lepton candidate, and no other tracks.
\item $x_{\mathrm{t}} > 0.16$
\end{itemize}

In order to veto events which may fake the missing transverse momentum
signature, events are rejected if they contain activity in
ECAL, GC, FD, SW, muon endcap or MIP-plug which is back-to-back with 
the observed
lepton.

\subsection{Selection II}
\label{sec-sele-graham}

Selection II 
%(see~\cite{grahamtn} for a detailed description) 
is
optimised to select high visible energy events 
typical of the \dWW\ process.  A low multiplicity preselection 
is applied such that the events contain at least one track 
but no more than 8.  Also the sum of the number of tracks 
plus the number of ECAL clusters is required to be less than 16. 

A cone-based jet finding algorithm~\cite{CONE} is applied 
requiring a minimum jet energy of 2.5~GeV and a cone half angle 
of 20$^\circ$.  Events are required to contain 1, 2 or 3 jets, and 
a separate selection is used for each value of $n_{\mathrm{jet}}$.  The
majority (about 90\%) of \dWW\ events have $n_{\mathrm{jet}}=2$.  One-jet
events are usually those where the decay products from one of the 
W bosons are poorly reconstructed (for example if the lepton is travelling
close to the beam pipe).  
Three-jet events can occur if there is an energetic 
photon in the event.

Electron and muon identification, similar to that in selection I, 
is applied to the most energetic
track in each jet.  Jets not identified as electrons or
muons are classified as hadronic tau decays. 

The most important cuts for each $n_{\mathrm{jet}}$ class are summarised below.

\subsubsection{Di-jet Selection}

\begin{enumerate}
\item $\theta_{\mathrm{acol}} > 5^\circ$, 
where $\theta_{\mathrm{acol}}$ is the
      acolinearity angle between the two jets (defined as
      the supplement of the three-dimensional opening angle).
\item $x_{\mathrm{t}} > 0.05$. It is further
      required that the significance by which 
$x_{\mathrm{t}}$ 
exceeds 0.05 be 
      greater than 1 standard deviation.
\item For events with $\acop > \pi/2$
      it is required that the direction of the missing
      momentum satisfy $|\cos\theta_p^{\mathrm{miss}}|<0.95$.  For events
      with $\acop < \pi/2$ it is required that 
      $\staxic >0.022$ and that $\sin \athet > 0.06$.
\end{enumerate}

Further cuts are made on the quality of each jet, and for background
rejection (mainly for events at low 
$x_{\mathrm{t}}$
):

\begin{itemize}
\item Events are rejected if there are any tracks which are not
      associated with either jet.
\item Events with low 
$x_{\mathrm{t}}$
($x_{\mathrm{t}} < 0.15$) are rejected if there is
      evidence of activity in the MIP-plug with
      azimuthal angle within $60^\circ$ of the missing transverse
      momentum direction. For events with medium 
$x_{\mathrm{t}}$
($0.15< x_{\mathrm{t}} <0.23)$,
      the veto uses only the outer MIP-plug scintillators.

\end{itemize}

\subsubsection{Tri-jet Selection}

For events classified as tri-jet, significant missing momentum is 
required:
\begin{itemize}
\item The sum of the opening angles among the three two-jet pairings
      should be less than $359^\circ$.
\item 
$x_{\mathrm{t}}$
of the three-jet system should exceed 0.05 with a significance
      exceeding 1 standard deviation.
\end{itemize}

Further requirements depend in part on the three-jet event topology 
in the transverse plane characterised by $\Delta\phi^{\mathrm{max}}$.
This is calculated by ordering the jets by increasing azimuth, and finding the 
maximum azimuthal di-jet separation angle 
when rotating clockwise from 
one jet to the next\footnote{This definition results in $\Delta\phi^{\mathrm{max}}$ being in the range from 120$^{\circ}$ to 
360$^{\circ}$.}.

The most important additional requirements are:
\begin{itemize}
\item For $\Delta\phi^{\mathrm{max}} > 185^\circ$, there should be no pair 
      of vertex drift chamber axial tracks with an azimuthal opening angle 
      exceeding 165$^{\circ}$. These tracks are reconstructed independently 
      from standard jet chamber based tracks and this criterion helps 
      to reduce cases where the wrong jet chamber 
      left-right ambiguity is chosen in the 
      standard tracking.

\item For $\Delta\phi^{\mathrm{max}} < 185^\circ$ it is explicitly required 
      that two of the jets have associated tracks (charged jet) and 
      the third jet has no associated tracks (neutral jet).
      For the events with $\Delta\phi^{\mathrm{max}} < 180^\circ$ 
      an axis in the transverse
      plane is defined using the most energetic charged jet.  The
      event is rejected if the transverse momentum of the neutral 
      jet with respect to this axis exceeds 80\% of the transverse
      momentum of the less energetic charged jet.  This cut is
      effective against $\tau^+\tau^-\gamma$ events.
\item Events are rejected if there is evidence for a particle passing
      through the MIP-plug (similar to di-jet MIP-plug veto 
      but without the
      directional requirement).
\end{itemize}

\subsubsection{Single-jet Selection}

The single-jet selection applies to events where one lepton with 
high transverse
momentum is observed at wide angle $|\cos{\theta}| < 0.82$ 
with evidence for a
partially reconstructed lepton at small polar angle, or events where
the two leptons fall within the same cone (massive mono-jet) for 
$|\cos{\theta}| < 0.90$. 
In contrast to the single 
lepton category of selection I, it is required here that there be 
some activity in the forward 
region (endcap ECAL, GC, FD, SW, muon endcap, MIP-plug) when there 
is evidence for only one lepton at wide angle.

Backgrounds from cosmic rays are reduced by requiring in-time TOF hits
(within 20~ns) for tracks in the barrel region, and that the most
energetic track in the jet be associated with hits in the silicon
micro-vertex detector.

Three different minimum requirements on $x_{\mathrm{t}}$ 
are applied (0.20, 0.30 and 0.25) depending on whether the event
satisfies the massive mono-jet criteria and whether the forward activity 
is solely due to the MIP-plug or not, respectively.

\section{Comparison of Data with Standard Model Processes}
\label{sec:compare}
The numbers of events passing the general selection at each centre-of-mass
energy in the data are compared with the \smc\ predictions 
in Table~\ref{tab-samples}. 
The total number of events predicted by the  \sm\ is given, 
together with a breakdown into the
contributions from individual processes.
The data have been binned by centre-of-mass energy in bins which reflect the
predominant centre-of-mass energies where data was collected.
The number of observed candidates  is consistent with the 
expectation from \sm\ sources, which is dominated by 
the \llnunu\ final state arising mostly from \wpair\ production.
% in which both W's
%decay leptonically.

\begin{table}[htb]
\centering
\begin{tabular}{|rr|rr|rrrrr|}
\hline
$\sqrt{s}$ & $\cal{L}$
                & Data & SM         &  \llnunu   & \eell    & \llqq    & \llgam   & \nngbra \\
\hline 
  182.7 & 56.4  &   78 &  81.4\p0.8 &  77.5\p0.7 & 3.4\p0.5 &0.07\p0.03&0.31\p0.04& 0.06\p0.03 \\
  188.6 & 183.5 &  332 & 348.2\p1.9 & 337.2\p1.7 & 3.6\p0.7 & 1.6\p0.1 & 4.6\p0.3 & 1.1\p0.2   \\
  191.6 & 29.3  &   60 &  56.1\p0.6 &  54.2\p0.6 & 0.6\p0.1 &0.2\p0.04 &0.9\p0.2  & 0.10\p0.04 \\
  195.5 & 76.4  &  166 & 150.5\p1.2 & 144.7\p1.0 & 2.7\p0.4 &0.6\p0.06 &2.1\p0.4  & 0.42\p0.05 \\
  199.5 & 76.6  &  155 & 153.5\p0.9 & 148.7\p0.7 & 1.8\p0.3 &0.7\p0.06 &1.8\p0.2  & 0.45\p0.05 \\
  202.0 & 45.5  &  110 &  90.6\p0.7 &  87.6\p0.6 & 1.4\p0.2 &0.4\p0.04 &1.0\p0.1  & 0.19\p0.04 \\
  205.1 & 79.0  &  154 & 155.6\p1.2 & 150.8\p1.1 & 1.8\p0.3 &0.6\p0.07 &2.0\p0.3  & 0.38\p0.07 \\
  206.5 & 124.6 &  243 & 249.5\p1.4 & 241.4\p1.2 & 3.9\p0.6 &1.1\p0.1  &2.3\p0.3  & 0.64\p0.08 \\
  207.9 & 9.03  &   19 &  18.2\p0.2 &  17.7\p0.2 & 0.20\p0.04&0.08\p0.01&0.18\p0.03& 0.06\p0.01 \\ \hline
  All   & 680.4 & 1317 &1303.6\p3.3 &1259.8\p2.9 &19.6\p1.2 & 5.4\p0.2 & 15.2\p0.7& 3.4\p0.2   \\
\hline
\end{tabular}
\caption[]{\sl
%  \protect{\parbox[t]{15cm}{
Comparison between data and \mc\ of the 
number of events passing the general selection in each centre-of-mass
energy bin.
The total number of events predicted by the  \sm\ is given, 
together with a breakdown into the
contributions from individual processes.
The \mc\ statistical errors are shown. 
Also listed for each bin is the mean $\sqrt{s}$ (in GeV) 
and the integrated luminosity, $\cal{L}$, in pb$^{-1}$.
\label{tab-samples}
%}}
}
\end{table}

\subsection{Lepton Identification and Kinematic Distributions}
\label{sec:lept}

Information about the type of lepton found in the selected events and
measurement of the event kinematics was used to compare the data 
with expectations from Standard Model processes. 
Some of these distributions are then used as likelihood variables 
to distinguish between
Standard Model and new physics sources
of acoplanar di-lepton events (section 5).
We examine the distributions of the likelihood variables and other 
related kinematics distributions,
as measured in the data, 
and compare to expectations from Standard Model processes.
The following distributions are used as the likelihood variables:
\begin{itemize}
\item Di-lepton identities;
\item Acolinearity of the event, defined as 
the supplement of the three-dimensional opening angle between the two leptons;
\item Momentum asymmetry of the event, ($|p_1 - p_2|/(p_1 + p_2)$), where $p_1$
and $p_2$ are the momenta of each lepton;
\item Scaled momentum of each lepton, ($p/E_{\mathrm{beam}}$) using the best estimate combining tracking and calorimetry information;
\item $-q \cos\theta$ of each lepton, where $q$ is the particle charge
and $\theta$ is the polar angle with respect to the electron beam direction.
\end{itemize}

%Information on lepton identification, the acolinearity
%and momentum asymmetry ($|p_1 - p_2|/(p_1 + p_2)$) of the event,
%the scaled momentum ($p/E_{\mathrm{beam}}$)
%and $-q \cos\theta$ of the observed lepton 
%candidates\footnote{where $q$ and $\theta$ are the charge and 
%polar production angle of
%the lepton} is used to distinguish between Standard Model and new physics 
%sources of acoplanar
%di-lepton events.

The lepton identification algorithm has been
improved using techniques developed for tau lepton decay analyses 
such as described in \cite{a_tau-paper}. The efficiencies
for correctly identifying leptons as electrons, muons or
hadronically decaying tau leptons are shown in Table~\ref{tab-leptid_eff}
for a sample of leptons representative of the geometrical and kinematical
acceptance of the general selection.
Note that we have six separate mutually exclusive 
lepton identification decisions:
electron (e), muon ($\mu$), hadronically decaying tau (h), 
electron-hadron ambiguous (e/h), 
muon-hadron ambiguous ($\mu$/h) and ``rest-of-event'' (x).
The ambiguous categories correspond to cases where the lepton candidate
has properties which do not permit clear identification between the
two lepton types, while the rest-of-event category arises mostly from
tau leptons which fail to pass basic lepton identification criteria.

%(Table 5.1 from Tom's thesis)

\begin{table}
\begin{center}
\begin{tabular}{|c|c|c|c|}
\hline
           &\multicolumn{3}{|c|}{ True Identity of Lepton} \\
\cline{2-4}
 Lepton ID & e or $\tau \ra\ $ e & $\mu$ or $\tau \ra\ \mu$ & $\tau \ra\ $ h \\
\hline
%           &\multicolumn{3}{|c||}{ }\\
 e         &       96.3 \% &  0.1 \%  &  4.7 \%  \\
$\mu$      &        0.1 \% & 98.2 \%  &  3.5 \%  \\
 h         &        1.2 \% &  0.8 \%  & 86.9 \%  \\
e/h        &        2.2 \% &  0.1 \%  &  2.8 \%  \\
$\mu$/h    &        0.2 \% &  0.8 \%  &  1.8 \%  \\
 x         &        0.0 \% &  0.0 \%  &  0.4 \%  \\
\hline
\end{tabular}
\end{center}
\caption[Lepton ID performance using TP]{\sl Lepton identification 
performance, calculated using {\sc KoralW} four-lepton Monte 
Carlo events at $\sqrt{s}=$189-206~GeV. 
The efficiency is evaluated for leptons generated with
$|\cos{\theta}| < 0.95$ and $p/E_{\mathrm{beam}} > 0.02$.
\label{tab-leptid_eff}}
\end{table}

\begin{figure}[htbp]
 \epsfxsize=\textwidth
% \epsffile{figures/fig-dileptonid.eps}
% \epsffile{/home/graham/acopll/kumacs/leptid_edited.eps}
\epsffile{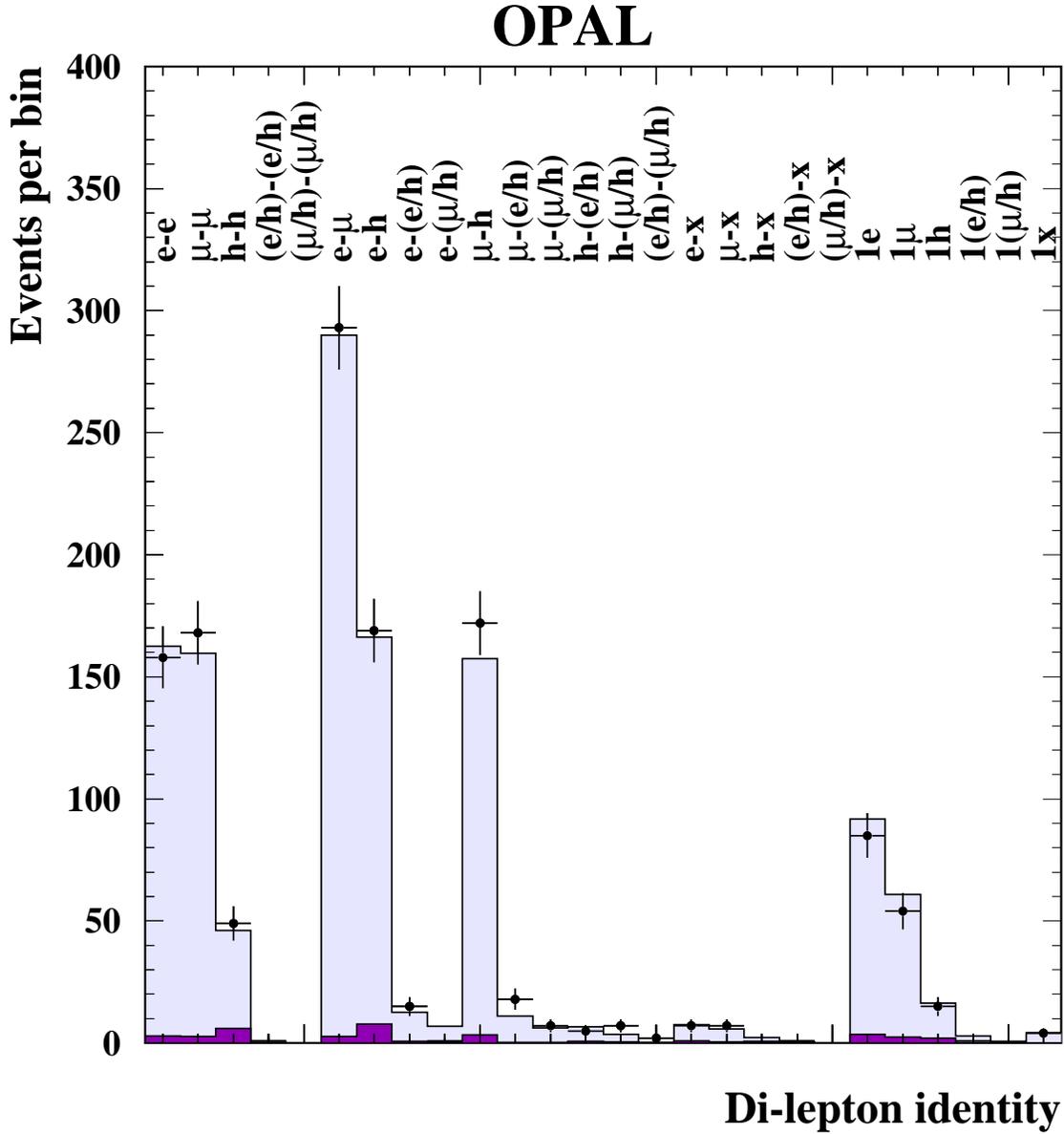}
 \caption{\sl Di-lepton identities
for the data at $\sqrt{s}=$189-208 GeV compared with Standard Model
expectations.
The data are shown as the points with error bars (statistical errors only).
The Standard Model \mc\ prediction dominated by 4-fermion processes with
genuine prompt missing energy and momentum (\llnunu ) is shown as the
lightly shaded histogram and the
background component, 
arising mainly from processes with four charged leptons in
the final state, is shown as the darkly shaded histogram.
The Standard Model Monte Carlo histograms are normalized 
to the integrated luminosity of the data.
The last six bins correspond to ``single-lepton events''.
\label{fig-leptid}
} 
\end{figure}

\begin{figure}[htbp]
 \epsfxsize=\textwidth
% \epsffile{pr290_01.eps}
%\epsffile{/home/graham/acopll/kumacs/evt_props_1_new.eps}
\epsffile{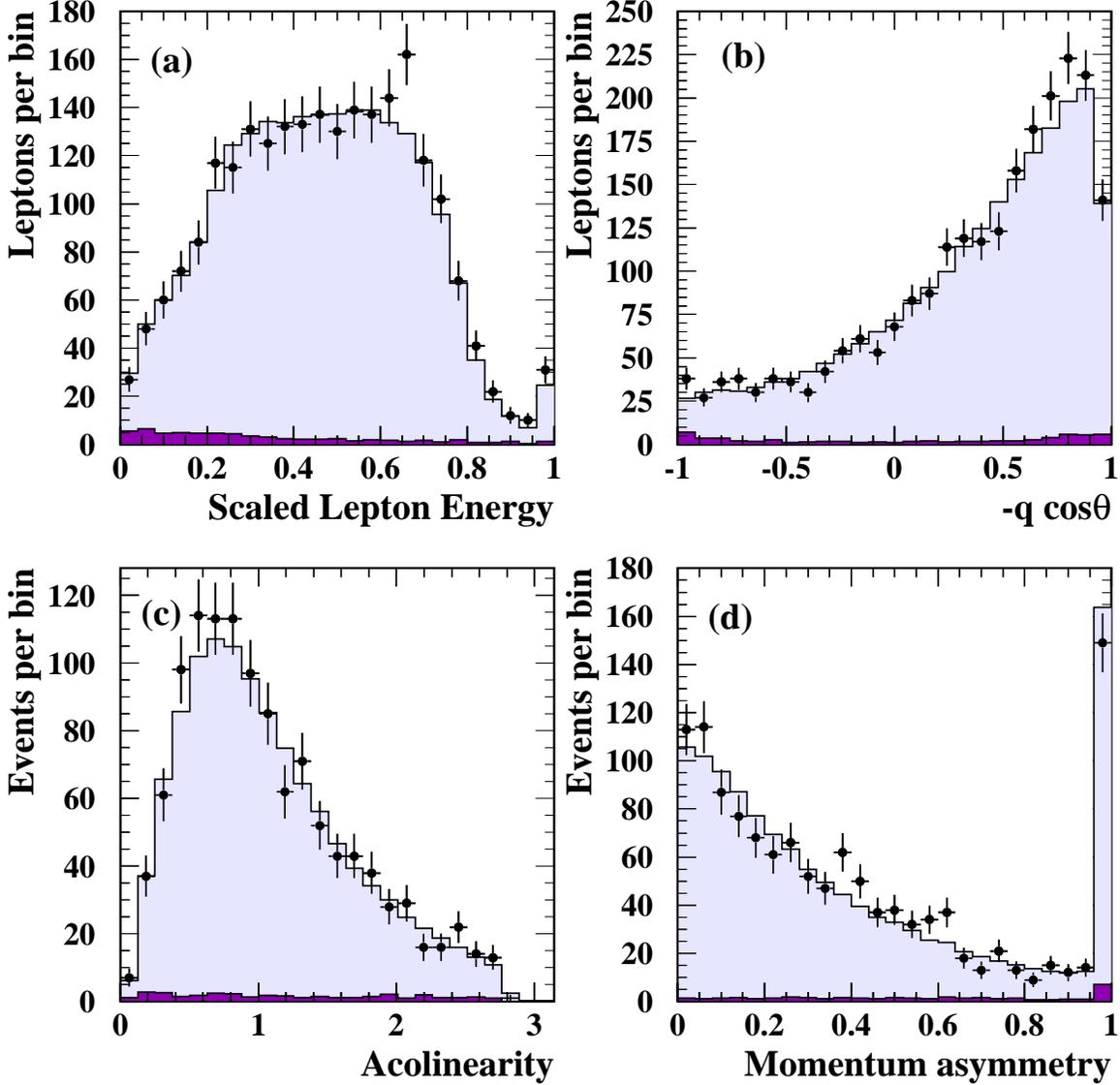}
 \caption
{\sl Distributions of (a) the lepton momentum divided by the beam 
energy, (b) $-q \cos\theta$, (c) acolinearity (in radians) and (d) 
momentum asymmetry for 
the event sample produced by the general selection 
at $\sqrt{s}=$189-208 GeV.
The data are shown as the points with error bars (statistical errors only).
The Standard Model \mc\ prediction dominated by 4-fermion processes with
genuine prompt missing energy and momentum (\llnunu ) is shown as the
lightly shaded histogram and the
background component, 
arising mainly from processes with four charged leptons in
the final state, is shown as the darkly shaded histogram.
The Standard Model Monte Carlo histograms are normalized to the 
integrated luminosity of the data.
In (a) and
(b) there are two entries per event
for events containing two identified leptons.
\label{fig-kine}
} 
\end{figure}

\begin{figure}[htbp]
 \epsfxsize=\textwidth
% \epsffile{pr290_01.eps}
%\epsffile{/home/graham/acopll/kumacs/evt_props_2_new.eps}
\epsffile{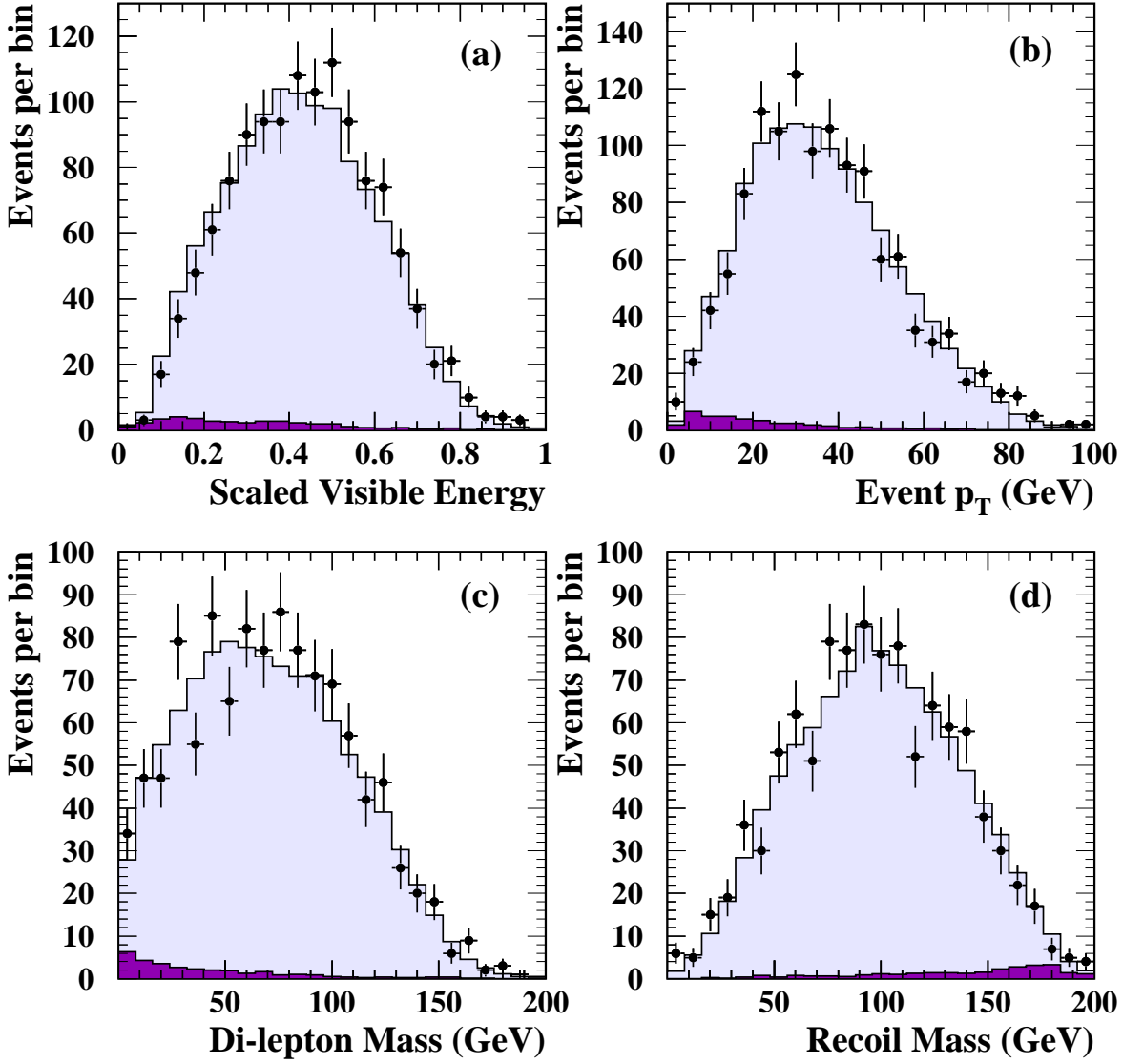}
 \caption
{\sl Distributions of (a) the total visible energy divided by the 
centre-of-mass energy, (b) the net transverse momentum of the event 
(c) di-lepton mass and (d) recoil mass to the di-lepton system
for 
the event sample produced by the general selection 
at $\sqrt{s}=$189-208 GeV.
The data are shown as the points with error bars (statistical errors only).
The Standard Model \mc\ prediction dominated by 4-fermion processes with
genuine prompt missing energy and momentum (\llnunu ) is shown as the
lightly shaded histogram and the
background component, 
arising mainly from processes with four charged leptons in
the final state, is shown as the darkly shaded histogram.
The Standard Model Monte Carlo histograms are normalized to the integrated 
luminosity of the data.
\label{fig-kine2}
} 
\end{figure}

We show in Figure~\ref{fig-leptid} the observed di-lepton identities
and the above kinematic variables are shown in Figure~\ref{fig-kine}.
Additional kinematic variables,
which are not directly used in the
likelihood, are also shown in Figure~\ref{fig-kine2}.
Reasonable agreement is observed between the data and the \smc\ in
all these distributions as quantified using $\chi^2$ tests.

\section{Likelihood Method for New Physics Search}
\label{sec:search}

\subsection{Method}

The method used in the second stage of the analysis, in which we 
distinguish between 
\sm\ and new physics sources of lepton pair
events with missing momentum, is essentially 
described in~\cite{paper189}.
Discrimination is provided by information on the 
lepton identification, the acolinearity and momentum asymmetry of the event, 
the scaled momentum
and $-q \cos\theta$ of the observed lepton candidates.
These variables are combined using a likelihood technique.
The discriminating quantity used is the relative likelihood, \LR\ , defined by:
\begin{center}
$$\LR = \frac{\LS}{\LS+\LB},$$
\end{center}
where \LS\ is the 
likelihood of the event being consistent 
with the signal hypothesis (a particular new physics scenario) 
and \LB\ is the likelihood of the event being consistent with
the background hypothesis  (\sm\ sources of lepton pair events with
missing transverse momentum).
The distributions depend on \dm\ and on the parent particle mass, $m$.  

In our previous publications we justified the inclusion
of most of these variables and described how they were used~\cite{paper189}.
In particular the $-q \cos\theta$ variable is not used
in the selectron and chargino searches 
because the $t$-channel neutralino (selectron search) and sneutrino
exchange (chargino search) contributions make 
this signal distribution model-dependent.

The new momentum asymmetry variable describes the correlation between
the momenta of the two leptons and is thus complementary to the 
two scaled momentum variables which are implemented
in the likelihood function
as
independent variables.
It is correlated with  
the acolinearity variable, and so the information was included
using a two-dimensional probability distribution of momentum asymmetry
versus acolinearity.
Sensitivity 
studies for the selectron and smuon searches justified this addition.
However similar studies for the other channels (stau, chargino
and charged Higgs), where momentum distributions
are broader, showed no net gain in sensitivity
and so for these channels the acolinearity distribution alone was
used. Also for these channels, 
we removed the distinction between
leptons being identified as electrons and muons 
and merged them into one category since we found that there 
was no significant difference in search sensitivity.

For the selectron search, we required that at least one lepton be 
identified as an electron and that no lepton in the event 
be compatible with being a muon (i.e. the other lepton was
not identified as a muon and it was not muon-hadron ambiguous).
Similarly for the smuon search we required at least one identified
muon and required that no lepton be compatible with being an electron.
Some additional requirements were used to reject 
events from these searches if the second lepton was identified as 
a hadronically decaying tau or a ``rest-of-event'' candidate and if it had 
properties strongly incompatible with being 
an electron for the selectron search and with being a muon for the 
smuon search. These
additional requirements cut the background by 21\% and 14\%
while only reducing the signal efficiency by 2.1\% and 0.8\%
for the selectron and smuon searches respectively.

%=======================================================================
%\subsection{Likelihood Distributions}
%=======================================================================
%\label{sec:lb}

For each search channel, reference histograms were constructed for each
of the likelihood variables at each point in $m$ and \dm\ for which
signal \mc\ had been generated (each point corresponds to a 
table entry in tables 3-8). A smoothing algorithm~\cite{smooth}
was applied to the histograms to reduce the effects of statistical
fluctuations.  The reference histograms were then used to construct
\LR\ distributions. An interpolation procedure was developed which 
allowed us to interpolate among neighbouring signal \mc\ ($m$, \dm\ , 
$\sqrt{s}$) 
points to intermediate values of ($m$, \dm\ and $\sqrt{s}$). Further 
details are given in~\cite{tom-thesis}.

\begin{figure}[htbp]
 \epsfxsize=\textwidth 
% \epsffile{figures/fig-lrexample_v7_edited.eps}
\epsffile{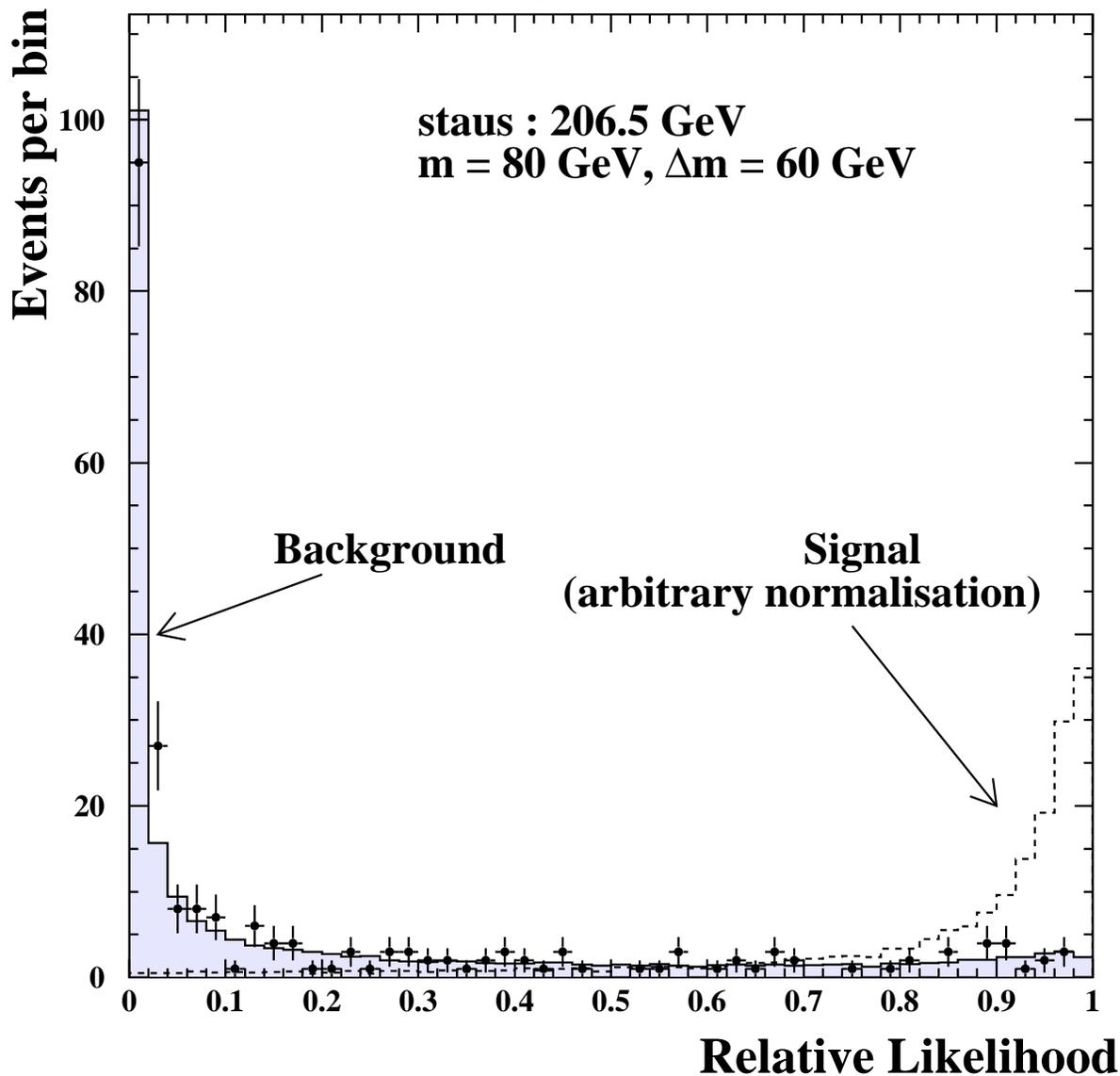}
 \caption{\sl Distributions of the relative likelihood, \LR , for
\smc\ (shaded histogram), signal (open
histogram) and data (points with error bars, statistical errors only), 
in the analysis for staus
with a mass of 80~GeV for a stau-neutralino mass difference of 60~GeV
for the $\sqrt{s}=$~206.5~GeV centre-of-mass energy bin.
The Standard Model Monte Carlo histogram is normalized to the 
integrated luminosity of the data.
\label{fig:lr}
} 
\end{figure}

\LR\ distributions for signal \mc , \smc\ and data are shown
in Figure~\ref{fig:lr} for the specific example of the analysis for
staus with a mass of 80~GeV and a stau-neutralino mass difference 
of 60~GeV.  There is considerable variation in the shapes of these
distributions with $m$ and \dm\ but agreement between data and
\smc\ is generally good.

%A check of consistency between data and the \sm\ can be performed without 
%reference
%to a particular signal by comparing the \LB\ distributions for data and \sm .
%Figure~\ref{fig:lb}(a) shows the \LB\ distributions for 
%the data compared with the Standard Model expectation for
%%the \sm\ (histogram)
%%and data (points) for 
%events passing the general selection.  All the likelihood
%variables are used.  Figures~\ref{fig:lb}(b) and (c) show 
%the same information 
%after making the initial lepton identification requirements 
%for the selectron and smuon searches respectively.
%In Figure~\ref{fig:lb}(b), only the variables used in the selectron analysis
%are used.  In each of the plots, the secondary peak at high \LB\ corresponds
%to events which have only one identified lepton and therefore fewer variables
%entering the likelihood\footnote{This effect cancels when the
%likelihood ratio \LR\ is calculated because the events will have high
%\LS\ for the same reason.}.  In all three plots, the data are in good
%agreement with the \sm\ expectation.

%=======================================================================
\subsection{Calculation of Cross-section Upper Limits}
%=======================================================================
\label{sec:limits}

%In~\cite{paper183} , the limit on the cross-section was calculated by finding
%an optimised cut on the value of \LR\ as a function of $m$ and \dm\ for each
%centre of mass energy, and applying
%this cut to signal \mc , \smc\ and data.  The resulting efficiencies,
%expected backgrounds and numbers of candidates were used to calculate 
%cross-section limits using the likelihood ratio method~\cite{LR} to
%combine the information.
For each search channel and for each
set of kinematic parameters ($m$ and \dm) 
we wish to test the consistency of the data
with the sum of background and an additional signal contribution.
The signal contribution depends on $\sigs$, 
the cross-section multiplied by branching ratio 
to leptonic final states
squared.
We use an extended maximum likelihood technique as described
in~\cite{paper189}
to measure $\sigs$ and, in the absence of a significant signal, to
set 95\% confidence level (CL) upper limits on $\sigs$.
We form a likelihood, $L(\sigs )$, of the set of
\LR\ values for the data being consistent with the expected \LR\ distribution
for \sm\ plus a signal contribution of $\sigs$.
The upper limit on the cross-section 
multiplied by branching ratio squared at 95\%
confidence level, \signine , is calculated as 
the value of $\sigs$ below which 95\% 
of the area under the likelihood function lies.
Details of the likelihood function and method 
as well as cross-checks of the method are described in~\cite{paper189}. 
Data from different centre-of-mass energies
are combined by weighting 
them using an assumed cross-section dependence on $\sqrt{s}$.
For this paper the reference centre-of-mass energy for the weighting 
is $\sqrt{s}=208$~GeV instead of 
$\sqrt{s}=189$~GeV.

We have also evaluated 
% ``counting-experiment'' 
results 
based on the number of events passing an optimised 
cut on the \LR\ value. These results are used as a cross-check, as
a direct comparison between data and background and as 
a simpler basis for combining this analysis with 
other experiments or other analyses.

%=======================================================================
\section{Constraints on New Physics}
%=======================================================================
\label{sec:results}

We present limits on the pair production
of charged scalar leptons, leptonically decaying charged Higgs bosons and 
charginos that decay to produce a charged lepton and invisible particles.

The 95\% CL upper limit on  new particle production at $\sqrt{s}=208$~GeV
obtained by combining the data at 
different centre-of-mass energies is 
calculated at each
kinematically allowed point on a 1~GeV by 1~GeV grid of $m$ and \dm , 
using
the \LR\ distributions for signal and background, the \LR\ values of 
the data events,
and the efficiency of the general selection at that point as input.

In addition to the \mc\ statistical error on the signal efficiency, we 
assess a 10\% relative
systematic error on the estimated selection efficiency to take into
account 
deficiencies in the \mc\ generators and the detector simulation (5\%), 
uncertainties in the interpolation procedure (5\%), 
effect of tau polarisation in stau decay\cite{Nojiri} (5\%),
fluctuations in the shape of the signal \LR\ distribution (2\%), 
uncertainties from lepton identification efficiency (2\%), 
detector occupancy (1\%), trigger efficiency ($<$ 1\%), 
and luminosity measurement (0.5\%). 

At high values of \dm\ the dominant background to 
the searches for new physics results from \wpair\ production.
High statistics \mc\ samples for this process are available and 
these samples describe the OPAL data well~\cite{wwpaper}.
In addition to the \mc\ statistical error, we assess a 5\% relative
systematic error on the estimated background to take into account uncertainties
in the shapes of the \LR\ distributions and reference 
histograms, in the 
interpolation procedure, and deficiencies in the \mc\ detector simulation.
At low values of \dm\ the dominant background results from two-photon \eell\
events.
The background uncertainty at low \dm\ is dominated by the limited \mc\
statistics; the
uncertainty is typically 20--80\% at low \dm .

We have examined distributions of the cut variables and 
background enriched samples in assessing these systematic errors.
The systematic errors on the selection efficiency and the 
estimated background are considered as global errors applicable to 
each search and for all values of $m$ and \dm\ .
In setting limits, the \mc\ statistical errors and the 
systematic uncertainties on the efficiency and the background 
expectation are taken into account using numerical convolution.
%~\cite{cousins}.

%=======================================================================
\subsection{Limits on {\boldmath$\sigs$ }}
%=======================================================================
\label{sec:crosssec}

\begin{figure}[htbp]
 \epsfxsize=\textwidth 
% \epsffile[0 0 580 600]{figures/fig-limit-sele.eps}
% \epsffile[0 0 580 600]{/home/graham/acopll/kumacs/limit_selectron_new.eps}
\epsffile[0 0 580 600]{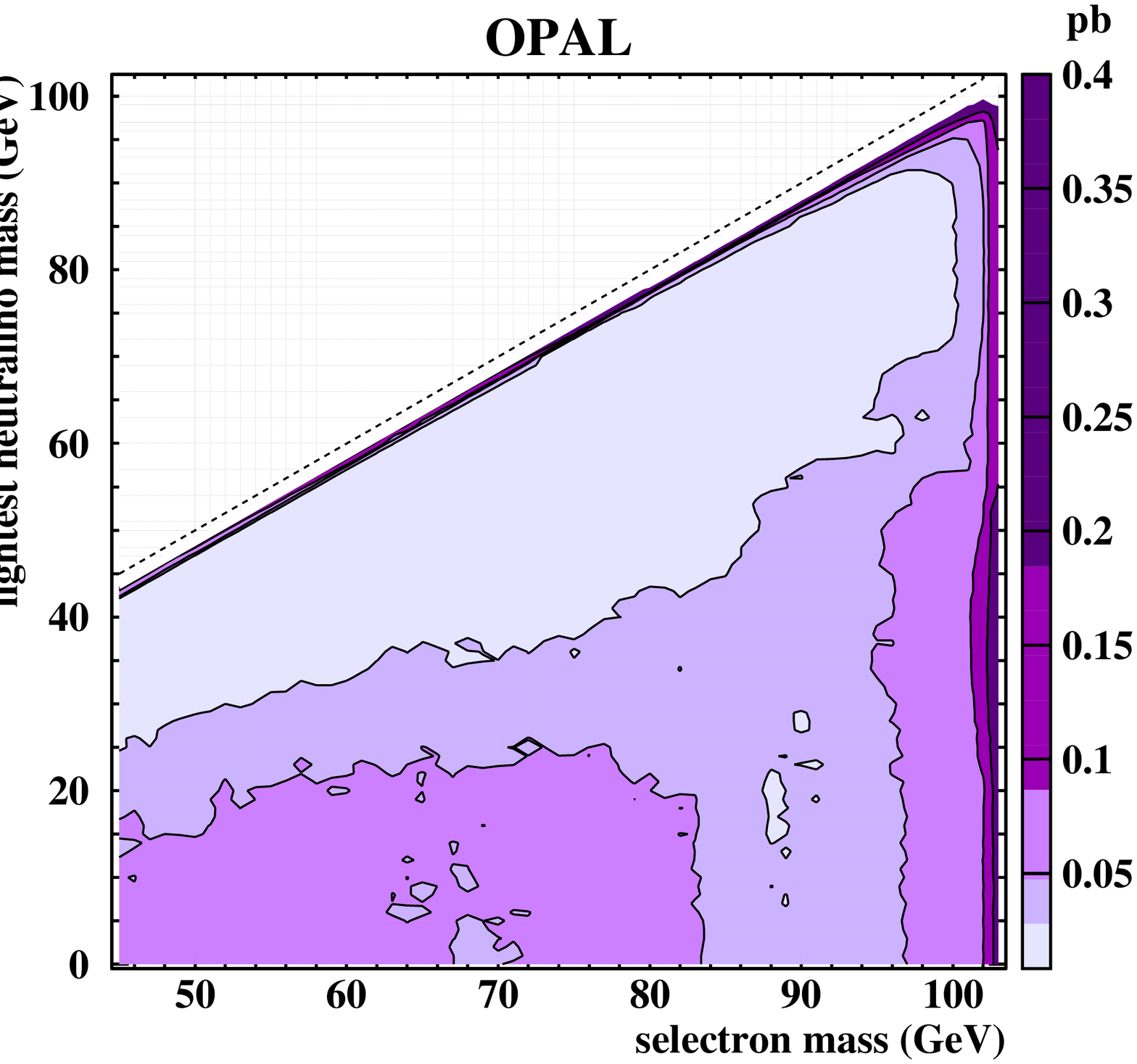}
 \caption{\sl
Contours of the 95\% CL upper limits on the selectron pair
cross-section times $BR^2(\sele \rightarrow \mathrm{e} \nt_1)$
at 208 GeV
based on combining the 183 - 208 GeV data-sets 
assuming a $\beta^3/s$ dependence of the cross-section.
The kinematically allowed region lies below the dashed line.  The
unshaded region at very low \dm\ is experimentally inaccessible in
this search.
} 
\label{fig:limit_1}
\end{figure}

\begin{figure}[htbp]
 \epsfxsize=\textwidth 
% \epsffile[0 0 580 600]{figures/fig-limit-chiggs.eps}
% \epsffile[0 0 580 600]{/home/graham/acopll/kumacs/chiggs_new.eps}
\epsffile[0 0 580 600]{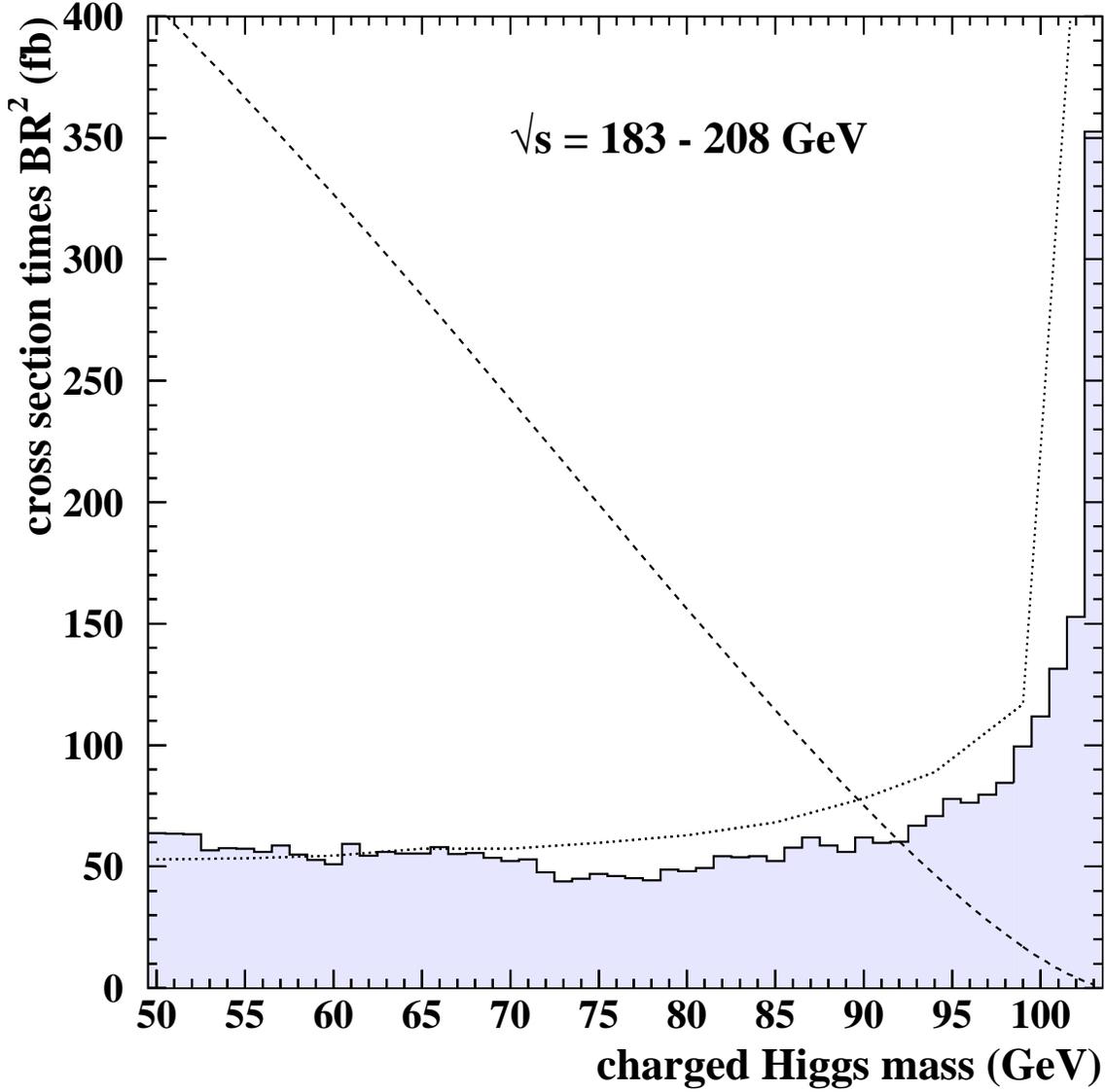}
%***  \epsffile[0 0 580 600]{higgs2.ps}
 \caption{\sl 
The solid histogram shows the 95\% CL upper limit on the 
charged Higgs pair production 
cross-section times branching ratio squared for the decay \dH\
at $\protect\sqrt{s}$~=~208~GeV.
The limit is obtained by combining the 183 - 208~GeV data-sets 
assuming a $\beta^3/s$ dependence of the cross-section.
For comparison, the dashed curve
shows the prediction from {\sc HZHA}
at $\protect\sqrt{s}$~=~208~GeV
assuming a 100\% branching ratio for the decay \dH .
The expected limit calculated from Standard Model \mc\ alone is indicated by
the dotted line.
} 
\label{fig:limit_5}
\end{figure}

Limits on $\sigs$, the production cross-section 
for new physics processes multiplied by the 
branching ratio squared, are presented in a manner intended to 
minimise the number of model assumptions.
The 95\% CL upper limits at $\sqrt{s}=208$~GeV shown in 
Figures~\protect\ref{fig:limit_1}~--~\protect\ref{fig:limit_5} are
obtained by combining the data at the various centre-of-mass 
energies using
the assumption that the
cross-section varies as $\beta^3/s$ for sleptons and charged Higgs 
and $\beta/s$ for charginos, where $\beta$ is the particle's velocity
in units of $c$.
The chosen functional forms are used for simplicity in presenting the
data and represent an approximation, most importantly 
for processes in which $t$-channel exchange may be important, namely
selectron pair and chargino pair production.
In these cases the cross-section
dependence on centre-of-mass energy 
is model dependent, depending on the mass of the exchanged particles 
and the couplings of the neutralinos and charginos.  The selectron
\mc\ events were generated with $\mu = -200$~GeV and
$\tan{\beta} = 1.5$ using {\sc Susygen}.
We have found by varying $\mu$ and $\tan{\beta}$, 
over the range $100 < |\mu| < 1000$~GeV and $1 < \tan{\beta} < 50$,  
that the above choice
gives a conservative estimate of the selection efficiency for selectrons.

\begin{figure}[htbp]
 \epsfxsize=\textwidth 
% \epsffile[0 0 580 600]{figures/fig-limit-smuon.eps}
% \epsffile[0 0 580 600]{/home/graham/acopll/kumacs/limit_smuon_new.eps}
\epsffile[0 0 580 600]{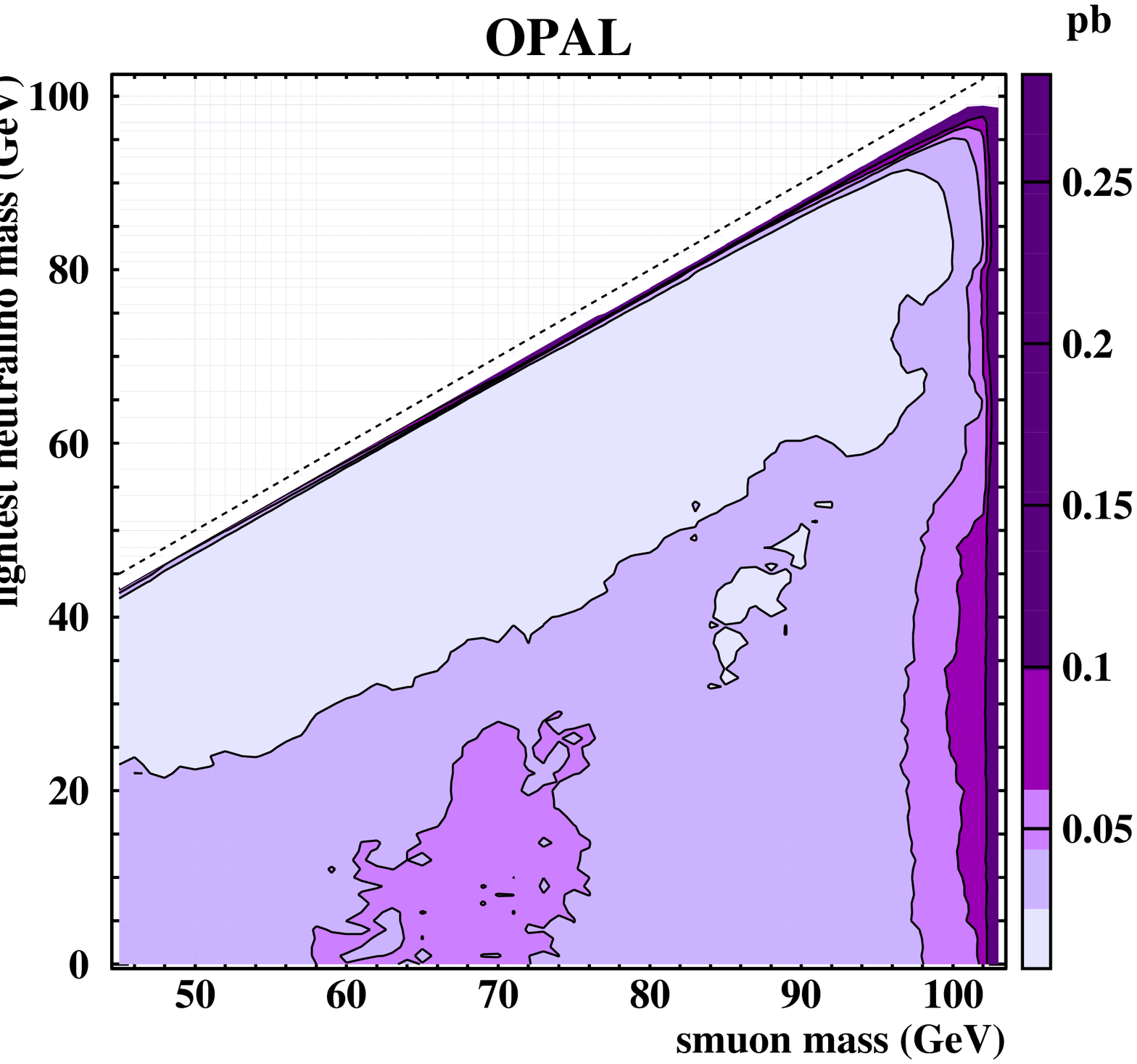}
 \caption{\sl
Contours of the 95\% CL upper limits on the smuon pair
cross-section times $BR^2(\smu \rightarrow \mu \nt_1)$
at 208 GeV
based on combining the 183 - 208 GeV data-sets 
assuming a $\beta^3/s$ dependence of the cross-section.
The kinematically allowed region lies below the dashed line.  The
unshaded region at very low \dm\ is experimentally inaccessible in
this search.
} 
\label{fig:limit_2}
\end{figure}

\begin{figure}[htbp]
 \epsfxsize=\textwidth 
% \epsffile[0 0 580 600]{figures/fig-limit-stau.eps}
% \epsffile[0 0 580 600]{/home/graham/acopll/kumacs/limit_stau_new.eps}
\epsffile[0 0 580 600]{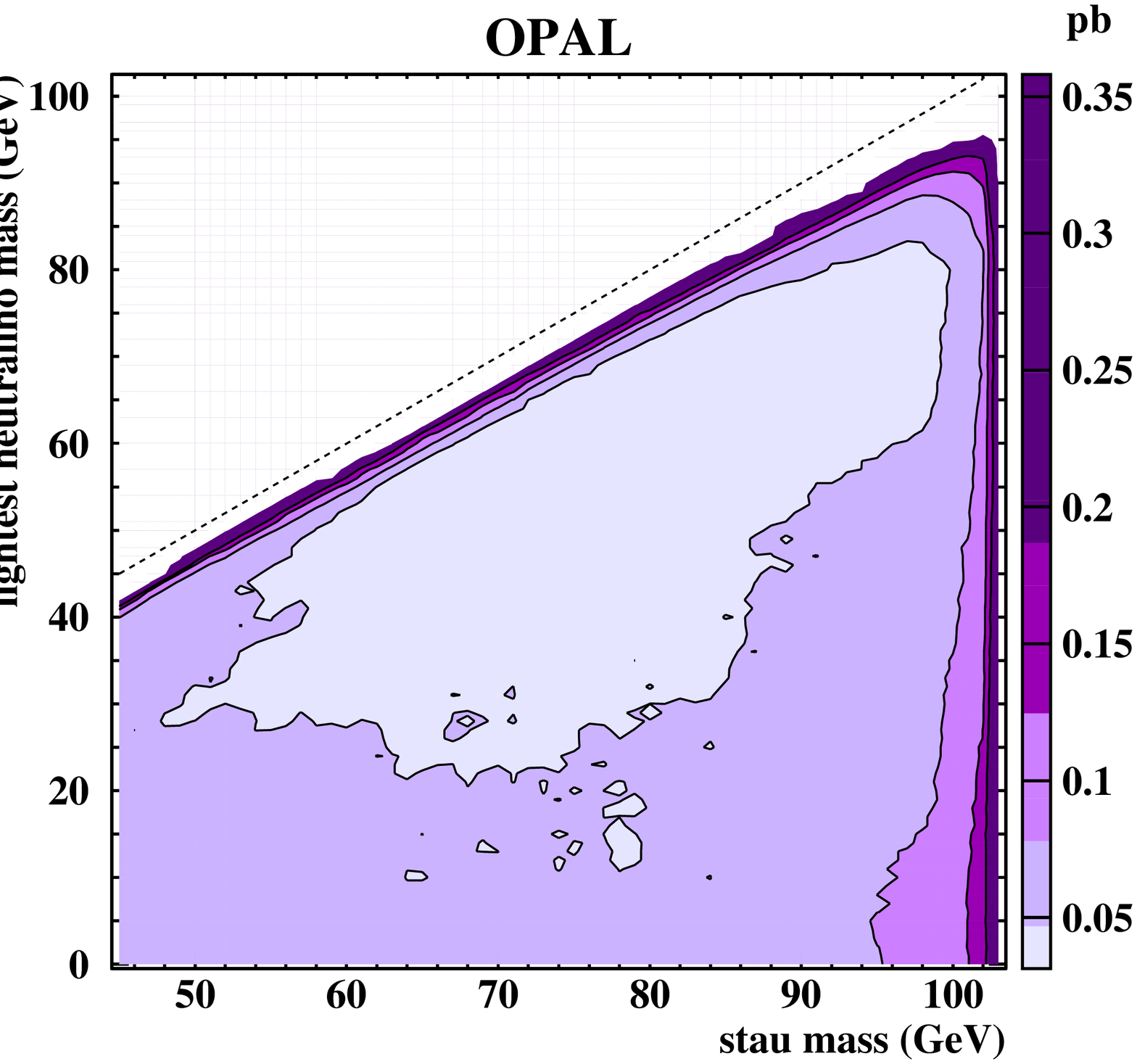}
 \caption{\sl
Contours of the 95\% CL upper limits on the stau pair
cross-section times $BR^2(\stau \rightarrow \tau \nt_1)$
at 208 GeV
based on combining the 183 - 208 GeV data-sets 
assuming a $\beta^3/s$ dependence of the cross-section.
The kinematically allowed region lies below the dashed line.  The
unshaded region at very low \dm\ is experimentally inaccessible in
this search.
} 
\label{fig:limit_3}
\end{figure}

Upper limits at  95\% CL on the selectron pair cross-section 
at~\sten\ multiplied by the
branching ratio squared for the decay \dsele\
are shown in Figure~\ref{fig:limit_1} as a function of selectron mass 
and lightest neutralino mass.
These limits are applicable to
$\tilde{\mathrm e}^+_{\mathrm L}\tilde{\mathrm e}^-_{\mathrm L}$ and 
$\tilde{\mathrm e}^+_{\mathrm R}\tilde{\mathrm e}^-_{\mathrm R}$ production.
The corresponding plots for  the smuon and stau pair searches are shown in  
Figures~\ref{fig:limit_2} and~\ref{fig:limit_3}, respectively.
Note that if the  LSP is the effectively massless 
gravitino, $\tilde{G}$ , then for 
prompt slepton decays to a lepton and a gravitino the experimental signature 
would be the same as that
for  \dslept\ with a massless neutralino.
In this case the limits given in 
Figures~\protect\ref{fig:limit_1}~--~\protect\ref{fig:limit_3}
for $m_{\tilde{\chi}^0_1} = 0$ may be interpreted as limits on the
decay $\tilde{\ell}^-\rightarrow \ell^-\tilde{G}$.

\begin{figure}[htbp]
 \epsfxsize=\textwidth 
% \epsffile[0 0 580 600]{figures/fig-limit-charg2.eps}
% \epsffile[0 0 580 600]{/home/graham/acopll/kumacs/limit_c2_new.eps}
\epsffile[0 0 580 600]{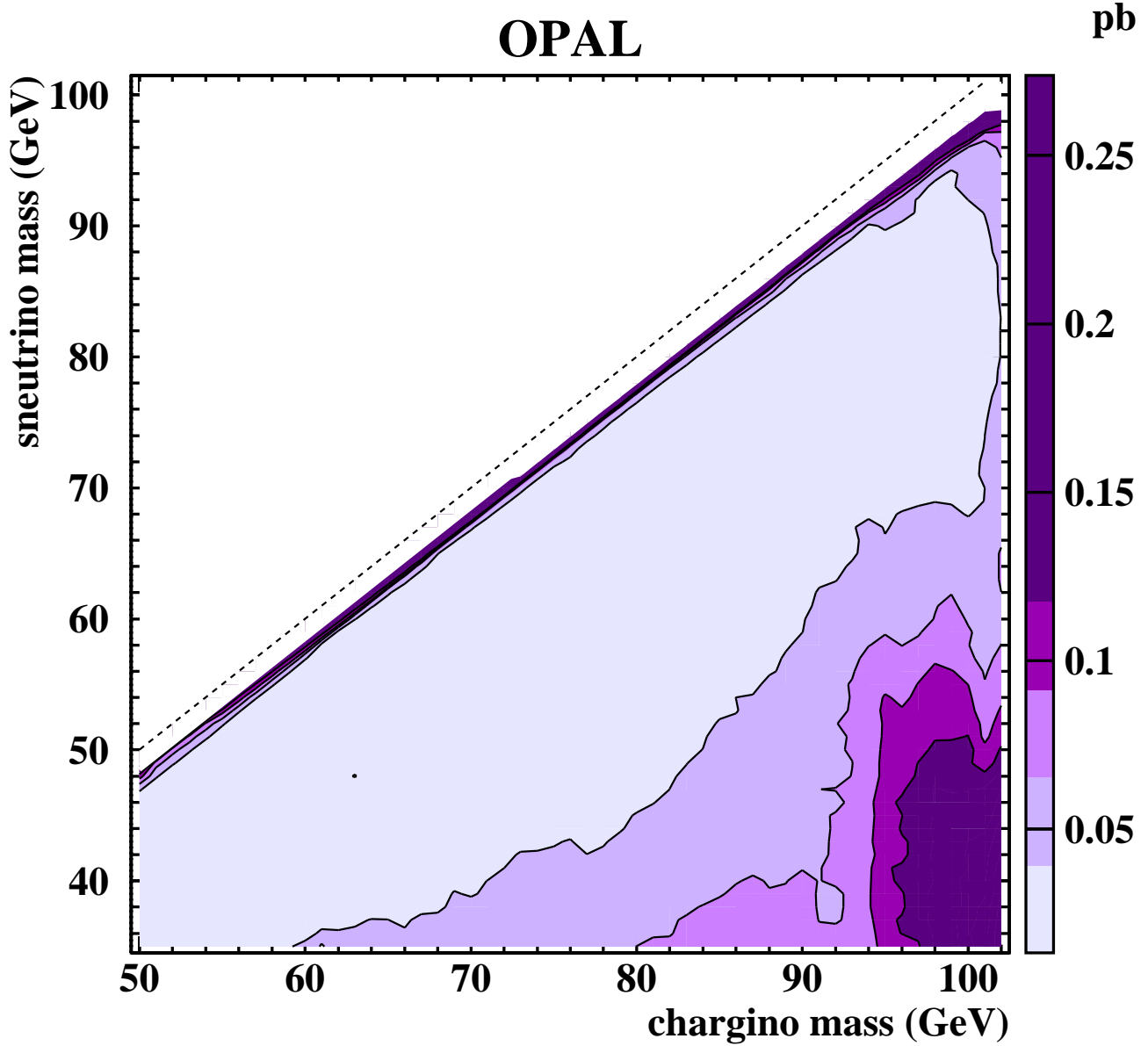}
 \caption{\sl 
Contours of the 95\% CL upper limits on the chargino pair
cross-section times branching ratio squared for 
$\chpm \rightarrow \ell^\pm \snu$ (2-body decay)
at $\protect\sqrt{s}$~=~208~GeV.
The  limits have been calculated for the 
case where the three sneutrino 
generations are mass degenerate. Only sneutrino masses above 35 GeV have 
been considered given constraints from the $\mathrm{Z}^0$ line-shape.
The limit is obtained by combining the 183 - 208~GeV data-sets 
assuming a $\beta/s$ dependence of the cross-section.
The kinematically allowed region lies below the dashed line.  The
unshaded region at very low \dm\ is experimentally inaccessible in
this search.
} 
\label{fig:limit_8}
\end{figure}

\begin{figure}[htbp]
 \epsfxsize=\textwidth 
% \epsffile[0 0 580 600]{figures/fig-limit-charg3.eps}
% \epsffile[0 0 580 600]{/home/graham/acopll/kumacs/limit_c3_new.eps}
\epsffile[0 0 580 600]{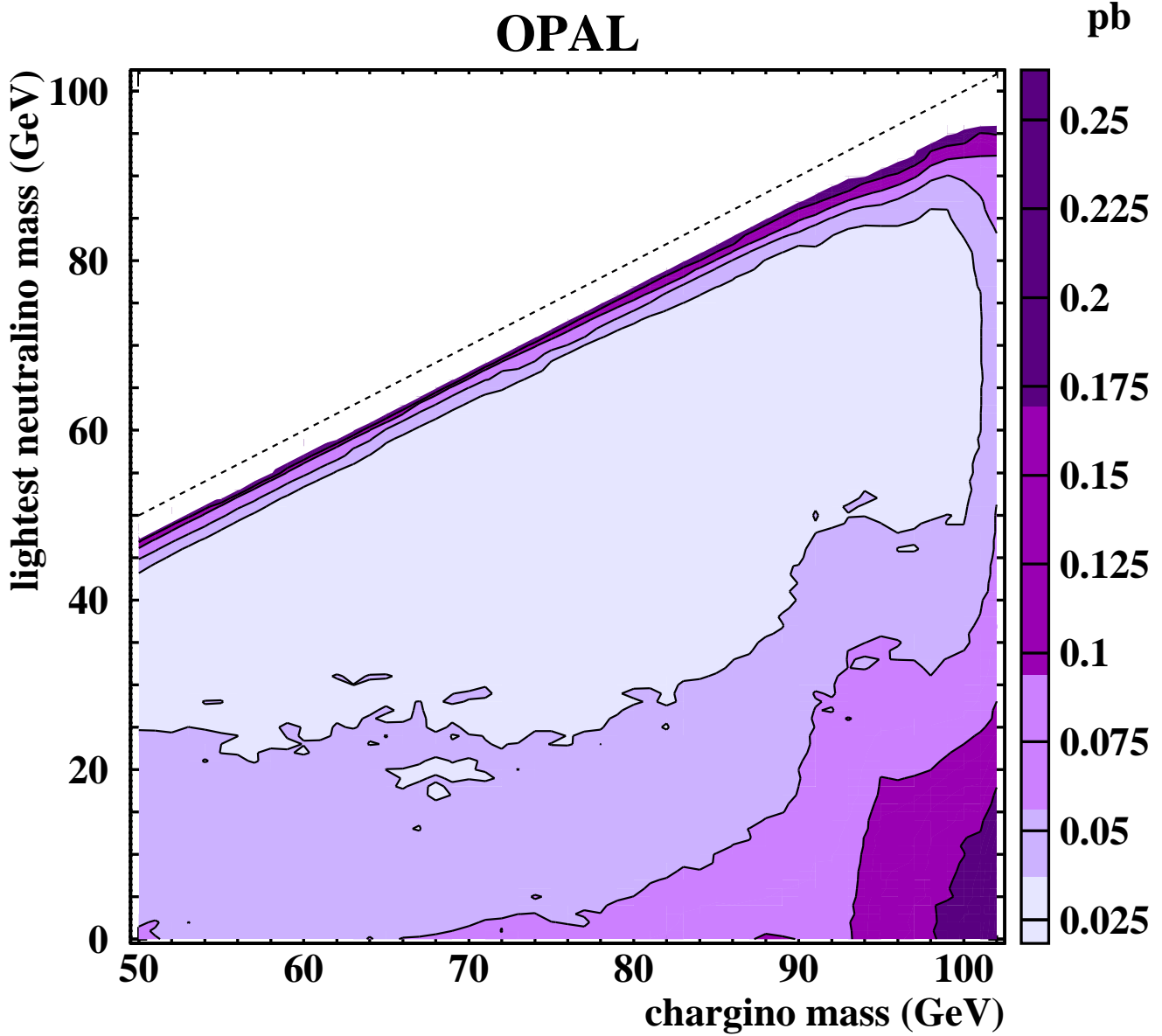}
 \caption{\sl 
Contours of the 95\% CL upper limits on the chargino pair
cross-section times branching ratio squared for 
$\chpm \rightarrow \ell^\pm \nu \chz$
 (3-body decay) at $\protect\sqrt{s}$~=~208~GeV.
The limit is obtained by combining the 183 - 208~GeV data-sets 
assuming a $\beta/s$ dependence of the cross-section.
The kinematically allowed region lies below the dashed line.  The
unshaded region at very low \dm\ is experimentally inaccessible in
this search.
} 
\label{fig:limit_4}
\end{figure}

The upper limit at 95\% CL on the chargino pair production 
cross-section multiplied by the branching
ratio squared for the decay \dchtwo\  (2-body decay)
is shown in Figure \ref{fig:limit_8}. 
The  limit has been calculated for the 
case where the three sneutrino 
generations are mass degenerate.
The upper limit at 95\% CL on the chargino pair production 
cross-section multiplied by the branching
ratio squared for the decay \dchthree\ (3-body decay)
is shown in Figure~\ref{fig:limit_4}. 

The upper limit at 95\% CL on the charged Higgs boson pair production 
cross-section multiplied by the branching ratio squared for the decay \dH\
is shown as a function of \mH\ as the solid histogram in Figure \ref{fig:limit_5}. 
%The limit is obtained by combining the data-sets 
%assuming the \mH\ and $\sqrt{s}$ 
%dependence of the cross-section predicted
%by {\sc HZHA}, which takes into account the effect of 
%initial state radiation.
The branching ratio for the decay \dH\ may be the dominant one for 
the charged Higgs masses explored with this data-set.
The dashed  line in Figure \ref{fig:limit_5} 
shows the prediction from {\sc HZHA}
at $\protect\sqrt{s}$~=~208~GeV
for a 100\% branching ratio for the decay \dH .
With this assumption we set a lower limit  at 95\% CL 
on \mH\ of 92.0~GeV.

%=======================================================================
\subsection{Expected Limits and Consistency with Expectation}
%=======================================================================
\label{sec:expect}
For each search, we provide a table showing quantitatively 
the signal efficiencies and the agreement 
of the data  
with the Standard Model background expectations for a number 
of values of $m$ and \dm\ .
Tables~\ref{tab-CL1},~\ref{tab-CL2},~\ref{tab-CL3},~\ref{tab-CL8}
,~\ref{tab-CL4} and~\ref{tab-CL5} 
give the values of the following quantities in the searches 
for selectrons, smuons, staus, charginos with two-body decay, 
charginos with three-body decay and charged Higgs, respectively:

\begin{enumerate}

\item
The signal efficiency for the general selection at 208~GeV with 
statistical error (the efficiencies 
at lower energies differ by less than 3\%).

\item
The 95\% CL upper limit on the 
cross-section multiplied by the 
branching ratio squared at 208~GeV, obtained by
combining the data from each centre-of-mass energy.

\item
The expected 95\% CL upper limit on the cross-section multiplied 
by the branching 
ratio squared in the absence of signal, \expsig .
This is calculated using an ensemble of 1000 \mc\ experiments to 
simulate the data. 
In each \mc\ experiment, the total number of candidates is
drawn from a Poisson distribution with mean equal to the number of events 
expected from the \sm . For each candidate, 
a value of $\LR$ is assigned, chosen randomly according to the 
expected $\LR$ distribution for \smp .
The expected limit at a given point in $m$ and \dm\ is the mean value of the
limit for the ensemble of simulated experiments.

\item
The confidence
level for consistency with the \sm , calculated as the fraction of the
simulated experiments for which the upper limit on the cross-section 
multiplied by the 
branching ratio squared is greater than or equal to the value calculated
using the actual data.  In the absence of signal, a CL of 
50\% is expected on average\footnote{Values of 100\% correspond
to points where there are no candidate events with non-zero \LR\ in the OPAL data.  In this
case, all toy \mc\ experiments will have a value of \signine\ equal to
or (if there are \mc\ candidates with non-zero \LR ) greater 
than the value for 
the data.}.

\end{enumerate}

Note that the number of candidate events with high values of 
relative likelihood varies greatly from search to search and 
with $m$ and \dm\ .
For the selectron and smuon searches where particular di-lepton identities 
are required and the kinematics of the lepton from the 
slepton decay are
measured directly (in contrast to events with tau leptons), relatively 
few events may be present at high relative likelihood.
For searches where the signal di-lepton identities are not that 
dissimilar to \dWW\ and a range of lepton momenta are expected, 
hundreds of events may potentially have high values of relative likelihood.

For some points in $m$ and \dm\ in these 
tables, 
the
confidence level for consistency with the \sm\ is relatively 
small (of order 1\%).  
The probability of getting a low confidence level for one or more points in 
$m$ and \dm\ for one or more of the search channels depends on
the degree of correlation among the different ($m$, \dm) points
and among the different channels.  The correlation between
adjacent points is strong when the momentum distributions for those points are
similar.  The momentum distributions vary slowly with
both $m$ and \dm\ when \dm\ is high (hence the clustering of low
confidence level values in Table~\ref{tab-CL2}), but vary considerably
with \dm\ when \dm\ is low.

This effect was investigated
by generating 1000 Monte Carlo experiments with the Standard Model 
Monte Carlo as described above.
For each experiment and each point on the ($m$, \dm) grid at 
which signal Monte Carlo has been generated, the CL was calculated.
It was found that 56\% of the experiments had at least one point with
a CL of 0.9\% or less; this is the lowest value observed in 
Tables~\ref{tab-CL1} to~\ref{tab-CL5}.

%calculating the cross-section limits for each of 1000 \mc\ 
%experiments in which 
%the data is simulated by randomly selected \smc\ events.  For 
%each experiment, 
%the number of events taken from a
%\mc\ sample simulating a given process is drawn from a Poisson
%distribution with mean equal to the number of events expected 
%for that process.
%For each experiment, the confidence level at each ($m$, \dm) point at
%which signal \mc\ has been generated was 
%calculated as already described, and the number
%of experiments for which a confidence level of 0.9\%
%\footnote{This is the lowest value of the confidence level in 
%Tables~\ref{tab-CL1} 
%to~\ref{tab-CL5}.} or less is obtained for 
%at least one point in $m$ and \dm\ in at least one search channel was 
%determined.  This was found to be the case for 56\% of the 1000 \mc\ 
%experiments.
%As a cross-check, taking the mean of the ensemble of limits obtained at %each 
%($m$, \dm) point for these simulated experiments was used as 
%an alternative to 
%the method described above to obtain \expsig .  The results were found 
%to be consistent.

%=======================================================================
\subsection{Limits on Slepton Masses}
%=======================================================================
\label{sec:mass}

\begin{figure}[htbp]
 \epsfxsize=\textwidth 
% \epsffile[0 0 550 550]{/home/graham/acopll/kumacs/smuon_br_new.eps}
\epsffile[0 0 550 550]{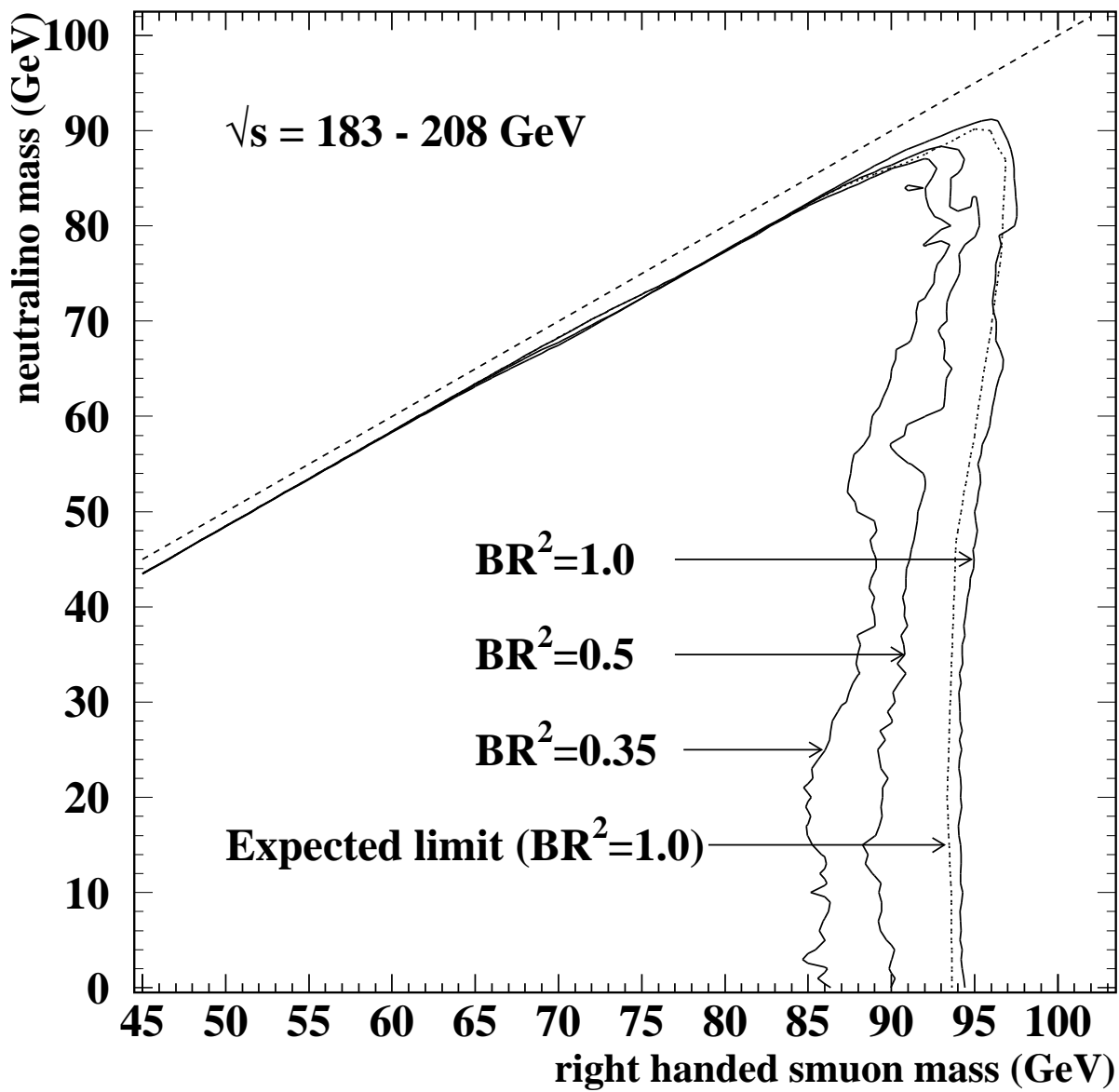}
 \caption{\sl 
95\% CL exclusion region for right-handed smuon pair 
production obtained by combining the $\protect\sqrt{s} = $ 183 - 208~GeV 
data-sets.
The limits are calculated for several values of
the branching ratio squared for 
$\smu^\pm_R \rightarrow  {\mu^\pm} \nt_1$ that are indicated in the figure.
Otherwise they have no supersymmetry model assumptions.
The kinematically allowed region lies below the dashed line.  The
expected limit for BR$^2$~=~1.0, calculated from \mc\ alone, is indicated by
the dash-dotted line.
} 
\label{fig-mssm_2}
\end{figure}

\begin{figure}[htbp]
 \epsfxsize=\textwidth 
% \epsffile[0 0 550 550]{/home/graham/acopll/kumacs/stau_br_new.eps}
\epsffile[0 0 550 550]{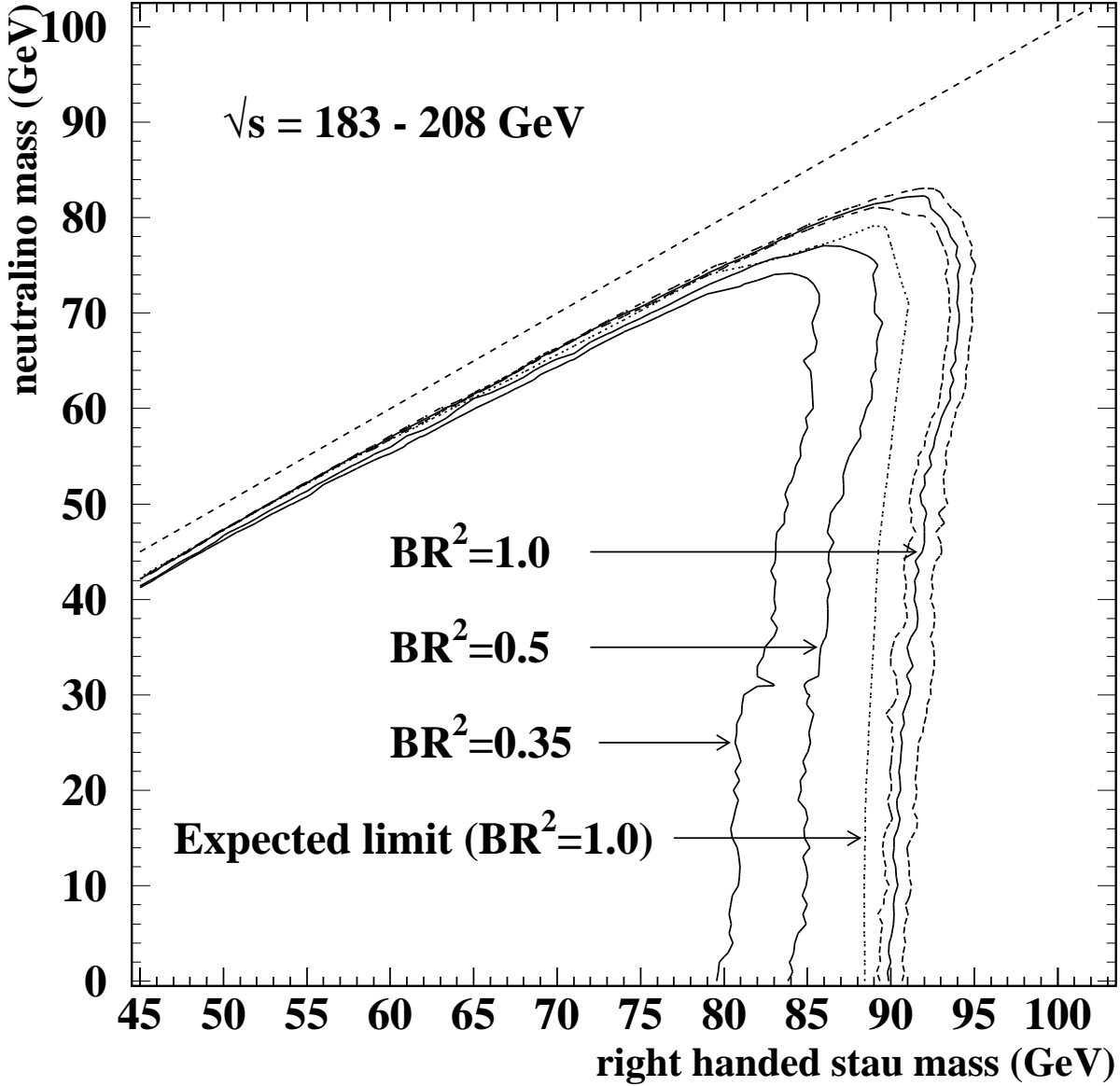}
 \caption{\sl 
95\% CL exclusion region for right-handed stau pair 
production obtained by combining the $\protect\sqrt{s} = $ 183 - 208~GeV 
data-sets.
The limits are calculated for several values of
the branching ratio squared for 
$\stau^\pm_R \rightarrow  {\tau^\pm} \nt_1$.
The selection efficiency for \staupair\ is calculated for the case
that the decay \dstau\ produces unpolarised $\tau^\pm$.
Otherwise the limits have no supersymmetry model assumptions.
The two broken lines adjacent to the limit for BR$^2$~=~1.0 
show the region in which this limit can
vary if stau mixing occurs (see text).
The kinematically allowed region is shown by the dashed
line.  The expected limit for BR$^2$~=~1.0, 
calculated from \mc\ alone, is represented by
the indicated broken line.
} 
\label{fig-mssm_3}
\end{figure}

We can use our data to set limits on the masses of right-handed 
sleptons\footnote{
The right-handed slepton is expected to be lighter
than the left-handed slepton. The
right-handed one tends (not generally valid for selectrons)
to
have a lower pair production cross-section, and so
conventionally limits are given for this (usually) conservative case.}
based on the expected right-handed slepton pair production 
cross-sections and
branching ratios.
The cross-sections were calculated using {\sc Susygen}
at each centre-of-mass energy 
and take into account initial state radiation.
In Figure~\ref{fig-mssm_2} we show  limits on right-handed smuons
 as a function of smuon mass and lightest
neutralino mass for several assumed values of
the branching ratio squared for $\smu^\pm_R \rightarrow  {\mu^\pm} \nt_1$.
The expected limit, calculated using Standard Model \mc\ only, is also 
shown 
for a branching ratio
of 100\%.
For a branching ratio $\smu^\pm_R \rightarrow  {\mu^\pm} \nt_1$ of
100\% and for a smuon-neutralino mass difference exceeding 4~GeV,
right-handed smuons are excluded at 95\% CL for 
masses below 94.0~GeV.
The  95\% CL upper limit on the  production of
right-handed \staupair\ multiplied by the
 branching ratio squared for $\stau^\pm_R \rightarrow  {\tau^\pm} \nt_1$
is shown in Figure~\ref{fig-mssm_3}.  The expected limit for a branching ratio
of 100\% is also shown.
For a branching ratio $\stau^\pm_R \rightarrow  {\tau^\pm} \nt_1$ of
100\% and for a stau-neutralino mass difference exceeding 8~GeV,
right-handed staus are excluded at 95\% CL for 
masses below 89.8~GeV.  No mixing 
between $\stau_L$ and $\stau_R$ is assumed.
%, but the cross-section limits are applicable to any degree of stau mixing.
However, the cross-section ratio 
$\sigma_{\stau_1^+ \stau_1^-}/\sigma_{\stau_R^+ \stau_R^-}$ at
$\roots \approx$ 208~GeV varies 
between 0.90 and 1.17 
%{\bf GWW need to check, think it is less different 
%at 208 GeV cf 189 GeV}, 
depending only on the mixing angle.  Using this
information, the limits shown in 
Figure~\ref{fig-mssm_3} can be applied to any degree of stau mixing by 
multiplying the predicted cross-section for $\stau_R^+ \stau_R^-$ by the 
value of $\sigma_{\stau_1^+ \stau_1^-}/\sigma_{\stau_R^+ \stau_R^-}$ 
corresponding to the mixing angle considered.  The two broken 
lines in 
Figure~\ref{fig-mssm_3} show the range of possible positions of the line 
defining the excluded region for a branching ratio 
$\stau^\pm_1 \rightarrow  {\tau^\pm} \nt_1$ of
100\% for any degree of stau mixing.

%{\bf GWW Need to revise} 
For the case of a massless 
neutralino (or gravitino) and 100\%
branching ratio, right-handed smuons and staus are excluded at 95\% CL for 
masses below 94.3~GeV and 89.8~GeV, respectively, and $\stau^\pm_1$ is
excluded at 95\% CL for masses below 88.7~GeV, for any degree of stau mixing.

\begin{figure}[htbp]
 \epsfxsize=\textwidth 
% \epsffile[0 0 550 550]{/home/graham/acopll/kumacs/mssm_selectron_new.eps}
\epsffile[0 0 550 550]{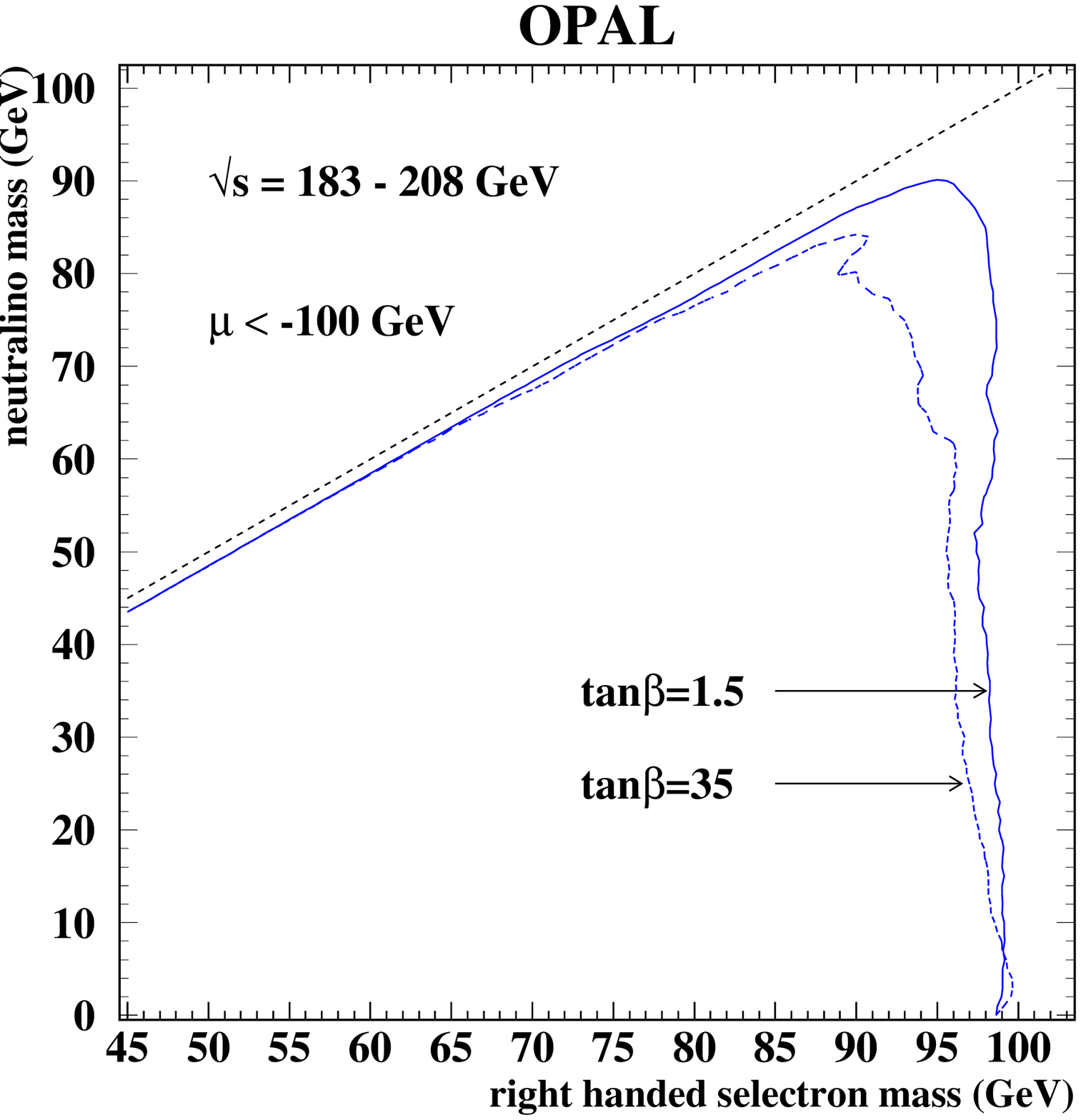}
 \caption{\sl 
For two values of $\tan{\beta}$ and $\mu < -100$~GeV,
95\% CL exclusion regions for right-handed selectron pairs within the MSSM, 
obtained by combining the $\protect\sqrt{s} = $ 183 - 208~GeV data-sets.
The excluded regions are calculated 
taking into account the 
predicted branching ratio for 
$\sele^\pm_R \rightarrow  {\mathrm{e}^\pm} \nt_1$.
The  gauge unification relation,
$M_1 =  \frac{5}{3} \tan^2 \theta_W M_2$, is assumed in calculating the
MSSM cross-sections and branching ratios.
The kinematically allowed region lies below the dashed line.
} 
\label{fig-mssm_1}
\end{figure}

\begin{figure}[htbp]
 \epsfxsize=\textwidth 
% \epsffile[0 0 550 550]{/home/graham/acopll/kumacs/mssm_smuon_new.eps}
\epsffile[0 0 550 550]{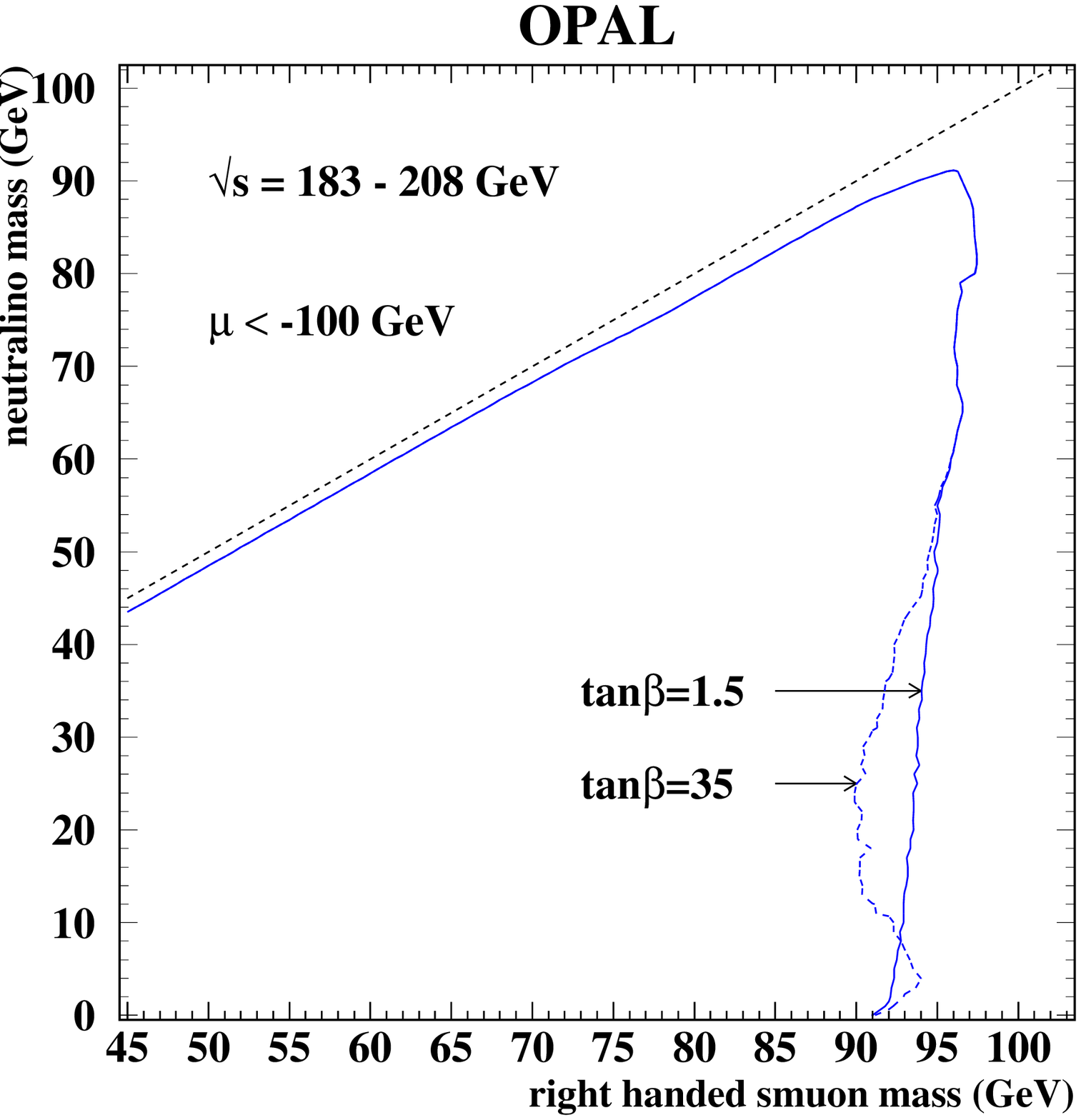}
 \caption{\sl 
For two values of $\tan{\beta}$ and $\mu < -100$~GeV,
 95\% CL exclusion regions for right-handed smuon pairs  within the MSSM,
obtained by combining the $\protect\sqrt{s} = $ 183 - 208~GeV data-sets.
The excluded regions are calculated 
taking into account the 
predicted branching ratio for $\smu^\pm_R \rightarrow  {\mu^\pm} \nt_1$.
The  gauge unification relation,
$M_1 =  \frac{5}{3} \tan^2 \theta_W M_2$, is assumed in calculating the
MSSM branching ratios.
The kinematically allowed region lies below the dashed line.
} 
\label{fig-mssm_2a}
\end{figure}

\begin{figure}[htbp]
 \epsfxsize=\textwidth 
% \epsffile[0 0 550 550]{/home/graham/acopll/kumacs/mssm_stau_new.eps}
\epsffile[0 0 550 550]{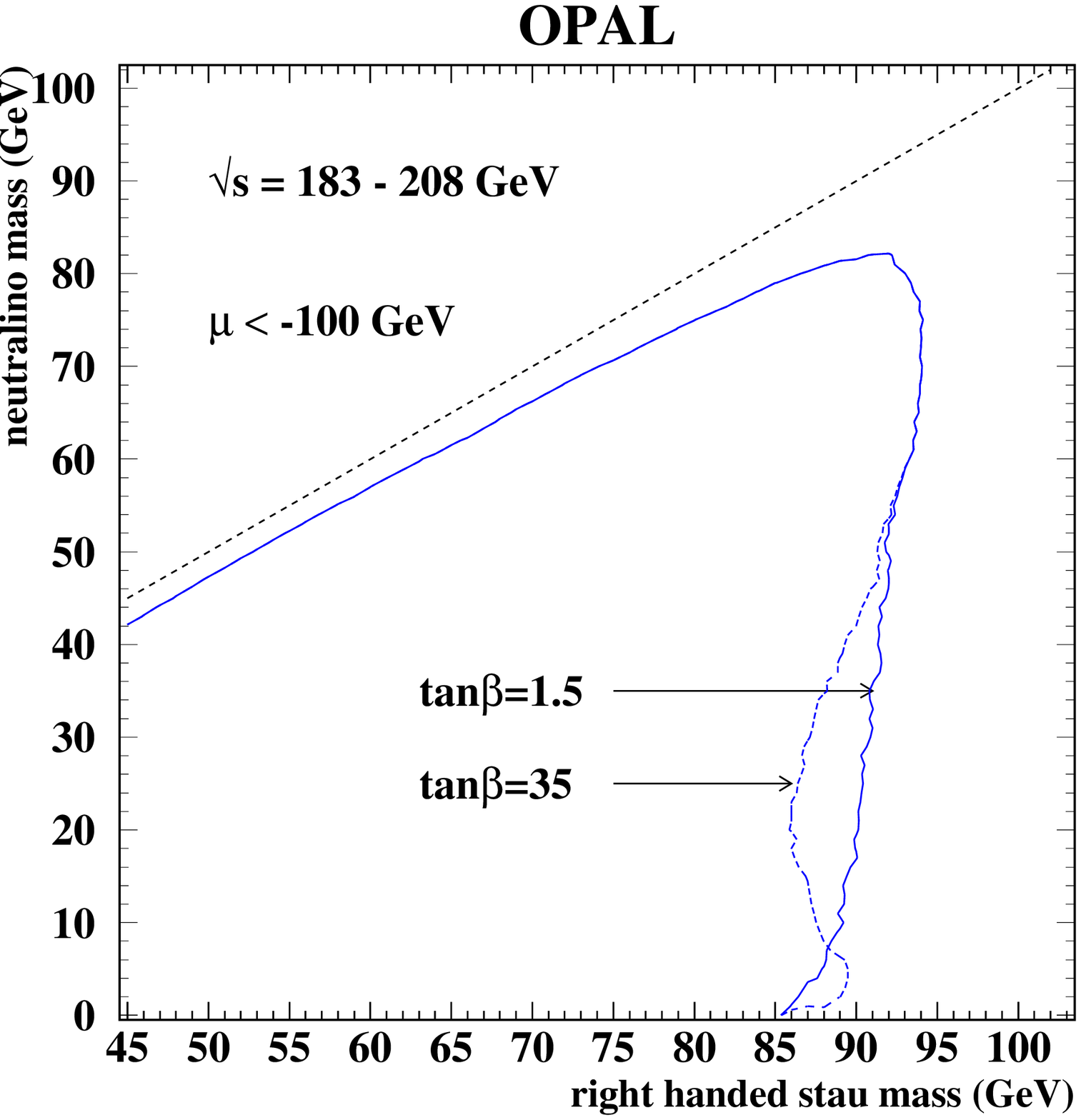}
 \caption{\sl 
For two values of $\tan{\beta}$ and $\mu < -100$~GeV,
 95\% CL exclusion regions for right-handed stau pairs within the MSSM,
obtained by combining the $\protect\sqrt{s} = $ 183 - 208~GeV data-sets.
The excluded regions are calculated 
taking into account the 
predicted branching ratio for $\stau^\pm_R \rightarrow  {\tau^\pm}
\nt_1$.
The  gauge unification relation,
$M_1 =  \frac{5}{3} \tan^2 \theta_W M_2$, is assumed in calculating the
MSSM branching ratios.
The selection efficiency for \staupair\ is calculated for the case
that the decay \dstau\ produces unpolarised $\tau^\pm$.
The kinematically allowed region lies below the dashed line.
} 
\label{fig-mssm_3a}
\end{figure}

An alternative approach is to set limits
taking into account the 
predicted cross-section and  branching ratio 
for specific choices of the parameters within the 
Minimal Supersymmetric Standard Model (MSSM)\footnote{
In particular 
 regions of the MSSM parameter space, the branching ratio for 
$\sell^\pm \rightarrow  {\ell^\pm} \nt_1$  
can be essentially zero 
as a result of competing cascade decays
and so 
it is not possible to provide general limits on sleptons within the MSSM
on the basis of this search alone.
The predicted cross-sections and  branching ratios within the MSSM 
are obtained using {\sc Susygen}
and are calculated with the gauge unification relation,
$M_1 =  \frac{5}{3} \tan^2 \theta_W M_2$.}.
For $\mu < -100$~GeV and for two
values of $\tan{\beta}$ (1.5 and 35),
Figures~\ref{fig-mssm_1},~\ref{fig-mssm_2a} and~\ref{fig-mssm_3a}
 show 95\% CL exclusion regions 
in the ($m_{\tilde{\ell}^\pm_{\mathrm{R}}}$, $m_{\nt_1}$) 
plane
for right-handed selectrons, smuons and staus, respectively.
For $\mu < -100$~GeV and $\tan{\beta}=1.5$, right-handed sleptons are 
excluded at
95\% CL as follows:
selectrons with masses below 97.5~GeV for \mbox{$\msele - \mchz > 11$}~GeV;
smuons with masses below 91.0~GeV for \mbox{$\msmu - \mchz > 3$}~GeV;
and staus with masses below 85.2~GeV for \mbox{$\mstau - \mchz > 6$}~GeV.

%=======================================================================
\subsection{Search for Unequal Mass Particle Production}
%=======================================================================
\label{sec:unequal}

All the search results described above are for the pair production
of new particles with equal mass.
New physics signals with unequal mass, 
such as the production of 
$\sele_L \sele_R$ in which the two selectrons have different mass and
each selectron decays to $\mathrm{e} \nt_1 $, 
can potentially be observed in the acoplanar di-lepton event sample.

We have examined the data
for di-electron, di-muon and arbitrary di-lepton identity events 
consistent with unequal mass particle production.
The search hypothesis is $\ee \rightarrow \mathrm{X} \mathrm{Y}$,
with subsequent decays, $\mathrm{X} \rightarrow \ell^{\pm} \mathrm{N}$ 
and $\mathrm{Y} \rightarrow \ell^{\mp} \mathrm{N}$, where X 
and Y are two massive charged particles, and N is an 
invisible particle such as the LSP.
The free parameters in the search are the 
masses of particles X, Y and N. The kinematics of 
producing two particles with different mass
constrain the lepton momenta to be within well defined ranges
depending on the new particle masses.
The data have been examined by scanning the three particle masses in
steps of 10 GeV for the three different di-lepton identity hypotheses 
and requiring the measured lepton momenta to be within the 
ranges specified by kinematics.

The consistency of the data with the Standard Model expectations is then 
examined. 
The lowest probability mass hypothesis occurs for di-leptons of arbitrary flavour
with particle mass 
$m_{\mathrm{X}} =$~130 GeV, 
$m_{\mathrm{Y}} =$~60 GeV and
$m_{\mathrm{N}} =$~20 GeV, where 242 events 
are observed with 197.3 expected from Standard Model sources (Poisson
probability of 0.7\%).
From an ensemble of 1000 Standard Model Monte Carlo experiments it is 
found that 12.5\% of the experiments had at least one mass hypothesis with a Poisson
probability of 0.7\% or lower.
In general, the agreement with expectation is good. In conclusion, we have 
explored this potential weakness of the standard equal mass pair production 
search and found no significant evidence for unequal 
mass particle production.

%=======================================================================
\section{Summary and Conclusions}
%=======================================================================
 
A selection of di-lepton events with significant missing transverse momentum 
is performed using a data sample with an integrated luminosity 
of 680~pb$^{-1}$ collected 
at e$^+$e$^-$ centre-of-mass energies ranging from 183 to 208~GeV.
The observed number of events, 1317, 
the dependence on centre-of-mass energy and the event properties
are consistent with expectations from Standard Model processes, 
dominantly arising from \wpair\ production with both W bosons 
decaying
leptonically.

Discrimination techniques are employed to search for the pair 
production of charged scalar leptons, leptonically decaying charged Higgs 
bosons and charginos that decay to produce a charged lepton and invisible 
particles.
No evidence for new phenomena is apparent. 
%largely 
%model independent 
Upper limits 
on the production cross-section multiplied by the branching ratio squared
for each new physics process are presented in a manner intended
to minimise the number of model assumptions.

Assuming a 100\% branching ratio for the decay
$\sell^\pm_R \rightarrow  {\ell^\pm} \nt_1$, we exclude at 95\% CL:
right-handed smuons with masses below 94.0~GeV for 
\mbox{$\msmu - \mchz > 4$}~GeV and
right-handed staus with masses below 89.8~GeV for 
\mbox{$\mstau - \mchz > 8$}~GeV.
Right-handed selectrons are excluded at 95\% CL for 
masses below 97.5~GeV for \mbox{$\msele - \mchz > 11$}~GeV
within the framework of the
MSSM assuming
$\mu < -100$~GeV and $\tan{\beta} = 1.5$.
Charged Higgs bosons are excluded at 95\%~CL for masses below 92.0~GeV,
assuming a 100\% branching ratio for the decay \dH .

%The cross-section times branching ratio squared limits 
%from the selectron, smuon and two-body chargino searches
%presented here 
%are used in the interpretation of the results of \cite{OPAL_chargino_189}
%in terms of mass limits on charginos and neutralinos.

\bigskip\bigskip
%=======================================================================
\noindent {\Large\bf Acknowledgements}
%=======================================================================
\par
We particularly wish to thank the SL Division for the efficient operation
of the LEP accelerator at all energies and for their close cooperation with
our experimental group.  In addition to the support staff at our own
institutions we are pleased to acknowledge the  \\
Department of Energy, USA, \\
National Science Foundation, USA, \\
Particle Physics and Astronomy Research Council, UK, \\
Natural Sciences and Engineering Research Council, Canada, \\
Israel Science Foundation, administered by the Israel
Academy of Science and Humanities, \\
Benoziyo Center for High Energy Physics,\\
Japanese Ministry of Education, Culture, Sports, Science and
Technology (MEXT) and a grant under the MEXT International
Science Research Program,\\
Japanese Society for the Promotion of Science (JSPS),\\
German Israeli Bi-national Science Foundation (GIF), \\
Bundesministerium f\"ur Bildung und Forschung, Germany, \\
National Research Council of Canada, \\
Hungarian Foundation for Scientific Research, OTKA T-038240, 
and T-042864,\\
The NWO/NATO Fund for Scientific Research, the Netherlands.\\

%=======================================================================
%       References
%=======================================================================

\clearpage

\begin{center}
\begin{table}[htb]
\renewcommand{\baselinestretch}{1.0}
\begin{center}
\hspace*{-2cm}
\vskip 6cm
\rotatebox{90}{
\begin{minipage}[htb]{\textwidth}
{\scriptsize
\begin{tabular}{|l||c|c|c|c|c|c|c|c|c|c|c|c|c|}
\hline
 \dm  & \multicolumn{13}{c|}{\msele\ (GeV)\ } \\
\cline{2-14}
(GeV)& 45 & 50 & 55 & 60 & 65 & 70 & 75 & 80 & 85 & 90 & 94 & 99 & 103\\
\hline\hline  \multicolumn{14}{||l||}{General Selection Efficiency ( \% )} \\
\hline
2     
&  5.0$\pm$0.7
&  3.8$\pm$0.6
&  4.5$\pm$0.7
&  2.2$\pm$0.5
&  2.0$\pm$0.4
&  1.6$\pm$0.4
&  1.2$\pm$0.3
&  0.8$\pm$0.3
&  0.6$\pm$0.2
&  0.5$\pm$0.2
&  0.5$\pm$0.2
&  0.1$\pm$0.1
&  0.0$\pm$0.0
\\
2.5   
&  5.0$\pm$0.7
&  3.8$\pm$0.6
&  4.5$\pm$0.7
&  2.2$\pm$0.5
&  2.0$\pm$0.4
&  1.6$\pm$0.4
&  1.2$\pm$0.3
&  0.8$\pm$0.3
&  0.6$\pm$0.2
&  0.5$\pm$0.2
&  0.5$\pm$0.2
&  0.1$\pm$0.1
&  0.0$\pm$0.0
\\
5     
& 40.7$\pm$1.6
& 42.9$\pm$1.6
& 41.9$\pm$1.6
& 44.5$\pm$1.6
& 45.0$\pm$1.6
& 43.5$\pm$1.6
& 43.1$\pm$1.6
& 39.9$\pm$1.5
& 41.6$\pm$1.6
& 44.1$\pm$1.6
& 43.1$\pm$1.6
& 42.3$\pm$1.6
& 42.4$\pm$1.6
\\
10    
& 60.9$\pm$1.5
& 64.8$\pm$1.5
& 65.4$\pm$1.5
& 66.4$\pm$1.5
& 67.7$\pm$1.5
& 67.2$\pm$1.5
& 65.9$\pm$1.5
& 67.8$\pm$1.5
& 66.0$\pm$1.5
& 67.4$\pm$1.5
& 67.4$\pm$1.5
& 66.6$\pm$1.5
& 67.8$\pm$1.5
\\
20    
& 74.8$\pm$1.4
& 76.3$\pm$1.3
& 77.4$\pm$1.3
& 79.1$\pm$1.3
& 81.2$\pm$1.2
& 82.4$\pm$1.2
& 82.5$\pm$1.2
& 81.1$\pm$1.2
& 81.4$\pm$1.2
& 81.3$\pm$1.2
& 82.1$\pm$1.2
& 83.3$\pm$1.2
& 80.2$\pm$1.3
\\
$m$/2 
& 75.0$\pm$1.4
& 78.6$\pm$1.3
& 79.4$\pm$1.3
& 83.2$\pm$1.2
& 85.9$\pm$1.1
& 84.4$\pm$1.1
& 85.5$\pm$1.1
& 87.8$\pm$1.0
& 88.2$\pm$1.0
& 87.8$\pm$1.0
& 88.5$\pm$1.0
& 88.5$\pm$1.0
& 89.4$\pm$1.0
\\
$m$-20
& 74.9$\pm$1.4
& 81.7$\pm$1.2
& 82.1$\pm$1.2
& 83.6$\pm$1.2
& 84.2$\pm$1.2
& 84.9$\pm$1.1
& 88.4$\pm$1.0
& 86.0$\pm$1.1
& 88.1$\pm$1.0
& 89.9$\pm$1.0
& 90.9$\pm$0.9
& 90.2$\pm$0.9
& 90.8$\pm$0.9
\\
$m$-10
& 74.6$\pm$1.4
& 79.3$\pm$1.3
& 79.8$\pm$1.3
& 82.0$\pm$1.2
& 85.2$\pm$1.1
& 85.2$\pm$1.1
& 89.1$\pm$1.0
& 89.2$\pm$1.0
& 88.7$\pm$1.0
& 90.7$\pm$0.9
& 90.1$\pm$0.9
& 90.3$\pm$0.9
& 90.5$\pm$0.9
\\
$m$   
& 69.5$\pm$1.5
& 77.3$\pm$1.3
& 80.8$\pm$1.2
& 83.1$\pm$1.2
& 85.0$\pm$1.1
& 88.2$\pm$1.0
& 87.3$\pm$1.1
& 89.2$\pm$1.0
& 89.5$\pm$1.0
& 88.5$\pm$1.0
& 91.1$\pm$0.9
& 90.4$\pm$0.9
& 91.1$\pm$0.9
\\

\hline\hline
\multicolumn{14}{|l|}{95\% CL upper limit on cross-section times
 $BR^2$(\selem$\rightarrow$e$\tilde{\chi}^0_1$) (fb)} \\
\hline
2     
&     82.1
&     86.6
&     99.4
&    139
&    113
&    160
&    294
&    412
&   1230
&   2530
&   5020
& --
& --
\\
2.5   
&     34.0
&     33.7
&     36.3
&     39.6
&     43.5
&     35.4
&     43.7
&     58.6
&     85.9
&    136
&    256
&    975
& --
\\
5     
&     13.8
&     11.6
&     12.1
&     12.3
&      9.6
&     15.0
&     10.3
&     11.9
&     12.7
&     21.4
&     23.2
&     36.9
&    270
\\
10    
&     14.0
&     17.2
&     12.9
&      9.7
&     10.2
&     10.3
&     10.0
&     10.2
&     10.6
&     11.1
&     14.9
&     23.9
&    174
\\
20    
&     26.9
&     23.4
&     21.8
&     19.1
&     16.6
&     21.0
&     21.5
&     16.5
&     13.2
&     13.0
&     14.1
&     22.6
&    162
\\
$m$/2 
&     34.4
&     35.0
&     37.2
&     34.6
&     34.2
&     27.9
&     25.7
&     31.0
&     31.2
&     37.5
&     40.3
&     75.6
&    282
\\
$m$-20
&     39.6
&     43.1
&     49.7
&     45.7
&     47.6
&     53.7
&     60.8
&     47.7
&     35.2
&     30.2
&     34.3
&     60.1
&    243
\\
$m$-10
&     48.3
&     56.3
&     56.1
&     56.5
&     49.4
&     48.1
&     61.7
&     54.8
&     42.2
&     33.9
&     40.3
&     52.9
&    228
\\
$m$   
&     56.0
&     59.3
&     61.3
&     52.0
&     51.2
&     46.7
&     53.9
&     62.9
&     39.7
&     36.5
&     37.9
&     54.8
&    229
\\

\hline\hline
\multicolumn{14}{|l|}{expected upper limit on cross-section times
 $BR^2$(\selem$\rightarrow$e$\tilde{\chi}^0_1$) (fb)} \\
\hline
2     
&     78.1
&     83.2
&     98.5
&    122
&    143
&    192
&    342
&    474
&   1360
&   2730
&   5280
& --
& --
\\
2.5   
&     34.4
&     38.1
&     39.8
&     42.0
&     43.4
&     46.8
&     56.0
&     72.8
&     98.9
&    150
&    275
&   1010
& --
\\
5     
&     15.6
&     15.4
&     15.5
&     15.4
&     14.4
&     15.0
&     15.6
&     15.9
&     16.4
&     20.0
&     27.0
&     41.1
&    274
\\
10    
&     18.0
&     15.6
&     14.4
&     13.4
&     12.7
&     12.6
&     12.1
&     12.4
&     13.2
&     16.0
&     19.9
&     29.4
&    188
\\
20    
&     29.9
&     26.7
&     24.1
&     22.1
&     21.3
&     18.6
&     17.6
&     17.2
&     17.7
&     19.3
&     21.8
&     27.3
&    171
\\
$m$/2 
&     34.0
&     32.9
&     32.2
&     32.8
&     31.9
&     32.9
&     33.4
&     35.3
&     38.9
&     42.1
&     43.7
&     45.9
&    191
\\
$m$-20
&     36.0
&     37.1
&     37.8
&     39.1
&     40.2
&     42.6
&     44.8
&     48.4
&     48.3
&     49.1
&     50.8
&     54.9
&    219
\\
$m$-10
&     38.0
&     38.3
&     37.2
&     38.5
&     40.3
&     44.6
&     46.4
&     49.5
&     48.8
&     47.7
&     51.1
&     55.0
&    218
\\
$m$   
&     38.9
&     37.5
&     37.9
&     38.9
&     40.8
&     43.0
&     47.5
&     49.7
&     47.7
&     48.8
&     49.7
&     53.9
&    216
\\

\hline\hline
\multicolumn{14}{|l|}{CL for consistency with SM (\%)} \\
\hline
2     
&  33.4
&  34.1
&  39.3
&  24.0
& 100.0
& 100.0
& 100.0
& 100.0
& 100.0
& 100.0
& 100.0
& --
& --
\\
2.5   
&  42.6
&  57.2
&  53.4
&  46.4
&  42.8
&  63.1
& 100.0
& 100.0
& 100.0
& 100.0
& 100.0
& 100.0
& --
\\
5     
&  57.7
&  73.5
&  72.8
&  68.0
&  85.2
&  44.5
&  87.2
&  73.9
&  71.2
&  35.4
& 100.0
& 100.0
& 100.0
\\
10    
&  72.9
&  32.2
&  56.0
&  81.0
&  67.8
&  66.8
&  63.3
&  69.0
&  68.6
&  88.9
& 100.0
&  66.3
& 100.0
\\
20    
&  54.3
&  59.4
&  54.8
&  58.4
&  69.9
&  30.6
&  22.8
&  47.6
&  76.0
&  87.4
&  95.3
&  64.2
&  40.1
\\
$m$/2 
&  42.2
&  36.2
&  26.4
&  37.2
&  36.1
&  63.4
&  74.5
&  58.4
&  67.4
&  55.1
&  52.5
&   5.3
&   7.2
\\
$m$-20
&  32.3
&  27.5
&  17.5
&  28.1
&  24.9
&  19.6
&  14.9
&  45.6
&  79.5
&  91.4
&  85.1
&  32.3
&  27.1
\\
$m$-10
&  18.8
&   8.4
&   7.3
&   9.4
&  20.9
&  34.5
&  14.2
&  32.4
&  62.0
&  79.6
&  70.4
&  45.9
&  33.6
\\
$m$   
&   9.9
&   5.3
&   4.4
&  16.3
&  21.5
&  33.7
&  30.5
&  18.8
&  65.6
&  75.5
&  73.6
&  42.0
&  34.2
\\

\hline
\end{tabular}
\renewcommand{\baselinestretch}{1.5}
%\caption[Selectron results summary table]{
%General Selection efficiency, 95~\% CL upper limit cross-section times
%braching ratio at $\sqrt{s}=$~208~GeV, expected upper limit, and CL
%for consistency with 
%\sm\ processes in the search for \selepair\ production at 
%$\sqrt{s}$ = 183-208~GeV, for different values of \msele\ and \dm\ .
%\label{tab-results-sele}
%}
}
\end{minipage}}
\caption{Selectron search results.\label{tab-CL1}}
\end{center}
\end{table}
\end{center}

\clearpage

\begin{center}
\begin{table}[htb]
\renewcommand{\baselinestretch}{1.0}
\begin{center}
\hspace*{-2cm}
\vskip 6cm
\rotatebox{90}{
\begin{minipage}[htb]{\textwidth}
{\scriptsize
\begin{tabular}{|l||c|c|c|c|c|c|c|c|c|c|c|c|c|}
\hline
 \dm  & \multicolumn{13}{c|}{\msmu\ (GeV)\ } \\
\cline{2-14}
(GeV)& 45 & 50 & 55 & 60 & 65 & 70 & 75 & 80 & 85 & 90 & 94 & 99 & 103\\
\hline\hline  \multicolumn{13}{|l|}{General Selection Efficiency ( \% )} \\
\hline
2     
&  8.9$\pm$0.9
&  8.5$\pm$0.9
&  7.0$\pm$0.8
&  4.3$\pm$0.6
&  5.1$\pm$0.7
&  2.7$\pm$0.5
&  2.0$\pm$0.4
&  0.5$\pm$0.2
&  0.6$\pm$0.2
&  0.2$\pm$0.1
&  0.3$\pm$0.2
&  0.1$\pm$0.1
&  0.0$\pm$0.0
\\
2.5   
&  8.9$\pm$0.9
&  8.5$\pm$0.9
&  7.0$\pm$0.8
&  4.3$\pm$0.6
&  5.1$\pm$0.7
&  2.7$\pm$0.5
&  2.0$\pm$0.4
&  0.5$\pm$0.2
&  0.6$\pm$0.2
&  0.2$\pm$0.1
&  0.3$\pm$0.2
&  0.1$\pm$0.1
&  0.0$\pm$0.0
\\
5     
& 53.1$\pm$1.6
& 53.3$\pm$1.6
& 50.7$\pm$1.6
& 52.0$\pm$1.6
& 51.3$\pm$1.6
& 50.6$\pm$1.6
& 47.0$\pm$1.6
& 48.2$\pm$1.6
& 48.4$\pm$1.6
& 48.9$\pm$1.6
& 47.4$\pm$1.6
& 47.4$\pm$1.6
& 46.4$\pm$1.6
\\
10    
& 71.1$\pm$1.4
& 73.4$\pm$1.4
& 70.6$\pm$1.4
& 71.8$\pm$1.4
& 70.1$\pm$1.4
& 69.6$\pm$1.5
& 70.3$\pm$1.4
& 65.9$\pm$1.5
& 73.3$\pm$1.4
& 70.9$\pm$1.4
& 69.9$\pm$1.5
& 70.7$\pm$1.4
& 70.0$\pm$1.4
\\
20    
& 82.6$\pm$1.2
& 82.7$\pm$1.2
& 84.4$\pm$1.1
& 83.8$\pm$1.2
& 84.9$\pm$1.1
& 83.4$\pm$1.2
& 85.7$\pm$1.1
& 82.1$\pm$1.2
& 84.7$\pm$1.1
& 83.2$\pm$1.2
& 83.3$\pm$1.2
& 83.4$\pm$1.2
& 83.5$\pm$1.2
\\
$m$/2 
& 84.0$\pm$1.2
& 85.8$\pm$1.1
& 86.5$\pm$1.1
& 87.8$\pm$1.0
& 90.3$\pm$0.9
& 88.2$\pm$1.0
& 89.2$\pm$1.0
& 91.5$\pm$0.9
& 92.1$\pm$0.9
& 91.3$\pm$0.9
& 91.3$\pm$0.9
& 92.7$\pm$0.8
& 92.5$\pm$0.8
\\
$m$-20
& 83.4$\pm$1.2
& 87.1$\pm$1.1
& 89.4$\pm$1.0
& 89.6$\pm$1.0
& 90.0$\pm$0.9
& 91.3$\pm$0.9
& 92.6$\pm$0.8
& 93.4$\pm$0.8
& 92.4$\pm$0.8
& 93.5$\pm$0.8
& 93.3$\pm$0.8
& 92.9$\pm$0.8
& 93.3$\pm$0.8
\\
$m$-10
& 88.6$\pm$1.0
& 87.9$\pm$1.0
& 89.2$\pm$1.0
& 90.0$\pm$0.9
& 91.8$\pm$0.9
& 91.3$\pm$0.9
& 93.0$\pm$0.8
& 90.1$\pm$0.9
& 93.2$\pm$0.8
& 93.3$\pm$0.8
& 93.5$\pm$0.8
& 94.2$\pm$0.7
& 94.6$\pm$0.7
\\
$m$   
& 88.6$\pm$1.0
& 88.8$\pm$1.0
& 88.1$\pm$1.0
& 91.3$\pm$0.9
& 92.1$\pm$0.9
& 93.3$\pm$0.8
& 91.6$\pm$0.9
& 93.5$\pm$0.8
& 93.6$\pm$0.8
& 92.8$\pm$0.8
& 92.1$\pm$0.9
& 93.5$\pm$0.8
& 94.5$\pm$0.7
\\

\hline\hline
\multicolumn{14}{|l|}{95\% CL upper limit on cross-section times
 $BR^2$(\smum$\rightarrow\mu\tilde{\chi}^0_1$) (fb)} \\
\hline
2     
&     49.6
&     43.4
&     52.6
&     64.5
&     88.0
&    152
&    242
&    522
&   1380
&   3410
&   4820
& --
& --
\\
2.5   
&     28.2
&     23.0
&     23.6
&     26.7
&     30.1
&     34.3
&     39.6
&     52.7
&     61.1
&    108
&    206
&    584
&   8850
\\
5     
&     11.9
&     12.1
&     10.0
&     10.1
&     10.4
&     10.9
&     11.6
&     10.0
&     11.6
&     14.9
&     20.2
&     33.3
&    245
\\
10    
&     14.4
&     11.1
&      8.4
&      7.8
&      7.1
&      7.8
&      6.9
&      8.3
&     11.3
&     14.6
&     21.2
&     23.7
&    173
\\
20    
&     21.2
&     16.4
&     13.3
&     12.4
&     12.2
&     11.6
&     11.3
&     10.1
&     14.3
&     16.4
&     19.6
&     21.9
&    146
\\
$m$/2 
&     25.3
&     20.6
&     18.9
&     28.2
&     25.2
&     34.6
&     29.7
&     34.5
&     22.1
&     26.1
&     30.6
&     49.6
&    200
\\
$m$-20
&     30.4
&     28.1
&     31.9
&     36.5
&     38.7
&     49.7
&     41.0
&     36.1
&     35.2
&     35.6
&     36.0
&     53.9
&    261
\\
$m$-10
&     28.5
&     35.5
&     36.7
&     43.0
&     45.3
&     50.1
&     43.9
&     34.6
&     35.7
&     34.9
&     35.2
&     50.8
&    241
\\
$m$   
&     31.7
&     33.8
&     37.0
&     43.8
&     44.0
&     46.4
&     36.2
&     37.9
&     32.3
&     32.3
&     34.0
&     48.4
&    222
\\

\hline\hline
\multicolumn{14}{|l|}{expected upper limit on cross-section times
 $BR^2$(\smum$\rightarrow\mu\tilde{\chi}^0_1$) (fb)} \\
\hline
2     
&     44.3
&     46.5
&     56.1
&     67.2
&     93.0
&    145
&    229
&    494
&   1600
&   3580
&   4970
& --
& --
\\
2.5   
&     25.5
&     25.1
&     27.0
&     28.8
&     32.1
&     36.0
&     43.1
&     52.3
&     72.8
&    124
&    231
&    601
&   8870
\\
5     
&     12.9
&     12.8
&     12.8
&     12.1
&     12.6
&     12.9
&     12.9
&     13.5
&     15.8
&     19.7
&     25.1
&     38.6
&    267
\\
10    
&     13.3
&     12.7
&     11.8
&     11.5
&     11.2
&     10.7
&     10.6
&     11.0
&     11.7
&     13.6
&     17.3
&     26.1
&    177
\\
20    
&     20.8
&     19.7
&     18.9
&     17.7
&     17.2
&     16.6
&     15.9
&     15.8
&     16.0
&     17.6
&     20.2
&     26.8
&    160
\\
$m$/2 
&     22.6
&     22.9
&     23.4
&     24.6
&     24.7
&     26.8
&     27.6
&     29.6
&     32.6
&     36.4
&     38.6
&     44.5
&    196
\\
$m$-20
&     23.3
&     25.1
&     25.9
&     26.8
&     28.6
&     29.4
&     32.1
&     35.2
&     37.3
&     39.9
&     42.2
&     47.9
&    209
\\
$m$-10
&     24.0
&     24.5
&     24.8
&     25.8
&     27.8
&     29.4
&     32.9
&     35.2
&     35.5
&     39.0
&     40.2
&     46.6
&    222
\\
$m$   
&     23.8
&     23.2
&     24.3
&     26.0
&     27.3
&     29.5
&     32.0
&     34.9
&     35.8
&     37.6
&     40.3
&     46.5
&    218
\\

\hline\hline
\multicolumn{14}{|l|}{CL for consistency with SM (\%)} \\
\hline
2     
&  29.5
&  51.4
&  55.8
&  53.4
&  52.9
&  44.5
&  41.7
&  43.5
& 100.0
& 100.0
& 100.0
& --
& --
\\
2.5   
&  32.3
&  59.9
&  63.4
&  61.7
&  51.1
&  60.6
&  59.8
&  47.7
& 100.0
& 100.0
& 100.0
& 100.0
& 100.0
\\
5     
&  52.0
&  47.7
&  76.7
&  71.4
&  71.8
&  70.8
&  61.9
&  68.7
&  66.5
& 100.0
& 100.0
& 100.0
& 100.0
\\
10    
&  31.9
&  57.9
&  80.2
&  84.9
&  91.3
&  77.9
&  92.0
&  71.2
&  49.8
&  35.9
&  19.7
& 100.0
& 100.0
\\
20    
&  40.9
&  62.6
&  81.9
&  83.7
&  82.1
&  83.0
&  80.2
&  87.4
&  54.5
&  51.1
&  44.3
&  70.4
&  75.4
\\
$m$/2 
&  30.6
&  54.6
&  68.9
&  29.1
&  41.7
&  18.4
&  34.7
&  26.9
&  83.5
&  80.0
&  69.4
&  29.7
&  39.2
\\
$m$-20
&  17.2
&  29.3
&  23.3
&  13.5
&  14.5
&   4.4
&  18.7
&  42.6
&  49.7
&  56.1
&  61.9
&  30.1
&  17.1
\\
$m$-10
&  25.8
&   9.9
&   8.1
&   3.4
&   4.9
&   3.0
&  15.5
&  45.9
&  42.4
&  56.5
&  59.3
&  31.6
&  29.8
\\
$m$   
&  16.6
&   9.6
&   7.6
&   3.5
&   5.6
&   6.9
&  29.0
&  34.7
&  55.6
&  59.1
&  63.2
&  36.5
&  39.5
\\

\hline
\end{tabular}
\renewcommand{\baselinestretch}{1.5}
%\caption[Smuon results summary table]{
%General Selection efficiency, 95~\% CL upper limit cross-section times
%braching ratio at $\sqrt{s}=$~208~GeV, expected upper limit, and CL for consistency with 
%\sm\ processes in the search for \smupair\ production at 
%$\sqrt{s}$ = 183-208~GeV, for different values of \msmu\ and \dm\ .
%\label{tab-results-smu}
%}
}
\end{minipage}}
\caption{Smuon search results.\label{tab-CL2}}
\end{center}
\end{table}
\end{center}

\clearpage

\begin{center}
\begin{table}[htb]
\renewcommand{\baselinestretch}{1.0}
\begin{center}
\hspace*{-2cm}
\vskip 6cm
\rotatebox{90}{
\begin{minipage}[htb]{\textwidth}
{\scriptsize
\begin{tabular}{|l||c|c|c|c|c|c|c|c|c|c|c|c|c|}
\hline
 \dm  & \multicolumn{13}{c|}{\mstau\ (GeV)\ } \\
\cline{2-14}
(GeV)& 45 & 50 & 55 & 60 & 65 & 70 & 75 & 80 & 85 & 90 & 94 & 99 & 103\\
\hline\hline  \multicolumn{12}{|l|}{General Selection Efficiency ( \% )} \\
\hline
2     
&  0.1$\pm$0.0
&  0.0$\pm$0.0
&  0.0$\pm$0.0
&  0.0$\pm$0.0
&  0.0$\pm$0.0
&  0.0$\pm$0.0
&  0.0$\pm$0.0
&  0.0$\pm$0.0
&  0.0$\pm$0.0
&  0.0$\pm$0.0
&  0.0$\pm$0.0
&  0.0$\pm$0.0
&  0.0$\pm$0.0
\\
2.5   
&  0.1$\pm$0.0
&  0.0$\pm$0.0
&  0.0$\pm$0.0
&  0.0$\pm$0.0
&  0.0$\pm$0.0
&  0.0$\pm$0.0
&  0.0$\pm$0.0
&  0.0$\pm$0.0
&  0.0$\pm$0.0
&  0.0$\pm$0.0
&  0.0$\pm$0.0
&  0.0$\pm$0.0
&  0.0$\pm$0.0
\\
5     
& 11.0$\pm$0.4
& 10.5$\pm$0.4
&  9.4$\pm$0.4
&  8.5$\pm$0.4
&  7.6$\pm$0.4
&  7.6$\pm$0.4
&  6.7$\pm$0.4
&  6.4$\pm$0.3
&  5.5$\pm$0.3
&  4.5$\pm$0.3
&  4.3$\pm$0.3
&  4.0$\pm$0.3
&  3.9$\pm$0.3
\\
10    
& 33.4$\pm$0.7
& 33.8$\pm$0.7
& 34.8$\pm$0.7
& 32.5$\pm$0.7
& 32.0$\pm$0.7
& 31.8$\pm$0.7
& 30.2$\pm$0.6
& 30.9$\pm$0.7
& 28.8$\pm$0.6
& 26.6$\pm$0.6
& 26.6$\pm$0.6
& 25.0$\pm$0.6
& 25.1$\pm$0.6
\\
20    
& 54.4$\pm$0.7
& 56.0$\pm$0.7
& 55.3$\pm$0.7
& 56.4$\pm$0.7
& 56.1$\pm$0.7
& 54.8$\pm$0.7
& 57.7$\pm$0.7
& 55.5$\pm$0.7
& 55.4$\pm$0.7
& 54.8$\pm$0.7
& 54.0$\pm$0.7
& 55.0$\pm$0.7
& 54.8$\pm$0.7
\\
$m$/2 
& 57.0$\pm$0.7
& 60.0$\pm$0.7
& 62.7$\pm$0.7
& 65.5$\pm$0.7
& 67.3$\pm$0.7
& 69.4$\pm$0.7
& 68.9$\pm$0.7
& 70.3$\pm$0.6
& 72.1$\pm$0.6
& 74.0$\pm$0.6
& 74.0$\pm$0.6
& 74.6$\pm$0.6
& 75.5$\pm$0.6
\\
$m$-20
& 59.4$\pm$0.7
& 62.7$\pm$0.7
& 66.5$\pm$0.7
& 68.9$\pm$0.7
& 71.1$\pm$0.6
& 74.1$\pm$0.6
& 75.7$\pm$0.6
& 76.9$\pm$0.6
& 76.4$\pm$0.6
& 78.1$\pm$0.6
& 79.3$\pm$0.6
& 79.7$\pm$0.6
& 79.5$\pm$0.6
\\
$m$-10
& 63.7$\pm$0.7
& 67.1$\pm$0.7
& 68.4$\pm$0.7
& 72.5$\pm$0.6
& 73.4$\pm$0.6
& 74.8$\pm$0.6
& 76.4$\pm$0.6
& 76.9$\pm$0.6
& 77.8$\pm$0.6
& 77.5$\pm$0.6
& 78.1$\pm$0.6
& 79.4$\pm$0.6
& 80.2$\pm$0.6
\\
$m$   
& 66.5$\pm$0.7
& 68.8$\pm$0.7
& 70.1$\pm$0.6
& 71.9$\pm$0.6
& 75.1$\pm$0.6
& 75.3$\pm$0.6
& 76.9$\pm$0.6
& 76.6$\pm$0.6
& 78.9$\pm$0.6
& 78.6$\pm$0.6
& 79.9$\pm$0.6
& 80.2$\pm$0.6
& 80.2$\pm$0.6
\\

\hline\hline
\multicolumn{14}{|l|}{95\% CL upper limit on cross-section times
 $BR^2$(\staum$\rightarrow\tau\tilde{\chi}^0_1$) (fb)} \\
\hline
2     
&   5120
&   6260
& --
& --
& --
& --
& --
& --
& --
& --
& --
& --
& --
\\
2.5   
&   2130
&   2750
&   2700
&   3020
&   4410
&   5060
&   5710
&   8190
& --
& --
& --
& --
& --
\\
5     
&     78.5
&     69.1
&     85.8
&    100
&    106
&    107
&    128
&    148
&    199
&    245
&    363
&    403
&   2960
\\
10    
&     52.6
&     49.5
&     44.2
&     39.2
&     33.7
&     35.2
&     35.2
&     34.9
&     39.6
&     48.0
&     56.0
&     81.4
&    474
\\
20    
&     50.3
&     42.5
&     42.8
&     40.3
&     40.8
&     36.9
&     36.7
&     39.0
&     37.2
&     34.8
&     37.5
&     42.3
&    235
\\
$m$/2 
&     51.4
&     51.6
&     46.6
&     45.7
&     42.0
&     44.3
&     41.9
&     42.8
&     44.7
&     48.6
&     53.3
&     61.4
&    256
\\
$m$-20
&     56.0
&     53.8
&     56.3
&     48.8
&     48.0
&     47.6
&     45.7
&     47.9
&     54.2
&     57.7
&     61.2
&     78.7
&    328
\\
$m$-10
&     63.6
&     62.8
&     55.7
&     49.7
&     46.1
&     48.3
&     49.1
&     48.5
&     52.1
&     58.7
&     70.0
&     91.8
&    356
\\
$m$   
&     67.4
&     64.4
&     53.7
&     49.8
&     53.7
&     52.1
&     51.3
&     53.7
&     57.1
&     65.8
&     77.3
&     92.4
&    345
\\

\hline\hline
\multicolumn{14}{|l|}{expected upper limit on cross-section times
 $BR^2$(\staum$\rightarrow\tau\tilde{\chi}^0_1$) (fb)} \\
\hline
2     
&   5900
&   6190
& --
& --
& --
& --
& --
& --
& --
& --
& --
& --
& --
\\
2.5   
&   1690
&   2440
&   2460
&   3330
&   4590
&   6350
&   6460
&  10100
& --
& --
& --
& --
& --
\\
5     
&     86.6
&     87.9
&     94.9
&    105
&    113
&    120
&    141
&    159
&    193
&    265
&    351
&    586
&   3570
\\
10    
&     44.0
&     42.0
&     41.0
&     42.6
&     43.0
&     45.0
&     49.0
&     50.3
&     56.3
&     68.0
&     78.0
&    114
&    646
\\
20    
&     45.5
&     45.0
&     43.5
&     43.3
&     42.3
&     42.9
&     43.4
&     43.9
&     47.0
&     55.1
&     62.2
&     79.1
&    405
\\
$m$/2 
&     47.0
&     48.5
&     48.0
&     50.0
&     49.3
&     50.7
&     51.9
&     55.2
&     61.0
&     71.4
&     82.2
&    110
&    505
\\
$m$-20
&     49.4
&     49.9
&     51.3
&     52.1
&     53.2
&     55.3
&     58.5
&     60.6
&     66.0
&     79.0
&     87.9
&    118
&    538
\\
$m$-10
&     52.8
&     52.5
&     53.1
&     53.8
&     56.0
&     56.9
&     59.4
&     61.5
&     68.6
&     79.2
&     92.1
&    119
&    535
\\
$m$   
&     53.6
&     52.8
&     53.2
&     55.3
&     55.8
&     58.1
&     61.2
&     62.9
&     69.0
&     78.7
&     91.1
&    118
&    535
\\

\hline\hline
\multicolumn{14}{|l|}{CL for consistency with SM (\%)} \\
\hline
2     
&  60.4
&  53.2
& --
& --
& --
& --
& --
& --
& --
& --
& --
& --
& --
\\
2.5   
&  17.6
&  28.4
&  31.7
&  54.2
&  49.2
&  80.4
&  65.9
&  61.2
& --
& --
& --
& --
& --
\\
5     
&  54.9
&  70.3
&  56.8
&  48.7
&  48.4
&  55.5
&  53.8
&  52.2
&  39.3
&  53.0
&  37.2
& 100.0
& 100.0
\\
10    
&  23.3
&  26.2
&  34.0
&  51.8
&  72.7
&  71.3
&  80.0
&  83.2
&  81.4
&  80.4
&  79.0
&  84.6
&  77.3
\\
20    
&  31.5
&  50.4
&  43.8
&  51.7
&  46.5
&  61.1
&  62.7
&  60.3
&  69.0
&  88.9
&  92.6
&  97.4
&  98.8
\\
$m$/2 
&  33.7
&  35.7
&  46.3
&  55.4
&  61.7
&  59.5
&  69.2
&  73.5
&  78.8
&  84.1
&  88.6
&  95.5
&  99.5
\\
$m$-20
&  30.4
&  35.7
&  34.5
&  51.8
&  54.6
&  60.4
&  72.9
&  70.9
&  66.1
&  78.6
&  82.2
&  86.3
&  92.5
\\
$m$-10
&  23.7
&  24.2
&  38.2
&  54.3
&  66.1
&  63.7
&  64.3
&  72.1
&  74.3
&  76.1
&  74.5
&  73.8
&  86.6
\\
$m$   
&  19.5
&  22.5
&  42.5
&  55.2
&  49.6
&  55.8
&  64.3
&  62.9
&  65.2
&  64.4
&  62.9
&  69.3
&  88.8
\\

\hline
\end{tabular}
\renewcommand{\baselinestretch}{1.5}
%\caption[Stau results summary table]{
%General Selection efficiency, 95~\% CL upper limit cross-section times
%braching ratio at $\sqrt{s}=$~208~GeV, expected upper limit, and CL for consistency with 
%\sm\ processes in the search for \staupair\ production at 
%$\sqrt{s}$ = 183-208~GeV, for different values of \mstau\ and \dm\ .
%\label{tab-results-stau}
%}
}
\end{minipage}}
\caption{Stau search results.\label{tab-CL3}}
\end{center}
\end{table}
\end{center}

\clearpage
\begin{center}
\begin{table}[htb]
\renewcommand{\baselinestretch}{1.0}
\begin{center}
\vskip 3cm
\rotatebox{90}{
\begin{minipage}[htb]{\textwidth}
{\scriptsize
\begin{tabular}{|l||c|c|c|c|c|c|c|c|c|}
\hline
 \dm  & \multicolumn{9}{c|}{\mch\ (GeV)\ } \\
\cline{2-10}
 (GeV) & 50 & 60 & 70 & 80 & 85 & 90 & 94 & 99 & 102 \\
\hline\hline  \multicolumn{10}{|l|}{General Selection Efficiency ( \% )} \\
\hline
2     
&  5.3$\pm$0.4
&  3.9$\pm$0.3
&  2.6$\pm$0.3
&  1.5$\pm$0.2
&  1.0$\pm$0.2
&  0.7$\pm$0.1
&  0.4$\pm$0.1
&  0.2$\pm$0.1
&  0.1$\pm$0.0
\\
3     
& 16.2$\pm$0.6
& 15.9$\pm$0.6
& 14.3$\pm$0.6
& 12.6$\pm$0.5
& 11.7$\pm$0.5
& 10.5$\pm$0.5
&  9.3$\pm$0.5
&  8.1$\pm$0.4
&  7.6$\pm$0.4
\\
4     
& 26.5$\pm$0.8
& 27.4$\pm$0.7
& 24.6$\pm$0.7
& 24.3$\pm$0.7
& 23.9$\pm$0.7
& 22.5$\pm$0.7
& 20.9$\pm$0.6
& 20.0$\pm$0.6
& 20.0$\pm$0.6
\\
5     
& 35.6$\pm$0.8
& 35.2$\pm$0.8
& 33.8$\pm$0.8
& 33.8$\pm$0.7
& 33.8$\pm$0.7
& 33.1$\pm$0.7
& 31.8$\pm$0.7
& 30.9$\pm$0.7
& 31.0$\pm$0.7
\\
10    
& 58.1$\pm$0.9
& 59.6$\pm$0.8
& 59.1$\pm$0.8
& 58.5$\pm$0.8
& 57.6$\pm$0.8
& 56.3$\pm$0.8
& 55.5$\pm$0.8
& 55.1$\pm$0.8
& 55.1$\pm$0.8
\\
20    
&  0.0$\pm$0.0
& 76.1$\pm$0.7
& 76.8$\pm$0.7
& 76.2$\pm$0.7
& 75.7$\pm$0.7
& 75.4$\pm$0.7
& 75.3$\pm$0.7
& 75.5$\pm$0.7
& 75.2$\pm$0.7
\\
\mf\  
&  0.0$\pm$0.0
& 77.6$\pm$0.7
& 81.4$\pm$0.6
& 84.1$\pm$0.6
& 84.9$\pm$0.6
& 85.2$\pm$0.6
& 85.8$\pm$0.6
& 87.2$\pm$0.5
& 88.1$\pm$0.5
\\
$m$-35
& 68.5$\pm$0.8
& 78.6$\pm$0.7
& 83.3$\pm$0.6
& 85.9$\pm$0.6
& 87.2$\pm$0.5
& 88.3$\pm$0.5
& 89.1$\pm$0.5
& 90.3$\pm$0.5
& 90.6$\pm$0.5
\\

\hline\hline
\multicolumn{10}{|l|}{95\% CL upper limit on cross-section times
$BR^2(\chpm \rightarrow \ell^\pm \snu)$ (fb)} \\
\hline
2     
&    104
&    133
&    218
&    473
&    835
&   1960
&   3390
&   8230
& --
\\
3     
&     41.7
&     41.1
&     42.0
&     52.7
&     63.9
&     73.4
&     88.1
&    135
&    350
\\
4     
&     26.0
&     24.5
&     29.1
&     25.6
&     28.8
&     29.4
&     36.1
&     52.3
&    129
\\
5     
&     20.2
&     19.5
&     19.7
&     21.3
&     22.7
&     24.6
&     34.1
&     34.5
&     84.1
\\
10    
&     29.3
&     24.2
&     20.2
&     17.1
&     16.2
&     17.0
&     21.2
&     25.6
&     48.5
\\
20    
& --
&     30.2
&     27.4
&     23.5
&     25.3
&     19.9
&     19.5
&     21.2
&     40.1
\\
\mf\  
& --
&     33.1
&     34.9
&     35.9
&     45.2
&     47.9
&     70.9
&     78.8
&     64.7
\\
$m$-35
&     65.1
&     42.0
&     45.3
&     65.6
&     73.0
&     69.3
&     77.3
&    156
&    273
\\

\hline\hline
\multicolumn{10}{|l|}{expected upper limit on cross-section times
$BR^2(\chpm \rightarrow \ell^\pm \snu)$ (fb)} \\
\hline
2     
&    111
&    156
&    250
&    503
&    791
&   1720
&   3000
&   9900
& --
\\
3     
&     49.0
&     47.6
&     50.3
&     56.5
&     61.9
&     74.8
&    106
&    174
&    401
\\
4     
&     34.6
&     33.2
&     34.4
&     32.4
&     34.3
&     38.5
&     51.0
&     72.9
&    150
\\
5     
&     30.5
&     28.6
&     28.2
&     27.5
&     28.0
&     29.6
&     37.0
&     53.1
&    105
\\
10    
&     35.7
&     29.7
&     26.3
&     23.5
&     23.3
&     24.1
&     28.4
&     36.8
&     71.4
\\
20    
& --
&     53.8
&     44.7
&     38.5
&     34.9
&     33.1
&     37.4
&     45.5
&     72.3
\\
\mf\  
& --
&     59.1
&     61.6
&     64.0
&     68.6
&     77.2
&     85.6
&     89.0
&    117
\\
$m$-35
&     98.9
&     66.1
&     77.8
&     90.2
&     98.9
&     97.5
&    100
&    101
&    136
\\

\hline\hline
\multicolumn{10}{|l|}{CL for consistency with SM (\%)} \\
\hline
2     
&  46.7
&  66.5
&  60.5
&  49.1
&  35.6
&  25.8
&  32.0
& 100.0
& --
\\
3     
&  61.3
&  60.6
&  66.3
&  54.1
&  37.8
&  43.6
&  75.4
& 100.0
& 100.0
\\
4     
&  76.7
&  76.5
&  64.7
&  68.7
&  63.6
&  77.0
&  81.4
& 100.0
& 100.0
\\
5     
&  85.7
&  83.9
&  83.4
&  72.3
&  68.3
&  65.3
&  54.8
&  86.2
& 100.0
\\
10    
&  66.0
&  66.6
&  72.9
&  78.0
&  81.6
&  81.2
&  76.4
&  83.5
&  92.2
\\
20    
& --
&  95.4
&  91.9
&  90.9
&  79.0
&  91.7
&  97.6
&  99.6
&  99.6
\\
\mf\  
& --
&  96.4
&  94.8
&  94.8
&  88.1
&  90.7
&  67.2
&  57.7
&  96.1
\\
$m$-35
&  86.3
&  91.1
&  94.6
&  79.5
&  78.9
&  82.9
&  72.8
&   7.5
&   0.9
\\

\hline
\end{tabular}
\renewcommand{\baselinestretch}{1.5}
%\caption[\chargtwo\ results summary table]{
%General Selection efficiency, 95~\% CL upper limit cross-section times
%braching ratio at $\sqrt{s}=$~208~GeV, expected upper limit, and CL for consistency with 
%\sm\ processes in the search for \chargtwo\ production at 
%$\sqrt{s}$ = 208~GeV, for different values of \mch\ and \dm\ .
%\label{tab-results-ch2}
%}
}
\end{minipage}}
\caption{Chargino (2-body decay) search results. \label{tab-CL8}}
\end{center}
\end{table}
\end{center}

\clearpage

\begin{center}
\begin{table}[htb]
\renewcommand{\baselinestretch}{1.0}
\begin{center}
\vskip 3cm
\rotatebox{90}{
\begin{minipage}[htb]{\textwidth}
{\scriptsize
\begin{tabular}{|l||c|c|c|c|c|c|c|c|c|c|}
\hline
 \dm  & \multicolumn{10}{c|}{\mch\ (GeV)\ } \\
\cline{2-11}
 (GeV) & 50 & 60 & 70 & 80 & 85 & 90 & 94 & 99 & 102 & 103 \\
\hline\hline  \multicolumn{11}{|l|}{General Selection Efficiency ( \% )} \\
\hline
3     
&  1.7$\pm$0.2
&  1.5$\pm$0.2
&  1.4$\pm$0.2
&  0.8$\pm$0.1
&  0.4$\pm$0.1
&  0.4$\pm$0.1
&  0.3$\pm$0.1
&  0.1$\pm$0.0
&  0.0$\pm$0.0
&  0.0$\pm$0.0
\\
5     
& 11.2$\pm$0.5
& 10.4$\pm$0.5
& 10.2$\pm$0.5
&  9.1$\pm$0.5
&  7.3$\pm$0.4
&  6.8$\pm$0.4
&  6.0$\pm$0.4
&  5.6$\pm$0.4
&  4.8$\pm$0.3
&  4.4$\pm$0.3
\\
10    
& 32.6$\pm$0.7
& 34.2$\pm$0.8
& 33.5$\pm$0.7
& 32.4$\pm$0.7
& 31.0$\pm$0.7
& 32.5$\pm$0.7
& 31.2$\pm$0.7
& 29.3$\pm$0.7
& 29.5$\pm$0.7
& 29.7$\pm$0.7
\\
20    
& 52.7$\pm$0.8
& 53.7$\pm$0.8
& 54.8$\pm$0.8
& 54.8$\pm$0.8
& 55.4$\pm$0.8
& 55.0$\pm$0.8
& 55.2$\pm$0.8
& 53.4$\pm$0.8
& 54.6$\pm$0.8
& 55.4$\pm$0.8
\\
$m$/2 
& 57.5$\pm$0.8
& 63.9$\pm$0.8
& 69.5$\pm$0.7
& 75.7$\pm$0.7
& 74.9$\pm$0.7
& 76.6$\pm$0.7
& 77.2$\pm$0.7
& 78.0$\pm$0.7
& 78.4$\pm$0.7
& 77.9$\pm$0.7
\\
$m$-20
& 62.2$\pm$0.8
& 71.9$\pm$0.7
& 75.8$\pm$0.7
& 79.2$\pm$0.6
& 82.4$\pm$0.6
& 82.9$\pm$0.6
& 84.6$\pm$0.6
& 87.2$\pm$0.5
& 88.5$\pm$0.5
& 88.8$\pm$0.5
\\
$m$-10
& 68.9$\pm$0.7
& 73.4$\pm$0.7
& 79.2$\pm$0.6
& 82.2$\pm$0.6
& 84.3$\pm$0.6
& 87.8$\pm$0.5
& 89.0$\pm$0.5
& 90.2$\pm$0.5
& 90.4$\pm$0.5
& 90.3$\pm$0.5
\\
$m$   
& 72.2$\pm$0.7
& 77.2$\pm$0.7
& 80.4$\pm$0.6
& 85.2$\pm$0.6
& 87.3$\pm$0.5
& 89.2$\pm$0.5
& 89.7$\pm$0.5
& 89.6$\pm$0.5
& 90.5$\pm$0.5
& 91.1$\pm$0.5
\\

\hline\hline
\multicolumn{11}{|l|}{95\% CL upper limit on cross-section times
 $BR^2(\chpm \rightarrow \ell^\pm \nu \chz)$ (fb)} \\
\hline
3     
&    190
&    249
&    374
&    979
&   1170
&   1810
&   4180
&  10400
& --
& --
\\
5     
&     59.0
&     62.9
&     55.9
&     80.0
&     86.1
&    106
&    151
&    195
&    565
&   7640
\\
10    
&     29.8
&     23.5
&     21.8
&     24.1
&     23.7
&     27.7
&     36.1
&     46.3
&     87.4
&   1130
\\
20    
&     33.6
&     27.9
&     25.3
&     19.7
&     19.1
&     19.5
&     22.7
&     26.8
&     54.3
&    605
\\
$m$/2 
&     37.2
&     32.1
&     33.2
&     27.8
&     30.1
&     34.2
&     45.4
&     37.5
&     56.5
&    441
\\
$m$-20
&     43.3
&     39.8
&     36.9
&     40.3
&     43.1
&     56.2
&     77.0
&    108
&    151
&   1050
\\
$m$-10
&     50.5
&     45.4
&     45.4
&     47.6
&     56.7
&     68.2
&     98.4
&    155
&    248
&   1230
\\
$m$   
&     58.4
&     52.6
&     58.7
&     66.6
&     92.5
&     93.5
&    105
&    175
&    264
&   1180
\\

\hline\hline
\multicolumn{11}{|l|}{expected upper limit on cross-section times
 $BR^2(\chpm \rightarrow \ell^\pm \nu \chz)$ (fb)} \\
\hline
3     
&    270
&    334
&    470
&    883
&   1360
&   2490
&   3960
&  12900
& --
& --
\\
5     
&     67.5
&     67.9
&     73.5
&     83.3
&     99.0
&    114
&    159
&    259
&    664
&   7670
\\
10    
&     36.4
&     33.5
&     32.5
&     32.5
&     33.0
&     35.8
&     43.0
&     60.2
&    117
&   1160
\\
20    
&     41.1
&     35.5
&     31.5
&     29.8
&     30.0
&     30.6
&     36.1
&     47.4
&     84.3
&    662
\\
$m$/2 
&     46.1
&     46.6
&     43.9
&     43.3
&     43.8
&     48.9
&     59.9
&     81.2
&    131
&    695
\\
$m$-20
&     50.9
&     54.3
&     56.2
&     61.5
&     70.0
&     80.9
&    102
&    128
&    189
&    833
\\
$m$-10
&     61.2
&     63.6
&     68.4
&     75.7
&     85.1
&    101
&    120
&    130
&    193
&    835
\\
$m$   
&     70.1
&     71.2
&     77.5
&     98.1
&    114
&    114
&    120
&    133
&    187
&    839
\\

\hline\hline
\multicolumn{11}{|l|}{CL for consistency with SM (\%)} \\
\hline
3     
&  83.8
&  77.6
&  72.4
&  29.0
&  59.5
&  86.6
&  33.5
& 100.0
& --
& --
\\
5     
&  59.6
&  50.9
&  74.3
&  47.3
&  60.5
&  49.8
&  47.3
& 100.0
& 100.0
& 100.0
\\
10    
&  66.3
&  82.8
&  86.0
&  76.4
&  80.2
&  71.8
&  61.6
&  73.0
& 100.0
& 100.0
\\
20    
&  66.3
&  72.6
&  67.2
&  86.3
&  88.3
&  88.4
&  90.6
&  95.7
&  90.7
& 100.0
\\
$m$/2 
&  70.3
&  81.2
&  73.8
&  88.1
&  82.8
&  81.2
&  72.5
&  99.5
&  99.8
&  96.3
\\
$m$-20
&  62.7
&  77.1
&  86.8
&  87.0
&  92.4
&  81.8
&  76.7
&  64.1
&  68.6
&  20.8
\\
$m$-10
&  64.4
&  80.3
&  86.9
&  90.6
&  88.0
&  86.5
&  68.5
&  23.5
&  17.3
&   8.8
\\
$m$   
&  64.5
&  75.9
&  76.7
&  86.2
&  69.9
&  67.8
&  59.3
&  15.6
&  11.3
&  10.9
\\

\hline
\end{tabular}
\renewcommand{\baselinestretch}{1.5}
%\caption[\chargthree\ results summary table]{
%General Selection efficiency, 95~\% CL upper limit cross-section times
%braching ratio at $\sqrt{s}=$~208~GeV, expected upper limit, and CL for consistency with 
%\sm\ processes in the search for \chargthree\ production at 
%$\sqrt{s}$ = 208~GeV, for different values of \mch\ and \dm\ .
%\label{tab-results-ch3}
%}
}
\end{minipage}}
\caption{Chargino (3-body decay) search results.\label{tab-CL4}}
\end{center}
\end{table}
\end{center}

\clearpage

\begin{center}
\begin{table}[htb]
% I changed this from 1.0 to 1.5 (graham)
\renewcommand{\baselinestretch}{1.5}
\begin{center}
\vskip 5cm
\rotatebox{90}{
\begin{minipage}[htb]{\textwidth}
{\scriptsize
\begin{tabular}{|c|c|c|c|c|c|c|c|c|c|c|c|}
\hline
 \multicolumn{12}{|c|}{\mH\ (GeV)\ } \\
\hline
 50 & 55 & 60 & 65 & 70 & 75 & 80 & 85 & 90 & 94 & 99 & 103 \\
\hline\hline  \multicolumn{12}{|l|}{General Selection Efficiency ( \% )} \\
\hline
  68.8$\pm$0.7
& 70.1$\pm$0.6
& 71.9$\pm$0.6
& 75.1$\pm$0.6
& 75.3$\pm$0.6
& 76.9$\pm$0.6
& 76.6$\pm$0.6
& 78.9$\pm$0.6
& 78.6$\pm$0.6
& 79.9$\pm$0.6
& 80.2$\pm$0.6
& 80.2$\pm$0.6
\\

\hline\hline
\multicolumn{12}{|l|}{95\% CL upper limit on cross-section times
 $BR^2(\dH)$ (fb)} \\
\hline
     63.8
&     57.4
&     50.9
&     55.4
&     52.2
&     47.0
&     48.1
&     52.4
&     61.9
&     70.8
&     99.5
&    352
\\

\hline\hline
\multicolumn{12}{|l|}{expected upper limit on cross-section times
 $BR^2(\dH)$ (fb)} \\
\hline
     53.0
&     53.5
&     54.4
&     57.3
&     57.4
&     59.7
&     62.9
&     68.1
&     78.1
&     88.9
&    117
&    539
\\

\hline\hline
\multicolumn{12}{|l|}{CL for consistency with SM (\%)} \\
\hline
  23.8
&  34.9
&  50.5
&  48.1
&  54.4
&  71.3
&  74.8
&  75.1
&  72.2
&  68.2
&  62.2
&  87.3
\\

\hline
\end{tabular}
\renewcommand{\baselinestretch}{1.5}
%\caption[Charged Higgs results summary table]{
%General Selection efficiency, 95~\% CL upper limit cross-section times
%$BR^2(\dH)$ at $\sqrt{s}=$~208~GeV, expected upper limit, and CL for consistency with 
%\sm\ processes in the search for $\mathrm{H}^+\mathrm{H}^-$ 
%production at $\sqrt{s}$~=~208~GeV, for different values of \mH .
%\label{tab-results-chiggs}
%}
}
\end{minipage}}
\caption{Charged Higgs search results.\label{tab-CL5}}
\end{center}
\end{table}
\end{center}

\clearpage

%============================================================================
%FIGURES
%============================================================================

\newpage

\clearpage

\clearpage

\clearpage

\clearpage

%\begin{figure}[htbp]
% \epsfxsize=\textwidth 
% \epsffile{pr290_05.eps_col}
% \caption{\sl {\bf GWW Still needs to be re-made for the full data-set.}
%Distributions of the background likelihood, \LB , for
%\smc\ (shaded histogram) and data 
%(points with error bars) for events passing the general selection, using all
%the likelihood variables (a).  (b) and (c) show the same information 
%after making the initial lepton identification requirements given in 
%Section~\ref{sec:lept} for the selectron and smuon searches respectively.
%In (b), only the variables used in the selectron analysis
%are used.
%\label{fig:lb}
%} 
%\end{figure}
%\clearpage

\clearpage

\clearpage

\clearpage

\clearpage

\clearpage

\clearpage

\clearpage

\clearpage

\clearpage

\clearpage

\clearpage

%%%%%%%%%%%%%%%%%%%%%%%%%%%%%%%%%%%%%%%%%%%%%%%%%%%%%%%%%%%%%%%%%%%%%%%%
\end{document}